\documentstyle{amltd2004}
\begin{document}
\annalsline{158}{2003}
 
\startingpage{253}
\def\bye{\end{document}}
 \font\tenrm=cmr10
\def\ritem#1{\item[{\rm #1}]}
\input amssym.def
\input amssym.tex
\def\eqref#1{(\ref{#1})}
\def\tfrac#1#2{{\textstyle\frac{#1}{#2}}}
\def\nhs{\hskip-6pt}
\def\binom#1#2{{#1\choose #2}}
\catcode`\@=11
\font\twelvemsb=msbm10 scaled 1100
\font\tenmsb=msbm10
\font\ninemsb=msbm10 scaled 800
\newfam\msbfam
\textfont\msbfam=\twelvemsb  \scriptfont\msbfam=\ninemsb
  \scriptscriptfont\msbfam=\ninemsb
\def\msb@{\hexnumber@\msbfam}
\def\Bbb{\relax\ifmmode\let\next\Bbb@\else
 \def\next{\errmessage{Use \string\Bbb\space only in math
mode}}\fi\next}
\def\Bbb@#1{{\Bbb@@{#1}}}
\def\Bbb@@#1{\fam\msbfam#1}
\catcode`\@=12

 \catcode`\@=11
\font\twelveeuf=eufm10 scaled 1100
\font\teneuf=eufm10
\font\nineeuf=eufm7 scaled 1100
\newfam\euffam
\textfont\euffam=\twelveeuf  \scriptfont\euffam=\teneuf
  \scriptscriptfont\euffam=\nineeuf
\def\euf@{\hexnumber@\euffam}
\def\frak{\relax\ifmmode\let\next\frak@\else
 \def\next{\errmessage{Use \string\frak\space only in math
mode}}\fi\next}
\def\frak@#1{{\frak@@{#1}}}
\def\frak@@#1{\fam\euffam#1}
\catcode`\@=12

\newcommand{\bbN}{{\Bbb N}}
\newcommand{\bbR}{{\Bbb R}}
\newcommand{\bbZ}{{\Bbb Z}}
\newcommand{\bbC}{{\Bbb C}}
\newcommand{\bbQ}{{\Bbb Q}}

\newcommand{\calG}{{\cal G}}
\newcommand{\calH}{{\cal H}}
\newcommand{\calI}{{\cal I}}


\newcommand{\dott}{\,\cdot\,}
 \newcommand{\no}{\nonumber}
\newcommand{\lb}{\label}
\newcommand{\f}{\frac}
\newcommand{\ul}{\underline}
\newcommand{\ol}{\overline}
\newcommand{\ti}{\tilde  }
\newcommand{\ess}{{\rm{ess}}}
\newcommand{\ac}{{\rm{ac}}}
\newcommand{\sing}{{\rm{sing}}}
\newcommand{\supp}{{\rm{supp}}}
\newcommand{\bi}{\bibitem}
\newcommand{\hatt}{\widehat}
\newcommand{\beq}{\begin{equation}}
\newcommand{\eeq}{\end{equation}}
\newcommand{\ba}{\begin{eqnarray}}
\newcommand{\ea}{\end{eqnarray}}
\newcommand{\veps}{\varepsilon}
\newcommand{\Tra}{\mathop{\rm Tr}\nolimits}
\newcommand{\Arg}{{\rm{Arg}}}
\newcommand{\Ran}{\mathop{\rm Ran}\nolimits}


\newcommand{\Real}{\mathop{\rm Re}\nolimits}
\newcommand{\Ima}{\mathop{\rm Im}\nolimits}
\newcommand{\diam}{\mathop{\rm diam}\nolimits}
\newcommand{\slim}{\mathop{\rm s-lim}\nolimits}
\newcommand{\wlim}{\mathop{\rm w-lim}\nolimits}
\newcommand{\simlim}{\sim}
\newcommand{\eqlim}{\mathop{=}}
\newcommand{\arrow}{\mathop{\rightarrow}}
 \newcommand{\abs}[1]{\left\vert#1\right\vert}


\title{Sum rules for Jacobi matrices \\ and their applications to spectral  theory}
\shorttitle{Sum rules for Jacobi matrices} 

 \acknowledgements{The first named author was supported in part by NSF grant
DMS-9729992.  The second named author was supported in part by 
NSF grant DMS-9707661.}
 \twoauthors{Rowan Killip}{Barry Simon}
\institutions{
California Institute of Technology,    
Pasadena, CA\\
{\eightpoint {\it E-mail addresses\/}:  killip@its.caltech.edu}\\
\hglue.97in {\eightpoint   bsimon@caltech.edu}\\ \vglue6pt
}
 
\vglue12pt
\centerline{\bf Abstract} 
\vglue12pt
We discuss the proof of and systematic application of Case's sum rules for Jacobi matrices. 
Of special interest is a linear combination of two of his sum rules which has strictly 
positive terms. Among our results are a complete classification of the spectral measures 
of all Jacobi matrices $J$ for which $J-J_0$ is Hilbert-Schmidt, and a proof of Nevai's 
conjecture that the Szeg\H{o} condition holds if $J-J_0$ is trace class.

\section{Introduction}\label{S1}

In this paper, we will look at the spectral theory of Jacobi matrices, that is, infinite 
tridiagonal matrices,
\begin{equation} \lb{1.1}
J= \left( \begin{array}{ccccc}
b_1 & a_1 & 0 & 0 & \cdots \\
a_1 & b_2 & a_2 & 0 & \cdots \\
0 & a_2 & b_3 & a_3 & \cdots \\
\vdots & \vdots & \vdots & \vdots & \ddots 
\end{array}\right)
\end{equation}
with $a_j>0$ and $b_j \in\bbR$.  We suppose that the entries of $J$ are bounded, that 
is, $\sup_n \abs{a_n}+\sup_n \abs{b_n}<\infty$ so that $J$ defines a bounded self-adjoint 
operator on $\ell^2 (\bbZ_+)=\ell^2 (\{1,2,\dots\})$. Let $\delta_j$ be the obvious vector
in $\ell^2 (\bbZ_+)$, that is, with components $\delta_{jn}$ which are $1$ if $n=j$ and $0$ if 
$n\neq j$.

The spectral measure we associate to $J$ is the one given by the spectral theorem for the
vector $\delta_1$.  That is, the measure $\mu$ defined by
\begin{equation} \label{1.2}
m_\mu (E)\equiv \langle\delta_1, (J-E)^{-1}\delta_1\rangle = \int\f{d\mu(x)}{x-E}\,.
\end{equation}

\pagebreak There is a one-to-one correspondence between bounded Jacobi matrices and 
unit measures whose support is both compact and contains an infinite number of points.
As we have described, one goes from $J$ to $\mu$ by the spectral theorem.
One way to find $J$, given $\mu$, is via orthogonal polynomials.
Applying the Gram-Schmidt process to $\{x^n\}_{n=0}^\infty$, one gets orthonormal 
polynomials $P_n(x) = \kappa_n x^n +\cdots$ with $\kappa_n >0$ and
\begin{equation} \label{1.3}
\int P_n (x) P_m(x) \, d\mu(x) = \delta_{nm}.
\end{equation}
These polynomials obey a three-term recurrence:
\begin{equation} \label{1.4}
xP_n(x) = a_{n+1} P_{n+1}(x) + b_{n+1} P_n(x) + a_n P_{n-1}(x),
\end{equation}
where $a_n, b_n$ are the Jacobi matrix coefficients of the Jacobi matrix with spectral measure
$\mu$ (and $P_{-1}\equiv 0$).

The more usual 
convention in the orthogonal polynomial literature is to start numbering of $\{a_n\}$ and 
$\{b_n\}$ with $n=0$ and then to have \eqref{1.4} with $(a_n, b_n, a_{n-1})$ instead of 
$(a_{n+1}, b_{n+1}, a_n)$. We made our choice to start numbering of $J$ at $n=1$ so that 
we could have $z^n$ for the free Jost function (well known in the physics literature with 
$z=e^{ik}$) and yet arrange for the Jost function to be regular at $z=0$. (Case's Jost 
function in \cite{C1,C2} has a pole since where we use $u_0$ below, he uses $u_{-1}$ because 
his numbering starts at $n=0$.)  There is, in any event, a notational conundrum which we 
solved in a way that we hope will not offend too many.

An alternate way of recovering $J$ from $\mu$ is the continued fraction expansion
for the function $m_\mu (z)$ near infinity,
\begin{equation} \label{1.5}
m_\mu(E) = \frac{1}{-E +b_1 - {\displaystyle\frac{a_1^2}{-E+b_2 +\cdots}}}.
\end{equation}

Both methods for finding $J$ essentially go back to Stieltjes' monumental paper \cite{St1}.
Three-term
recurrence relations appeared earlier in the work of Chebyshev and Markov but, of course,
Stieltjes was the first to consider general measures in this context.
While \cite{St1} does not have the continued fraction expansion given in \eqref{1.5}, Stieltjes 
did discuss \eqref{1.5} elsewhere. Wall \cite{Wall} calls \eqref{1.5} a $J$-fraction and
the fractions used in \cite{St1}, he calls $S$-fractions. This has been discussed in many
places, for example, \cite{GS}, \cite{S270}.

That every $J$ corresponds to a spectral measure is known in the orthogonal polynomial 
literature as Favard's theorem (after Favard \cite{Fav}).  As noted, it is a consequence 
for bounded $J$ of Hilbert's spectral theorem for bounded operators. This appears already 
in the Hellinger-Toeplitz encyclopedic article \cite{HT}. Even for the general unbounded 
case, Stone's book \cite{Stone} noted this consequence before Favard's work.

Given the one-to-one correspondence between $\mu$'s and $J$'s, it is natural to ask how  
properties of one are reflected in the other. One is especially interested in $J$'s 
``close" to the free matrix, $J_0$ with $a_n =1$ and $b_n=0$, that is,
\begin{equation} \label{1.6}
J_0 = \left( \begin{array}{ccccc} 
0 & 1 & 0 & 0 & \dots \\
1 & 0 & 1 & 0 & \dots \\
0 & 1 & 0 & 1 & \dots \\
0 & 0 & 1 & 0 & \dots 
\end{array}\right).
\end{equation}

In the orthogonal polynomial literature, the free Jacobi matrix is taken as $\f12$ of our  
$J_0$ since then the associated orthogonal polynomials are precisely Chebyshev polynomials 
(of the second kind). As a result, the spectral measure of our $J_0$ is supported by $[-2,2]$
and the natural parametrization is $E=2\cos\theta$.

Here is one of our main results:

\specialnumber{1}\proclaim{Theorem} \label{T1} Let $J$ be a Jacobi matrix and $\mu$ the corresponding spectral
measure. The operator  $J - J_0$ is Hilbert\/{\rm -}\/Schmidt{\rm ,} that is{\rm ,} 
\begin{equation} \label{1.7}
2 \sum_n (a_n -1)^2 + \sum b_n^2 < \infty
\end{equation}
if and only if $\mu$ has the following four properties\/{\rm :}\/
\vglue4pt
\noindent\hglue16pt\hangindent=34pt\hangafter=1
 {\rm(0)\enspace  (Blumenthal-Weyl Criterion)} The support of $\mu$ is $[-2,2]\cup 
\{E_j^+\}_{j=1}^{N_+}\cup\{E_j^-\}_{j=1}^{N_-}$ where $N_\pm$ are each zero{\rm ,} finite{\rm ,} or 
infinite{\rm ,} and $E_1^+ > E_2^+ >\cdots >2$ and $E_1^- < E_2^- <\cdots <-2$ and if $N_\pm$ is 
infinite{\rm ,} then\break $\lim_{j\to\infty} E_j^\pm =\pm2$.

\noindent\hglue16pt\hangindent=38pt\hangafter=1
 {\rm (1)\enspace (Quasi-Szeg\H{o} Condition)} Let $\mu_{\ac}(E)=f(E)\, dE$ where $\mu_{\ac}$ 
is the Lebesgue absolutely continuous component of $\mu$. Then

\begin{equation} \label{1.8}
\int_{-2}^2 \log[f(E)] \sqrt{4-E^2}\, dE >-\infty.
\end{equation}

\noindent\hglue16pt
 {\rm (2)\enspace   (Lieb-Thirring Bound)}
\begin{equation} \label{1.9}
\sum_{j=1}^{N_+} |E_j^+ -2|^{3/2} + \sum_{j=1}^{N_-} |E_j^- +2|^{3/2} <\infty.
\end{equation}

\noindent\hglue16pt\hangindent=38pt\hangafter=1
 {\rm (3)\enspace   (Normalization)} $\int d\mu(E)=1$.
\endproclaim

 {\it Remarks.}
1. Condition (0) is just a quantitative way of writing that the essential spectrum of $J$ is
the same as that of $J_0$, viz.~$[-2,2]$, consistent with the compactness of $J-J_0$. This is, 
of course, Weyl's invariance theorem \cite{Weyl}, \cite{RS4}. Earlier, Blumenthal \cite{Blu} proved 
something close to this in spirit for the case of orthogonal polynomials.

\vglue4pt 
2. Equation \eqref{1.9} is a Jacobi analog of a celebrated bound of Lieb and Thirring
\cite{LT1}, \cite{LT2} for Schr\"odinger operators. That it holds if $J-J_0$ is Hilbert-Schmidt 
has also been recently proven by Hundertmark-Simon \cite{HS}, although we do not use the 
$\f32$-bound of \cite{HS} below. We essentially reprove \eqref{1.9} if \eqref{1.7} holds.

\vglue4pt 
3. We call \eqref{1.8} the quasi-Szeg\H{o} condition to distinguish it from the Szeg\H{o} 
condition,
\begin{equation} \label{1.10}
\int_{-2}^2 \log[f(E)] (4-E^2)^{-1/2}\, dE> -\infty.
\end{equation}
This is stronger than \eqref{1.8} although the difference only matters if $f$ 
vanishes extremely rapidly at $\pm 2$.  For example, like
$\exp(-(2-\abs{E})^{-\alpha})$ with $\f12 \leq \alpha < \f32$. Such
behavior actually occurs for certain Pollaczek polynomials \cite{Chi}.

\vglue4pt 
4. It will often be useful to have a single sequence $e_1(J), e_2(J), \dots$  obtained from 
the numbers $\abs{E_j^\pm \mp 2}$ by reordering so $e_1(J)\geq e_2(J)\geq \cdots \to 0$.

\vglue4pt 
By property (1), for any $J$ with $J-J_0$ Hilbert-Schmidt, the essential support of the
a.c.~spectrum is $[-2,2]$.  That is, $\mu_{\ac}$ gives positive weight to any subset of $[-2,2]$
with positive measure.  This follows from \eqref{1.8} because $f$ cannot vanish on any such set.
This observation is the Jacobi matrix analogue of recent results which show that (continuous
and discrete) Schr\"odinger operators with potentials $V\in L^p$, $p\leq 2$, or
$|V(x)|\lesssim (1+x^2)^{-\alpha/2}$, $\alpha>1/2$, have a.c.~spectrum.  (It is known that the
a.c. spectrum can disappear once $p>2$ or $\alpha\leq 1/2$.)  Research in this direction began
with Kiselev
\cite{Kis} and culminated in the work of Christ-Kiselev \cite{CK}, Remling \cite{Rem},
Deift-Killip \cite{DK}, and Killip \cite{Kil}.
Especially relevant here is the work of Deift-Killip who used sum rules for finite
range perturbations to obtain an {\it a~priori} estimate.  Our work differs from theirs (and the
follow-up papers of Molchanov-Novitskii-Vainberg \cite{MNV} and
Laptev-Naboko-Safronov \cite{LNS}) in two critical ways:  we deal
with the half-line sum rules so the eigenvalues are the ones for the problem of
interest and we show that the sum rules still hold in the limit.  These developments are
particularly important for the converse direction (i.e., if $\mu$ obeys
(0--3) then $J-J_0$
is Hilbert-Schmidt).

In Theorem 1, the only restriction on the singular part of $\mu$ on $[-2,2]$
is in terms of its total mass.  Given any singular measure $\mu_{\sing}$ supported on $[-2,2]$
with total mass less than one, there is a Jacobi matrix $J$ obeying \eqref{1.7} for which
this is the singular part of the spectral measure.  In particular, there exist Jacobi matrices
$J$ with $J-J_0$ Hilbert-Schmidt for which $[-2,2]$ simultaneously supports dense point spectrum,
dense singular continuous spectrum and absolutely continuous spectrum.
Similarly, the only restriction on the norming
constants, that is, the values of $\mu(\{E_j^\pm\})$, is that their sum must be less than one.

In the related setting of Schr\"odinger operators on $\bbR$, Denisov \cite{Den}
has constructed an $L^2$ potential which gives rise to embedded singular continuous
spectrum.  In this vein see also Kiselev \cite{KisSC}.  
We realized that the key to Denisov's result was a sum rule, not the particular method he used
to construct his potentials. We decided to focus first on the discrete case
where one avoids certain technicalities, but are turning to the continuum case.

While \eqref{1.8} is the natural condition when $J-J_0$ is Hilbert-Schmidt, 
we have a one-directional result for the Szeg\H{o} condition. We prove the following 
conjecture of Nevai \cite{Nev1}: 

\specialnumber{2}\proclaim{Theorem} \lb{T2} If $J-J_0$ is in trace class{\rm ,} that is, 
\begin{equation} \label{1.11}
\sum_n \, \abs{a_n -1} + \sum_n \, \abs{b_n} <\infty,
\end{equation}
then the Szeg{\rm \H{\it o}} condition {\rm \eqref{1.10}} holds. 
\endproclaim

{\it Remark.} Nevai \cite{Nev2} and Geronimo-Van Assche \cite{GVA} prove the Szeg\H{o} 
condition holds under the slightly stronger hypothesis $$\sum_n (\log n) \abs{a_n-1} + \sum_n 
(\log n)\abs{b_n}<\infty.$$

\vglue6pt

We will also prove
\specialnumber{3}\proclaim{Theorem}  \lb{T3} If $J-J_0$ is compact and 
\vglue2pt
{\rm (i)}
\begin{equation} \label{1.12}
\sum_j \, \abs{E_j^+ -2}^{1/2} + \sum_j \, \abs{E_j^-+2}^{1/2}<\infty
\end{equation}

\vglue2pt {\rm (ii)} $\limsup_{N\to\infty} a_1 \dots a_N >0$
\vglue6pt\noindent 
then {\rm \eqref{1.10}} holds.
\endproclaim

We will prove Theorem~2 from Theorem~3 by using a $\f12$ power Lieb-Thirring 
inequality, as proven by Hundertmark-Simon \cite{HS}. 

For the special case where $\mu$ has no mass outside $[-2,2]$ (i.e., $N_+=N_-\break =0$), 
there are over seventy years of results related to Theorem~1 with important contributions 
by Szeg\H{o} \cite{Sz1}, \cite{Szb}, Shohat \cite{Sho}, Geronomius \cite{Ge}, Krein \cite{Kr}, 
and Kolmogorov \cite{Kol}. Their results are summarized by Nevai \cite{Nev1} as:

\proclaimtitle{Previously Known}
\specialnumber{4}
\proclaim{Theorem} \lb{T4} Suppose $\mu$ is a probability measure supported on 
$[-2,2]$. The Szeg{\rm \H{\it o}} condition {\rm \eqref{1.10}} holds if and only if
\vglue6pt
{\rm (i)} $J-J_0$ is Hilbert\/{\rm -}\/Schmidt.
\vglue6pt
{\rm (ii)} $\sum (a_n-1)$ and $\sum b_n$ are {\rm(}\/conditionally\/{\rm)} convergent.
\endproclaim

Of course, the major difference between this result and Theorem~1 is that we can handle 
bound states (i.e., eigenvalues outside $[-2,2]$) and the methods of Szeg\H{o}, Shohat, and 
Geronimus seem unable to. Indeed, as we will see below, the condition of no eigenvalues 
is very restrictive. A second issue is that we focus on the previously unstudied 
(or lightly studied; e.g., it is mentioned in \cite{Netal}) condition which we have called the 
quasi-Szeg\H{o} condition \eqref{1.8}, which is strictly weaker than the Szeg\H{o} condition 
\eqref{1.10}. Third, related to the first point, we do not have any requirement for 
conditional convergence of $\sum_{n=1}^N (a_n-1)$ or $\sum_{n=1}^N b_n$.
\vglue2pt
The Szeg\H{o} condition, though, has other uses (see Szeg\H{o} \cite{Szb}, Akhiezer \cite{Akh}), 
so it is a natural object independently of the issue of studying the spectral condition.

We emphasize that the assumption that $\mu$ has no pure points outside $[-2,2]$ is 
extremely strong. Indeed, while the Szeg\H{o} condition plus this assumption implies (i) and 
(ii) above, to deduce the Szeg\H{o} condition requires only a very small part of (ii). We 
\specialnumber{4'}
\proclaim{Theorem} \lb{T4'} If $\sigma(J)\subset [-2,2]$ and 
\vglue4pt {\rm (i)}
$
\limsup_N \sum_{n=1}^N \log (a_n)>-\infty,
$
\vglue4pt\noindent 
then the Szeg{\rm \H{\it o}} condition holds.  If $\sigma(J)\subset [-2,2]$ and 
either {\rm{(i)}} or the Szeg{\rm \H{\it o}} condition holds{\rm ,} then 
\begin{itemize}
\ritem{(ii)} $\sum_{n=1}^\infty (a_n-1)^2 + \sum_{n=1}^\infty b_n^2 <\infty,$
\ritem{(iii)} $\lim_{N\to\infty} \sum_{n=1}^N \log (a_n)$ exists {\rm{(}}\/and is 
finite\/{\rm{)},}
\ritem{(iv)} $\lim_{N\to\infty} \sum_{n=1}^N b_n$ exists {\rm{(}}\/and is finite\/{\rm{)}}.
\end{itemize}
In particular{\rm ,} if $\sigma(J)\subset [-2,2]${\rm ,} then {\rm{(i)}} implies {\rm{(ii)--(iv)}}.
\endproclaim

In Nevai \cite{Nevmem}, it is stated and proven (see pg.~124) that $\sum_{n=1}^\infty 
\abs{a_n-1}<\infty$ implies the Szeg\H{o} condition, but it turns out that his method of 
proof only requires our condition (i). Nevai informs us that he believes his result was 
probably known to Geronimus. 

The key to our proofs is a family of sum rules stated by Case in \cite{C2}.   
Case was motivated by Flaschka's calculation of the first integrals for the Toda 
lattice for finite \cite{Fla1} and doubly infinite Jacobi matrices \cite{Fla2}. 
Case's method of proof is partly patterned after that of Flaschka in \cite{Fla2}.

To state these rules, it is natural to change variables from $E$ to $z$ via
\begin{equation} \label{1.13}
E=z+\f{1}{z}\, .
\end{equation}
We choose the solution of \eqref{1.13} with $\abs{z}<1$, namely
\begin{equation} \label{1.14}
z={\textstyle \frac{1}{2}} \bigl[ E-\sqrt{E^2 -4}\,\bigr],
\end{equation}
where we \pagebreak take the branch of $\sqrt{\vphantom{\mu}\phantom{\mu}}$ with $\sqrt{\mu} >0$ for 
$\mu>0$. In this way, $E\mapsto z$ is the conformal map of $\{\infty\}\cup \bbC\backslash 
[-2,2]$ to $D\equiv\{z\mid \,\abs{z} <1\}$, which takes $\infty$ to $0$ and (in the limit) 
$\pm 2$ to $\pm 1$.  The points $E\in [-2,2]$ are mapped to $z=e^{\pm i\theta}$ where 
$E=2\cos\theta$.

The conformal map suggests replacing $m_\mu$ by 
\begin{equation} \label{1.15a}
M_\mu (z) = -m_\mu\big(E(z)\big) = -m_\mu \big(z+z^{-1}\big)
	= \int\f{z\,d\mu(x)}{1-xz+z^2}\,. \hskip.5in
\end{equation}
We have introduced a minus sign so that $\Ima M_\mu(z) >0$ when $\Ima z>0$.  Note that
$\Ima E>0 \Rightarrow m_\mu(E)>0$ but $E\mapsto z$ maps the upper half-plane to the
lower half-disk.

If $\mu$ obeys the Blumenthal-Weyl criterion, $M_\mu$ is meromorphic on $D$ with poles at 
the points $(\gamma_j^\pm)^{-1}$ where 
\begin{equation} \label{1.16}
\abs{\gamma_j}>1 \quad\hbox{and}\quad E_j^\pm =\gamma_j^\pm + (\gamma_j^\pm)^{-1}.
\end{equation}
As with $E_j^\pm$, we renumber $\gamma_j^\pm$ to a single sequence $\abs{\beta_1} \geq 
\abs{\beta_2} \geq\cdots\geq 1$. 

By general principles, $M_\mu$ has boundary values almost everywhere on the circle,
\begin{equation} \label{1.18}
M_\mu (e^{i\theta})=\lim_{r\uparrow 1}\, M_\mu (re^{i\theta})
\end{equation}
with $M_\mu (e^{-i\theta}) = \ol{M_\mu (e^{i\theta})}$ and $\Ima M_\mu (e^{i\theta})\geq 0$
for $\theta\in (0,\pi)$.

{}From the integral representation \eqref{1.2},
\begin{equation} \label{1.21}
\Ima m_\mu (E+i0) = \pi\f{d\mu_{\ac}}{dE}
\end{equation}
so using $dE = -2\sin\theta\, d\theta = - (4-E^2)^{1/2}\, d\theta$,
the quasi-Szeg\H{o} condition \eqref{1.8} becomes
$$
4\int_0^\pi \log[\Ima M_\mu (e^{i\theta})]\sin^2\theta\, d\theta >-\infty
$$
and the Szeg\H{o} condition \eqref{1.10} is
$$
\int_0^\pi \log[\Ima M_\mu (e^{i\theta})]\, d\theta > -\infty.
$$
Moreover, we have by \eqref{1.21} that
\begin{equation} \label{1.22}
\tfrac{2}{\pi}\int_0^\pi \Ima[M_\mu (e^{i\theta})]\sin\theta\, d\theta = 
	\mu_{\ac} (-2,2)\leq 1.
\end{equation}

With these notational preliminaries out of the way, we can state Case's sum rules.
For future reference, we give them names: 
\vglue6pt\noindent 
\ul{$\mathbold{C_0}${\bf :}}
\begin{equation} \label{1.23}
\f{1}{4\pi} \int_{-\pi}^\pi \log \biggl[ \f{\sin\theta}{\Ima M(e^{i\theta})}\biggr] 
d\theta = \sum_j \log \abs{\beta_j} - \sum_j \log \abs{a_j}
\end{equation}
and for $n=1,2,\dots$,\\
\vglue6pt\noindent 
\ul{$\mathbold{C_n}${\bf :}}
\begin{eqnarray} \lb{1.24}
&&-\f{1}{2\pi} \int_{-\pi}^\pi  \log\biggl[ \f{\sin\theta}{\Ima M(e^{i\theta})}\biggr] 
 \cos (n\theta)\, d\theta + \f{1}{n} \sum_j (\beta_j^n - \beta_j^{-n}) \\
&=& \f{2}{n} \, \Tra \Big\{ T_n \big(\tfrac{1}{2}J\big) - T_n \big( \tfrac{1}{2}J_0\big) \Big\} 
\nonumber
\end{eqnarray}
where $T_n$ is the $n^{\rm th}$ Chebyshev polynomial (of the first kind).

We note that Case did not have the compact form of the right side of \eqref{1.24}, but he 
used implicitly defined polynomials which he did not recognize as Chebyshev polynomials
(though he did give explicit formulae for small $n$). Moreover, his arguments are
formal. In an earlier paper, he indicates that the conditions he needs are
\begin{equation} \label{1.27a}
\abs{a_n -1} + \abs{b_n} \leq C (1+n^2)^{-1}
\end{equation}
but he also claims this implies $N_+ < \infty$, $N_- <\infty$, and, as Chihara \cite{Chi1}
noted, this is false. We believe that Case's  implicit methods could be made to work if
$\sum n [\abs{a_n-1} + \abs{b_n}] <\infty$ rather than \eqref{1.27a}. In any event, we
will provide explicit proofs of the sum rules---indeed, from two points of view.

One of our primary observations is the power of a certain combination of the 
Case sum rules, $C_0 + \f12 C_2$.  It says
\vglue4pt\noindent 
\ul{$\mathbold{P_2}${\bf :}}
\begin{eqnarray}\label{1.25}
&&\f{1}{2\pi} \int_{-\pi}^\pi
	\log\biggl(\f{\sin\theta}{\Ima M(\theta)}\biggr) \sin^2 \theta\, d\theta
	+ \sum_j [F(E_j^+)+F(E_j^-)] \\
& &\qquad\qquad= \f14 \sum_j b_j^2 + \f12 \sum_j G(a_j) 
\no
\end{eqnarray}
where $G(a) = a^2 - 1 -\log\abs{a}^2$ and
$
F(E) = \tfrac14\, [\beta^2 - \beta^{-2} -\log \abs{\beta}^4\, ],
$
with $\beta$ given by $E=\beta+\beta^{-1}$, $|\beta|>1$ (cf. \eqref{1.16}).

As with the other sum rules, the terms on the left-hand side are purely spectral---they 
can be easily found from $\mu$; those on the right depend in a simple way on the coefficients
of $J$.

The significance of \eqref{1.25} lies in the fact that each of its terms
is nonnegative.  It is not difficult to see (see the end of \S\ref{S3}) that $F(E)
\geq 0$ for $E\in\bbR\setminus[-2,2]$ and that $G(a)\geq 0$ for $a\in(0,\infty)$.  
To see that the integral is also nonnegative, we employ Jensen's inequality.  Notice that 
$y\mapsto -\log(y)$ is convex and $\f{2}{\pi}\int_0^\pi \sin^2 \theta\, d\theta =1$ so
\pagebreak

\begin{eqnarray}
&&\no\\
\noalign{\vskip-30pt}
 \lb{1.27} \qquad
\f{1}{2\pi}\int_{-\pi}^\pi \log \biggl[\f{\sin(\theta)}{\Ima M(e^{i\theta})}\biggr] 
\sin^2 \theta \, d\theta &\nhs =\nhs& \f12 \, \f{2}{\pi} \int_0^\pi -\log \biggl[ 
\f{\Ima M}{\sin\theta}\biggr] \sin^2(\theta)\, d\theta  \\[4pt]
&\nhs\geq\nhs& -\f12 \log \biggl[ \f{2}{\pi} \int_0^\pi (\Ima M) \sin(\theta)\, d\theta \biggr]\no \\[4pt]
&\nhs=\nhs& -\f12 \log[\mu_{\ac} (-2,2)]\geq 0 \no
\end{eqnarray}
by \eqref{1.22}.

The hard work in this paper will be to extend the sum rule to equalities or inequalities 
in fairly general settings. Indeed, we will prove the following: 

\specialnumber{5}\proclaim{Theorem} If $J$ is a Jacobi matrix for which the right-hand side of {\rm \eqref{1.25}} is
finite{\rm ,}  then the left\/{\rm -}\/hand side is also finite and\/ ${\rm LHS}\leq{\rm RHS}$. 
\endproclaim

\specialnumber{6}\proclaim{Theorem} \lb{T6} If $\mu$ is a probability measure that obeys the Blumenthal\/{\rm -}\/Weyl 
criterion and the left{\rm -}hand side of {\rm \eqref{1.25}} is finite{\rm ,} then the right{\rm -}hand side of
{\rm \eqref{1.25}} is also finite and ${\rm LHS}\geq{\rm RHS}$.
\endproclaim

In other words, the $P_2$ sum rule {\it always\/} holds although both sides may be infinite. 
We will see (Proposition~\ref{P3.4}) that $G(a)$ has a zero only at $a=1$ where $G(a)
=2(a-1)^2 + O((a-1)^3)$ so the RHS of \eqref{1.25} is finite if and only if $\sum b_n^2 + 
\sum (a_n-1)^2 <\infty$, that is, $J$ is Hilbert-Schmidt. On the other hand, we will see (see 
Proposition~\ref{P3.5}) that $F(E_j) = (\abs{E_j}-2)^{3/2} +O((\abs{E_j}-2)^2)$ so the 
LHS of \eqref{1.25} is finite if and only if the quasi-Szeg\H{o} condition \eqref{1.8} and 
Lieb-Thirring bound \eqref{1.9} hold. Thus, Theorems~5 and 6 imply Theorem~1.

The major tool in proving the Case sum rules is a function that arises in essentially 
four distinct guises: 

\vglue6pt \noindent\hskip12pt\hangindent=18pt\hangafter=1 
 (1)  The perturbation determinant defined as
\vglue-12pt
\begin{equation} \label{1.28}
L(z;J) = \det \big[(J-z-z^{-1})(J_0 -z-z^{-1})^{-1}\big].
\end{equation}

\vglue4pt \noindent\hskip12pt\hangindent=30pt\hangafter=1 (2)  The Jost function, $u_0(z;J)$ defined for suitable
$z$ and
$J$. The Jost  solution is the unique solution of 
\vglue-12pt
\begin{equation} \label{1.29}
a_n u_{n+1} + b_n u_n + a_{n-1} u_{n-1} = (z+z^{-1})u_n
\end{equation}

\noindent\hskip12pt\hangindent=30pt\hangafter=1 \phantom{(2)} $n\geq 1$ with $a_0\equiv 1$ which obeys
\vglue-12pt
\begin{equation} \label{1.30}
\lim_{n\to\infty} z^{-n} u_n =1.
\end{equation}

\noindent\hskip12pt\hangindent=30pt\hangafter=1 \phantom{(2)} The Jost function is $u_0(z;J)=u_0$.
\vglue4pt
\noindent\hskip12pt\hangindent=30pt\hangafter=1  (3)   Ratio asymptotics of the orthogonal polynomials
$P_n$, 

\vglue-6pt
\begin{equation} \label{1.32}
\lim_{n\to\infty} P_n (z+z^{-1}) z^n.
\end{equation}

\pagebreak
\noindent\hskip12pt\hangindent=18pt\hangafter=1 (4)   The Szeg\H{o} function, normally only defined when
$N_+=N_-=0$:

\vglue-12pt
\begin{equation} \label{1.33}
D(z) = \exp \biggl( \frac{1}{4\pi} \int \log\big|2\pi\sin(\theta)f(2\cos\theta)\big|\,
\f{e^{i\theta}+z}{e^{i\theta}-z}\, d\theta \biggr)
\end{equation}

\noindent\hskip12pt\hangindent=18pt\hangafter=1 \phantom{(2)} where $d\mu=f(E)dE+d\mu_{\sing}$.
\vglue6pt

These functions are not all equal, but they are closely related.
$L(z;J)$ is defined  
for $\abs{z}<1$ by the trace class theory of determinants \cite{GK}, \cite{STr} so long as 
$J-J_0$ is trace class. We will see in that case it has a continuation to $\{z\mid \,
\abs{z}\leq 1, \, z\neq\pm 1\}$ and, when $J-J_0$ is finite rank, it is a polynomial. 
The Jost function is related to $L$ by 
\begin{equation} \label{1.34}
u_0(z;J) = \biggl(\prod_1^\infty a_j\biggr)^{-1} L(z;J).
\end{equation}
Indeed, we will define all $u_n$ by formulae analogous to \eqref{1.34} and show that 
they obey \eqref{1.29}/\eqref{1.30}.  The Jost solution is normally constructed using
existence theory for the difference equation \eqref{1.29}.
We show directly that the limit in \eqref{1.32} is $u_0(J,z)/(1-z^2)$.
Finally, the connection of $D(z)$ to $u_0(z)$ is
\begin{equation} \label{1.35}
D(z) = (2)^{-1/2}\,(1-z^2)\,u_0 (z;J)^{-1}.
\end{equation}
Connected to this formula, we will prove that
\begin{equation} \label{1.36}
\abs{u_0(e^{i\theta})}^2 = \f{\sin\theta}{\Ima M_\mu(\theta)}\, ,
\end{equation}
from which \eqref{1.35} will follow easily when $J-J_0$ is nice enough.
The result for general trace class 
$J-J_0$ is obviously new since it requires Nevai's conjecture to even define $D$ in that 
generality. It will require the analytic tools of this paper.

In going from the formal sum rules to our general results like Theorems~4 and 5, we will 
use three technical tools:
\begin{itemize}
\item[(1)] That the map $\mu\mapsto\int_{-\pi}^\pi \log(\f{\sin\theta}{\Ima M_\mu}) 
\sin^2 \theta\, d\theta$ and the similar map with $\sin^2\theta\, d\theta$ replaced by 
$d\theta$ is weakly lower semicontinuous. As we will see, these maps are essentially the 
negatives of entropies and this will be a known upper semicontinuity of an entropy.
\item[(2)] Rather than prove the sum rules in one step, we will have a way to prove them 
one site at a time, which yields inequalities that go in the opposite direction from the 
semicontinuity in (1).
\item[(3)] A detailed analysis of how eigenvalues change as a truncation is removed.
\end{itemize}

\pagebreak
In Section~\ref{S2}, we discuss the construction and properties of the perturbation 
determinant and the Jost function. In Section~\ref{S3}, we give a proof of the Case sum 
rules for nice enough $J-J_0$ in the spirit of Flaschka's \cite{Fla1} and Case's \cite{C2}
papers, and in Section~\ref{S4}, a second proof implementing tool (2) above. Section~\ref{S5} 
discusses the Szeg\H{o} and quasi-Szeg\H{o} integrals as entropies and the associated 
semicontinuity, and Section~\ref{S6} implements tool (3). Theorem~5 is proven in  
Section~\ref{S7}, and Theorem~6 in Section~\ref{S8}.

Section~\ref{S9} discusses the $C_0$ sum rule and proves Nevai's conjecture.

The proof of Nevai's conjecture itself will be quite simple---the $C_0$ sum rule and 
semicontinuity of the entropy will provide an inequality that shows the Szeg\H{o} 
integral is finite. We will have to work quite a bit harder to show that the sum rule 
holds in this case, that is, that the inequality we get is actually an equality.

In Section~\ref{S10}, we turn to another aspect that the sum rules expose: the fact that 
a dearth of bound states forces a.c.~spectrum. For Schr\"odinger operators, there are 
many $V$'s which lead to $\sigma(-\Delta +V)=[0,\infty)$. This always happens, for 
example, if $V(x)\geq 0$ and $\lim_{\abs{x}\to\infty} V(x)=0$. But for discrete 
Schr\"odinger operators, that is, Jacobi matrices with $a_n\equiv 1$, this 
phenomenon is not widespread because $\sigma(J_0)$ has two sides. Making $b_n\geq 0$ 
to prevent eigenvalues in $(-\infty, -2)$ just forces them in $(2,\infty)$! We will 
prove two somewhat surprising results (the $e_n(J)$ are defined in Remark~6 after 
Theorem~1). 

\specialnumber{7} \proclaim{Theorem} \lb{T7} If $J$ is a Jacobi matrix with $a_n\equiv 1$ and $\sum_n 
\abs{e_n (J)}^{1/2}<\infty${\rm ,} then $\sigma_{\ac}(J)=[-2,2]$.
\endproclaim

\specialnumber{8}\proclaim{Theorem} \lb{T8} Let $W$ be a two\/{\rm -}\/sided Jacobi matrix 
with $a_n\equiv 1$ and no 
eigenvalues. Then $b_n=0${\rm ,} that is{\rm ,} $W=W_0${\rm ,} the free Jacobi matrix. 
\endproclaim

We emphasize that Theorem 8 does {\it not\/} presuppose any reflectionless condition.

\demo{Acknowledgments} We thank F.~Gesztesy, N.~Makarov, P.~Nevai,\break M. B.~Ruskai, and V.~Totik 
for useful discussions. R.K. would like to thank T.~Tombrello for the hospitality of 
Caltech where this work was initiated.
\enddemo
 
\section{Perturbation determinants and the Jost function}\label{S2}

In this section we introduce the perturbation determinant 
$$
  L(z;J)=\det\big[\big(J-E(z)\big)\,\big(J_0-E(z)\big)^{-1}\big];\qquad E(z)=z+z^{-1}
$$
and describe its analytic properties.  This leads naturally to a discussion 
of the Jost function commencing with the introduction of the Jost solution \eqref{2.62}.
The section ends with some remarks on the asymptotics of orthogonal polynomials.
We begin, however, with notation, the basic properties of $J_0$, and a brief review
of determinants for trace class and Hilbert-Schmidt operators.
The analysis of $L$ begins in earnest with Theorem~\ref{T2.4}.

Throughout, $J$ represents a matrix of the form \eqref{1.1} thought of as an operator on 
$\ell^2 (\bbZ_+)$.  The special case $a_n\equiv 1$, $b_n\equiv 0$ is denoted by $J_0$
and $\delta J=J-J_0$ constitutes the perturbation.
If $\delta J$ is finite rank (i.e., for 
large $n$, $a_n =1$ and $b_n=0$), we say that $J$ is {\it finite range}.

It is natural to approximate the true perturbation by one of finite rank.
We define $J_n$ as the semi-infinite matrix,
\begin{equation} \label{2.10}
J_n = \left( \begin{array}{ccccccc}
b_1 & a_1 & 0 & {} & {} & {} & {} \\
a_1 & b_2 & a_2 & {} & {} & {} & {} \\
{} & \dots & \dots & \dots & {} & {} & {} \\
{} & \dots & b_{n-1} & a_{n-1} & {} & {} & {} \\
{} & {} & a_{n-1} & b_n & 1 & {} & {} \\
{} & {} & {} & 1 & 0 & 1 & {} \\
{} & {} & {} & {} & 1 & 0 & \dots 
\end{array}\right)
\end{equation}
that is, $J_n$ has $b_m=0$ for $m>n$ and $a_m =1$ for $m>n-1$.  Notice that $J_n-J_0$
has rank at most $n$.

We write the $n\times n$ matrix obtained by taking the first $n$ rows and columns of $J$ 
(or of $J_n$) as $J_{n;F}$.  The $n\times n$ matrix formed from $J_0$ will be called
$J_{0;n;F}$.

A different class of associated objects will be the semi-infinite matrices $J^{(n)}$ 
obtained from $J$ by dropping the first $n$ rows and columns of $J$, that is,
\begin{equation} \label{2.10a}
J^{(n)} = 
\left( \begin{array}{ccccc} 
b_{n+1} & a_{n+1} & 0 & \dots \\
a_{n+1} & b_{n+2} & a_{n+2} & \dots \\
0 & a_{n+2} & b_{n+3} & \dots \\
\dots & \dots & \dots & \dots 
\end{array}\right).
\end{equation}

As the next preliminary, we need some elementary facts about $J_0$, the free Jacobi matrix. 
Fix $z$ with $\abs{z}<1$. Look for solutions of
\begin{equation} \label{2.11}
u_{n+1} + u_{n-1} = (z+z^{-1}) u_n, \qquad n\geq 2
\end{equation}
as sequences without any {\it a~priori} conditions at infinity or $n=1$. The solutions of 
\eqref{2.11} are linear combinations of the two ``obvious" solutions $u^\pm$ given by
\begin{equation} \label{2.12a}
u_n^\pm (z) = z^{\pm n}.
\end{equation}
Note that $u^+$ is $\ell^2$ at infinity since $\abs{z}<1$. The linear combination that obeys
$$
u_2 = (z+z^{-1})u_1
$$
as required by the matrix ending at zero is (unique up to a constant)
\begin{equation} \label{2.13a}
u_n^{(0)}(z) = z^{-n} - z^n.
\end{equation}
Noting that the Wronskian of $u^{(0)}$ and $u^+$ is $z^{-1}-z$, we see that $(J_0 -E(z))^{-1}$ has 
the matrix elements $-(z^{-1} -z)^{-1} u_{\min(n,m)}^{(0)} (z) u_{\max(n,m)}^+(z)$ either 
by a direct calculation or standard Green's function formula. We have thus proven that
\begin{eqnarray}
(J_0 -E(z))_{nm}^{-1}&=&-(z^{-1} -z)^{-1} [z^{\abs{m-n}} -z^{m+n}] \lb{2.12} \\[6pt]
&=&-\sum_{j=0}^{\min(m,n)-1} z^{1+\abs{m-n}+2j} \lb{2.13}
\end{eqnarray}
where the second comes from $(z^{-1}-z) (z^{1-n} + z^{3-n} + \cdots + z^{n-1}) = z^{-n} 
-z^n$ by telescoping. \eqref{2.13} has two implications we will need later:
\begin{equation} \label{2.14}
\abs{z}\leq 1 \Rightarrow \abs{(J_0 -E(z)_{nm}^{-1}}\leq \min(n,m) \abs{z}^{1+\abs{m-n}}
\end{equation} 
and that while the operator $(J_0 -E(z))^{-1}$ becomes singular as $\abs{z}\uparrow 1$, the 
matrix elements do not; indeed, they are polynomials in $z$.

We need an additional fact about $J_0$:

\proclaim{Proposition} \lb{P2.3} The characteristic polynomial of $J_{0;n;F}$ is 
\begin{equation} \label{2.15}
\det (E(z)-J_{0,n;F})=\f{(z^{-n-1}-z^{n+1})}{(z^{-1}-z)}=U_n\big({\textstyle \frac{1}{2}}\, E(z)\big) 
\end{equation}
where $U_n(\cos \theta)=\sin[(n+1)\theta]/\sin(\theta)$ is the Chebyshev polynomial
of the second kind.  In particular{\rm ,} 
\begin{equation} \label{2.17}
\lim_{n\to\infty}\, \frac{\det[E(z)-J_{0;n+j;F}]}{\det [E(z)-J_{0;n;F}]} =z^{-j}.
\end{equation}
\endproclaim

\demo{Proof} Let
\begin{equation} \lb{2.16}
g_n(z) = \det (E(z)-J_{0;n;F}).
\end{equation}
By expanding in minors
$$
g_{n+2}(z) = (z+z^{-1}) g_{n+1}(z) - g_n(z).
$$
Given that $g_1 = z+z^{-1}$ and $g_0 =1$, we obtain the first equality of \eqref{2.15}
by induction. The second equality and \eqref{2.17} then follow easily.
\enddemo

In Section~\ref{S4}, we will need

\proclaim{Proposition} \lb{P2.3a} Let $T_m$ be the Chebyshev polynomial {\rm{(}}\/of the first 
kind\/{\rm{):}}
\begin{equation} \lb{2.20a} 
T_m (\cos\theta) = \cos(m\theta).\pagebreak
\end{equation}
Then
\begin{equation} \lb{2.20b} 
\Tra\big[T_m \big(\tfrac{1}{2}J_{0,n;F}\big)\big] = 
\left\{ \begin{array}{ll} n & m=2\ell (n+1);\, \ell\in \bbZ\\
-\f12 - \f12 (-1)^m &\hbox{otherwise.} 
\end{array}\right.  \hskip.35in
\end{equation}
In particular{\rm ,} for $m$ fixed{\rm ,} once $n>\f12 m-1$ the trace is independent of $n$.
\endproclaim

\demo{Proof} As noted above, the characteristic polynomial of $J_{0,n;F}$ is $U_n(E/2)$.
That is, $\det[2\cos(\theta)-J_{0;n;F}]=\sin[(n+1)\theta]/\sin[\theta]$.
This implies that the eigenvalues of $J_{0;n;F}$ are given by 
\begin{equation} \lb{2.20c} 
E_n^{(k)} = 2\cos \biggl( \f{k\pi}{n+1}\biggr) \qquad k=1,\dots, n.
\end{equation}
So by \eqref{2.20a}, $T_m \big(\tfrac{1}{2}E_n^{(k)}\big) = \cos\big(\f{km\pi}{n+1}\big)$.
Thus,
\begin{eqnarray*}
\Tra \big[ T_m \big( \tfrac{1}{2}J_{0;n;F}\big) \big] &=& \sum_{k=1}^n \cos 
\biggl( \f{km\pi}{n+1}\biggr) \\
&=& -\tfrac{1}{2} - \tfrac{1}{2}\, (-1)^m + \tfrac{1}{2}\sum_{k=-n}^{n+1} \exp
	\biggl( \f{ikm\pi}{n+1}\biggr).
\end{eqnarray*}
The final sum is $2n+2$ if $m$ is a multiple of $2(n+1)$ and $0$ if it is not.
\enddemo

As a final preliminary, we discuss Hilbert space determinants \cite{GK}, \cite{S74}, \cite{STr}. 
Let $\calI_p$ denote the Schatten classes of operators with norm $\|A\|_p=\Tra(|A|^p)$ as 
described for example, in \cite{STr}.  In particular, $\calI_1$ and $\calI_2$ are the trace 
class and Hilbert-Schmidt operators, respectively.

For each $A\in\calI_1$, one can define 
a complex-valued function $\det (1+A)$ (see \cite{GK}, \cite{STr}, \cite{S74}), so that 
\begin{equation} \label{2.18}
\abs{\det(1+A)}\leq \exp (\|A\|_1)
\end{equation}
and $A\mapsto \det(1+A)$ is continuous; indeed \cite[pg.~48]{STr}, 
\begin{equation} \label{2.19}
\abs{\det (1+A) -\det (1+B)} \leq \|A-B\|_1 \exp (\|A\|_1 + \|B\|_1 +1). \hskip.25in
\end{equation}
We will also use the following properties: 
\begin{eqnarray}
&&A,B\in\calI_1\   \Rightarrow\ \det (1+A) \det (1+B) = \det (1+A+B+AB) \lb{2.20} \\
&&\phantom{A,B\in\calI_1\ } AB,BA\in\calI_1\ \Rightarrow\ \det (1+AB) =\det (1+BA) \lb{2.21}  \\
&&\phantom{A,B\in\calI} (1+A)\hbox{ is invertible if and only if } \det(1+A)\neq 0   \label{2.22} \\
&&\phantom{A,B\in\calI_1\ }z\mapsto A(z) \hbox{ analytic} \Rightarrow \det (1+A(z)) \hbox{ analytic}. \label{2.22a}
\end{eqnarray}

If $A$ is finite rank and $P$ is a finite-dimensional self-adjoint projection,  
\begin{equation} \label{2.22b}
PAP=A\quad \Rightarrow\quad \det (1+A) = \det\nolimits_{P\calH} (1_{P\calH}+PAP),
\end{equation}
where $\det_{P\calH}$ is the standard finite-dimensional determinant.

For $A\in\calI_2$, $(1+A) e^{-A}-1\in\calI_1$, so one defines 
(see \cite[pp.~106--108]{STr}) 
\begin{equation} \label{2.23}
\det\nolimits_2 (1+A) =\det ((1+A)e^{-A}).
\end{equation}
Then
\begin{eqnarray} 
\abs{\det\nolimits_2 (1+A)} &\nhs\leq \nhs&\exp (\|A\|_2^2) \label{2.24} \\
\qquad\abs{\det\nolimits_2 (1+A)-\det\nolimits_2 (1+B)} &\nhs\leq\nhs &\|A-B\|_2 \exp ((\|A\|_2 + 
\|B\|_2 + 1)^2) \lb{2.25}
\end{eqnarray}
and, if $A\in\calI_1$, 
\begin{equation} \label{2.24a}
\det\nolimits_2 (1+A) = \det (1+A)e^{-\Tra (A)}
\end{equation}
or
\begin{equation} \label{2.26}
\det (1+A) = \det\nolimits_2 (1+A) e^{\Tra (A)}.
\end{equation}

To estimate the $\calI_p$ norms of operators we use

\proclaim{Lemma} \lb{L2.1} If $A$ is a matrix and $\|\dott\|_p$ the Schatten $\calI_p$
norm {\rm \cite{STr},} then

{\rm{(i)}} 
\begin{equation} \label{2.3}
\|A\|_2^2 = \sum_{n,m}\, \abs{a_{nm}}^2,
\end{equation}

{\rm{(ii)}} 
\begin{equation} \label{2.4}
\|A\|_1 \leq \sum_{n,m}\, \abs{a_{nm}},
\end{equation}

{\rm{(iii)}}  For any $j$ and $p${\rm ,}
\begin{equation} \label{2.5}
\sum_n \, \abs{a_{n,n+j}}^p \leq \|A\|_p^p.
\end{equation}
\endproclaim

\demo{Proof} (i) is standard. (ii) follows from the triangle inequality for $\|\dott\|_1$
and the fact that a matrix which a single nonzero matrix element, $\alpha$, has trace norm
$\abs{\alpha}$. (iii) follows from a result of Simon \cite{STr}, \cite{S73} that
\vglue12pt
\hfill ${\displaystyle 
\|A\|_p^p =\sup \biggl\{ \sum_n \abs{\langle \varphi_n, A\psi_n\rangle}^p \biggm|
\{\varphi_n\}, \{\psi_n\} \hbox{ orthonormal sets}\biggr\}.
}$
\enddemo
\vglue6pt
 
The following factorization will often be useful. Define
$$
c_n =\max(\abs{a_{n-1} -1}, \abs{b_n}, \abs{a_n-1})
$$
which is the maximum matrix element in the $n^{\rm th}$ row and $n^{\rm th}$ column.
Let $C$ be the diagonal matrix with matrix elements $c_n$. Define $U$ by
\begin{equation} \label{2.6}
\delta J = C^{1/2} UC^{1/2}.
\end{equation}
Then $U$ is a tridiagonal matrix with matrix elements bounded by $1$ so
\begin{equation} \label{2.7}
\|U\|\leq 3.
\end{equation} 

One use of \eqref{2.6} is the following:

\proclaim{Theorem} \lb{T2.2} Let $c_n =\max(\abs{a_{n-1} -1}, \abs{b_n}, \abs{a_n-1})$.
For any $p\in [1,\infty)${\rm ,}
\begin{equation} \label{2.9}
\tfrac13 \biggl(\sum_n \abs{c_n}^p\biggr)^{1/p} \leq \|\delta J\|_p \leq
3\biggl(\sum_n \abs{c_n}^p\biggr)^{1/p}.
\end{equation}
\endproclaim 

\demo{Proof} The right side is immediate from \eqref{2.6} and H\"older's inequality for
trace ideals \cite{STr}. The leftmost inequality follows from \eqref{2.5} and
\vglue12pt
\hfill ${\displaystyle
\biggl( \sum_n \abs{c_n}^p\biggr)^{1/p} \leq \biggl( \sum_n \abs{b_n}^p \biggr)^{1/p}
+2\biggl(\sum_n \abs{a_n-1}^p\biggr)^{1/p}.
}$
\enddemo
\vglue6pt

With these preliminaries out of the way, we can begin discussing the perturbation
determinant $L$. For any $J$ with $\delta J\in\calI_1$ (by \eqref{2.9} this is equivalent 
to $\sum\abs{a_n-1} + \sum \abs{b_n}<\infty$), we define
\begin{equation} \label{2.27}
L(z;J) = \det\big[ \big(J-E(z)\big)\;\big(J_0 -E(z)\big)^{-1}\big] 
\end{equation}
for all $|z|<1$.  Since
\begin{equation} \label{2.28}
(J-E)(J_0 -E)^{-1} = 1+\delta J (J_0 -E)^{-1}, 
\end{equation}
the determinant in \eqref{2.27} is of the form $1+A$ with $A\in \calI_1$.

\proclaim{Theorem} \lb{T2.4} Suppose $\delta J\in\calI_1$.
\begin{itemize} 
\item[{\rm{(i)}}] $L(z;J)$ is analytic in $D\equiv\{z\mid\,\abs{z}<1\}$.
\item[{\rm{(ii)}}] $L(z;J)$ has a zero in $D$ only at points $z_j$ where 
$E(z_j)$ is an eigenvalue of~$J${\rm ,} and it has zeros at all such points. 
All zeros are simple.
\item[{\rm{(iii)}}] If $J$ is finite range{\rm ,} then $L(z;J)$ is a polynomial and so 
has an analytic continuation to all of $\bbC$.
\end{itemize}

\endproclaim 

\demo{Proof} (i) follows from \eqref{2.22a}. 

\vglue5pt
(ii) If $E_0=E(z_0)$ is not an eigenvalue of $J$, then $E_0\notin \sigma(J)$ since
$E:D\to\bbC\backslash [-2,2]$ and $\sigma_{\ess}(J)=[-2,2]$. Thus, $(J-E_0)/(J_0 -E_0)$ 
has an inverse (namely, $(J_0 -E_0)/(J-E_0)$), and so by \eqref{2.22}, $L(z;J)\neq 0$. 
If $E_0$ is an eigenvalue, $(J-E_0)/(J_0-E_0)$ is not invertible, so by \eqref{2.22}, 
$L(z_0;J)=0$. Finally, if $E(z_0)$ is an eigenvalue, eigenvalues of $J$ are simple 
by a Wronskian argument. That $L$ has a simple zero under these circumstances comes 
from the following.

If $P$ is the projection onto the eigenvector at $E_0 =E(z_0)$, then\break $(J-E)^{-1}(1-P)$ 
has a removable singularity at $E_0$. Define
\begin{equation} \label{2.29}
C(E) = (J-E)^{-1} (1-P)+P
\end{equation}
so
\begin{equation} \label{2.30}
(J-E)C(E) = 1- P+(E_0 -E)P.
\end{equation}

Define 
\begin{eqnarray}
D(E) &\equiv& (J_0 -E)C(E) \lb{2.31} \\
&= &-\delta JC(E) + (J-E)C(E) \no  \\
&=& 1-P + (E_0 -E)P - \delta JC(E)\no \\
&=&1 + \hbox{trace class}. \no 
\end{eqnarray}
Moreover, 
\begin{eqnarray*}
D(E) [ (J-E)/(J_0 -E)] &=& (J_0 -E) [1-P+(E_0 -E)P](J_0-E)^{-1} \\
&=& 1+(J_0 -E) [-P+(E_0 -E)P](J_0-E)^{-1}.
\end{eqnarray*}
Thus by \eqref{2.20} first and then \eqref{2.21}, 
\begin{eqnarray*}
\det(D(E(z))) L(z;J) &=& \det (1+(J_0-E)[-P+(E_0-E)P](J_0-E)^{-1}) \\
&=&\det (1-P +(E_0-E)P) \\
&=& E_0 -E(z),
\end{eqnarray*}
where we used \eqref{2.22b} in the last step. Since $L(z;J)$ has a zero at $z_0$ and $E_0 -E(z) 
=(z-z_0) [1-\f{1}{zz_0}]$ has a simple zero, $L(z;J)$ has a simple zero.

\vglue5pt
(iii) Suppose $\delta J$ has range $N$, that is, $N=\max\{n\mid\,|b_n|+|a_{n-1}-1|>0\}$ and
let $P^{(N)}$ be the projection onto the span of
$\{\delta_j\}_{j=1}^{N}$.  As $P^{(N)}\delta J=\delta J$,
$$
\delta J (J_0 -E)^{-1} = P^{(N)} P^{(N)} \delta J (J_0 -E)^{-1}.
$$
By \eqref{2.21},
$$
L(z;J) = \det \big(1+P^{(N)} \delta J \big(J_0-E(z)\big)^{-1} P^{(N)}\big).
$$
Thus by \eqref{2.13}, $L(z;J)$ is a polynomial if $\delta J$ is finite range.
\enddemo

 {\it Remarks.} 1. By this argument, if $\delta J$ has range $n$, $L(z;J)$ is the 
determinant of an $n\times n$ matrix whose $ij$ element is a polynomial of degree 
$i+j+1$. That implies that we have shown $L(z;J)$ is a polynomial of degree at most 
$2n(n+1)/2 + n =(n+1)^2$. We will show later it is actually a polynomial of degree 
at most $2n-1$.
\vglue5pt
2. The same idea shows that if $\sum_n \abs{(a_n-1)\rho^{2n}} + \abs{b_n\rho^{2n}}<\infty$ for 
some $\rho >1$, then $C^{1/2} (J_0 -z-z^{-1})^{-1} C^{1/2}$ is trace class for $\abs{z}<\rho$, and 
thus $L(z;J)$ has an analytic continuation to $\{z\mid\, \abs{z}<\rho\}$.
\vglue8pt

We are now interested in showing that $L(z;J)$, defined initially only on~$D$,
can be continued to $\partial D$ or part of $\partial D$. Our goal is to show:  
 
 \phantom{i}(i) \   If
\vglue-12pt
\begin{equation} \label{2.32}
\sum_{n=1}^\infty n [\abs{a_n -1} + \abs{b_n}] <\infty,
\end{equation}
 
\noindent\hglue42pt\hangindent=42pt\hangafter=1
 then $L(z;J)$ can be continued to all of $\bar D$, that is, extends
to a function continuous  on $\bar D$ and analytic in $D$.

\vglue4pt \hangindent=42pt\hangafter=1 (ii) For the general trace class situation, $L(z;J)$ has a continuation to 
$\bar D\backslash \{-1,1\}$.  

\vglue4pt \hangindent=42pt\hangafter=1 \hskip-3pt(iii) \hskip4pt  As $x$ real approaches $\pm 1$, $\abs{L(x;J)}$
is bounded by 
$\exp\{o(1)/(1-\abs{x})\}$.
\vglue6pt

We could interpolate between (i) and (iii) and obtain more information about cases where 
\eqref{2.32} has $n$ replaced by $n^\alpha$ with $0<\alpha<1$ or even $\log n$ (as is 
done in \cite{Nev2}, \cite{GVA}), but using the theory of Nevanlinna functions and (iii), we will 
be able to handle the general trace class case (in Section~\ref{S9}), so we forgo 
these intermediate results.

\proclaim{Lemma} \lb{L2.5} Let $C$ be  diagonal positive trace class matrix. For $\abs{z} 
<1${\rm ,} define 
\begin{equation} \label{2.33}
A(z) = C^{1/2} (J_0 -E(z))^{-1} C^{1/2}.
\end{equation}
Then{\rm ,} as a Hilbert\/{\rm -}\/Schmidt operator\/{\rm -}\/valued function{\rm ,} $A(z)$ extends continuously to
$\bar D\setminus \{-1, 1\}$. If 
\begin{equation} \label{2.34}
\sum_n n c_n <\infty,
\end{equation}
it has a Hilbert\/{\rm -}\/Schmidt continuation to $\bar D$.
\endproclaim

\demo{Proof} Let $A_{nm}(z)$ be the matrix elements of $A(z)$. It follows from $\abs{z}<1$ 
and \eqref{2.12}/\eqref{2.14} that 
\begin{eqnarray}
\abs{A_{nm}(z)} &\leq& 2c_n^{1/2} c_m^{1/2} \abs{z-1}^{-1} \abs{z+1}^{-1} \lb{2.35} \\ 
\abs{A_{nm}(z)} &\leq &\min(m,n) c_n^{1/2} c_m^{1/2}  \lb{2.36}
\end{eqnarray}

\noindent 
and each $A_{n,m}(z)$ has a continuous extension to $\bar D$. It follows from \eqref{2.35}, the 
dominated convergence theorem, and
$$
\sum_{n,m} (c_n^{1/2}c_m^{1/2})^2 = \biggl( \sum_n c_n\biggr)^2
$$
that so long as $z$ stays away from $\pm 1$, $\{A_{mn}(z)\}_{n,m}$ is continuous in the space
$\ell^2((1,\infty) \times (1,\infty))$ so $A(z)$ is Hilbert-Schmidt and continuous on $\bar D
\backslash \{-1, 1\}$. Moreover, \eqref{2.36} and 
$$
\sum_{n,m} \big[\min(m,n) c_n^{1/2} c_m^{1/2}\big]^2 \leq \sum_{mn} mn c_nc_m = 
\biggl( \sum_n nc_n\biggr)^2
$$
imply that $A(z)$ is Hilbert-Schmidt on $\bar D$ if \eqref{2.34} holds.
\enddemo

{\it Remark.} When  \eqref{2.34} holds---indeed, when 
\begin{equation} \label{2.37}
\sum n^\alpha c_n <\infty
\end{equation}
for some $\alpha >0$---we believe that one can show $A(z)$ has trace class boundary values 
on $\partial D\backslash \{-1, 1\}$ but we will not provide all the details since the 
Hilbert-Schmidt result suffices. To see this trace class result, we note that $\Ima A(z) 
= (A(z) - A^*(z))/2i$ has a rank 1 boundary value as $z\to e^{i\theta}$; explicitly, 
\begin{equation} \label{2.38}
\Ima A(e^{i\theta})_{mn} = -c_n^{1/2} c_m^{1/2}\,
 \f{(\sin m\theta)(\sin n\theta)}{(\sin\theta)}\,.
\end{equation}
Thus, $\Ima A(e^{i\theta})$ is trace class and is H\"older continuous in the trace norm
if \eqref{2.37} holds.  Now $\Real A(e^{i\theta})$ is the Hilbert transform of a H\"older 
continuous trace class operator-valued function and so trace class. This is because when a 
function is H\"older continuous, its Hilbert transform is given by a convergent integral, 
hence limit of Riemann sums. Because of potential singularities at $\pm 1$, the details 
will be involved.  

\proclaim{Lemma} \lb{L2.6} Let $\delta J$ be trace class. Then
\begin{equation} \label{2.39}
t(z) = \Tra ((\delta J)(J_0 -E(z))^{-1})
\end{equation}
has a continuation to $\bar D\backslash\{-1, 1\}$. If {\rm \eqref{2.32}} holds{\rm ,} $t(z)$ can be 
continued to $\bar D$.
\endproclaim

 {\it Remark.} We are only claiming $t(z)$ can be continued to $\partial D$, not 
that it equals the trace of $(\delta J)(J_0 -E(z))^{-1}$ since $\delta J(J_0 -E(z))^{-1}$ 
is not even a bounded operator for $z\in\partial D$!

\demo{Proof} $t(z) = t_1 (z) + t_2 (z) + t_3(z)$ where
\begin{eqnarray*}
t_1(z) &=& \sum b_n (J_0 -E(z))_{nn}^{-1} \\
t_2(z) &=& \sum (a_n -1)(J_0 -E(z))_{n+1, n}^{-1} \\
t_3(z) &=&\sum (a_n -1) (J_0 -E(z))_{n, n+1}^{-1}\,.
\end{eqnarray*}
Since, by \eqref{2.12}, \eqref{2.14},
\begin{eqnarray*} 
\abs{(J_0 -E(z))_{nm}^{-1}} &\leq &2\abs{z-1}^{-1} \abs{z+1}^{-1} \\
\abs{(J_0 -E(z))_{nm}^{-1}} &\leq& \min (n,m),
\end{eqnarray*}
the result is immediate. 
\enddemo

\proclaim{Theorem} \lb{T2.7} If $\delta J$ is trace class{\rm ,} $L(z;J)$ can be extended to a 
continuous function on $\bar D\backslash\{-1,1\}$ with
\begin{equation} \label{2.40}
\abs{L(z;J)} \leq \exp\Big\{ c\big[\|\delta J\|_1 + \|\delta J\|_1^2\big]\, 
\abs{z-1}^{-2}\abs{z+1}^{-2} \Big\} \hskip.5in
\end{equation}
for a universal constant{\rm ,} $c$. If {\rm \eqref{2.32}} holds{\rm ,} $L(z;J)$ can be extended to all 
of $\bar D$ with
\begin{equation} \label{2.41}
\abs{L(z;J)} \leq\exp\biggl\{\tilde c\Big[1+ \sum_{n=1}^\infty n\big[\abs{a_n-1} + \abs{b_n}\big] 
\Big]^2 \biggr\}
\end{equation}
for a universal constant{\rm ,} $\tilde c$.
\endproclaim 

\demo{Proof} This follows immediately from \eqref{2.23}, \eqref{2.24}, \eqref{2.24a}, and 
the last two lemmas and their proofs. 
\enddemo

While we cannot control $\|C^{1/2} (J_0 -E(z))^{-1} C^{1/2}\|_1$ for arbitrary $z$ with 
$\abs{z}\to 1$, we can at the crucial points $\pm 1$ if we approach along the real axis, 
because of positivity conditions.

\proclaim{Lemma} \lb{L2.8} Let $C$ be a positive diagonal trace class operator. Then
\begin{equation} \label{2.41a}
\lim_{{\abs{x}\uparrow 1 \atop x\ {\rm real} }} \, (1-\abs{x}) \|C^{1/2} 
(J_0 -E(x))^{-1} C^{1/2}\|_1 =0.
\end{equation}
\endproclaim

\demo{Proof} For $x<0$, $E(x)<-2$, and $J_0 - E(x)>0$, while for $x>0$, $E(x) >2$, so 
$J_0 -E(x) <0$. It follows that
\begin{eqnarray} 
\|C^{1/2} (J_0 -E(x))^{-1} C^{1/2}\|_1 &=& \abs{\Tra (C^{1/2} (J_0 -E(x))^{-1} C^{1/2})} 
 \label{2.42}\\
&\leq&\sum_n c_n \abs{(J_0 -E(x))_{nn}^{-1}}. \no
\end{eqnarray}
By \eqref{2.12},
$$
(1-\abs{x}) \abs{(J_0 -E(x))_{nn}^{-1}}\leq 1
$$
and by \eqref{2.13} for each fixed $n$, 
$$
\lim_{{ \abs{x}\uparrow 1 \atop x\ {\rm real} }} \, (1-\abs{x}) 
\abs{(J_0 -E(x))_{nn}^{-1}}=0.
$$
Thus \eqref{2.42} and the dominated convergence theorem proves \eqref{2.41a}.
\enddemo

\proclaim{Theorem} \lb{T2.9}
\begin{equation} \label{2.43}
\limsup_{{ \abs{x}\uparrow 1 \atop x \ {\rm  real} }} \, (1-\abs{x}) \log 
\abs{L(x;J)} \leq 0.
\end{equation}
\endproclaim  

\demo{Proof} Use \eqref{2.6} and \eqref{2.21} to write
$$
L(x;J) = \det (1+UC^{1/2} (J_0 -E(x))^{-1}C^{1/2})
$$
and then \eqref{2.18} and \eqref{2.7} to obtain 
\begin{eqnarray*}
\log \abs{L(x;J)} &\leq& \|UC^{1/2} (J_0 -E(x))^{-1} C^{1/2}\|_1 \\
&\leq& 3 \|C^{1/2} (J_0 -E(x))^{-1}C^{1/2}\|_1.
\end{eqnarray*}
The result now follows from the lemma. 
\enddemo

Next, we want to find the Taylor coefficients for $L(z;J)$ at $z=0$, which we will 
need in the next section.

\proclaim{Lemma} \lb{L2.10} For each fixed $h>0$ and $\abs{z}$ small{\rm ,} 
\begin{equation} \label{2.44}
\log\biggl(1- \f{h}{E(z)}\biggr) = \sum_{n=1}^\infty
	\tfrac{2}{n}\big[T_n(0)-T_n(\tfrac{1}{2}h)\big]z^n 
\end{equation}
where $T_n (x)$ is the $n^{\rm th}$ Chebyshev polynomial of the first kind\/{\rm :} 
$T_n(\cos\theta) = \cos (n\theta)$.  In particular{\rm ,} $T_{2n+1}(0)=0$ and $T_{2n}(0)=(-1)^n$.
\endproclaim

\demo{Proof} Consider the following generating function: 
\begin{equation}\label{ChebGF}
g(x,z)\equiv \sum_{n=1}^\infty T_n(x) \frac{z^n}{n} =  - \tfrac{1}{2}\log[1-2xz+z^2]. 
\end{equation}
The lemma now follows from
$$
\log\Big[1-\frac{2x}{z+z^{-1}}\Big] = 2[g(0,z)-g(x,z)] =
	\sum \tfrac{2}{n} \big[T_n(0)-T_n(x)\big]z^n
$$
by choosing $x=h/2$.  The generation function is well known (Abramowitz and
Stegun \cite[Formula 22.9.8]{AB} or Szeg\H{o} \cite[Equation 4.7.25]{Szb}) and
easily proved: for $\theta\in\bbR$ and $|z|<1$, 
\begin{eqnarray*}
\frac{\partial g}{\partial z}(\cos\theta,z) &= &\frac{1}{z}\sum_{n=1}^\infty \cos(n\theta) z^n \\
&=& \frac{1}{2z} \sum_{n=1}^\infty \Big[\big(ze^{i\theta}\big)^n + \big(ze^{-i\theta}\big)^n\Big] \\
&=& \frac{\cos(\theta)+z}{z^2-2z\cos\theta+1} \\
&= &-\tfrac{1}{2}\; \frac{\partial\phantom{z}}{\partial z}\; \log[1-2xz+z^2]
\end{eqnarray*}
at $x=\cos\theta$. Integrating this equation from $z=0$ proves \eqref{ChebGF} for $x\in[-1,1]$ 
and $|z|<1$. For more general $x$ one need only consider $\theta\in\bbC$ and require $|z|<\exp
\{-|\Ima \theta|\}$.
\enddemo

\proclaim{Lemma} \lb{L2.11} Let $A$ and $B$ be two self\/{\rm -}\/adjoint $m\times m$ matrices. Then
\begin{equation} \label{2.52}
\log\,\det\big[\big(A-E(z)\big)\;\big(B-E(z)\big)^{-1}\big] =\sum_{n=0}^\infty c_n (A,B)z^n 
\end{equation}
where 
\begin{equation} \label{2.53}
c_n (A,B) =-\tfrac{2}{n} \Tra \big[ T_n\big(\tfrac{1}{2}A\big) 
- T_n \big(\tfrac{1}{2}B\big)\big].
\end{equation}
\endproclaim

\demo{Proof} Let $\lambda_1, \dots, \lambda_m$ be the eigenvalues of $A$ and $\mu_1, \dots, 
\mu_m$ the eigenvalues of $B$. Then 
\begin{eqnarray*}
\det \biggl[ \f{A-E(z)}{B-E(z)}\biggr] &=& \prod_{j=1}^m 
\biggl[ \f{\lambda_j -E(z)}{\mu_j -E(z)}\biggr] \\
\hbox to 0mm{\hss$\Rightarrow$\quad}
\log\,\det \biggl[ \f{A-E(z)}{B-E(z)}\biggr]
&= &\sum_{j=1}^m\, \log[1-\lambda_j/E(z)] - \log[1-\mu_j/E(z)]
\end{eqnarray*}
so \eqref{2.52}/\eqref{2.53} follow from the preceding lemma.
\enddemo

\proclaim{Theorem} \lb{T2.12} If $\delta J$ is trace class{\rm ,} then for each $n${\rm ,} $T_n (J/2)- 
T_n (J_0/2)$ is trace class. Moreover{\rm ,} near $z=0${\rm ,}
\begin{equation} \label{2.54}
\log [L(z;J)] = \sum_{n=1}^\infty c_n (J) z^n 
\end{equation}
where
\begin{equation} \label{2.55}
c_n (J) = -\f{2}{n}\, \Tra \big[ T_n\big(\tfrac{1}{2}J\big) - T_n\big(\tfrac{1}{2}J_0\big)\big].
\end{equation}
In particular{\rm ,} 
\begin{eqnarray} 
c_1 (J)&=& -\Tra (J-J_0) = -\sum_{m=1}^\infty b_m \label{2.56} \\
c_2 (J)& =& -\tfrac{1}{2}\, \Tra (J^2 -J_0^2) = -\tfrac{1}{2} 
\sum_{m=1}^\infty [b_m^2 + 2(a_m^2 -1)]. \lb{2.57}
\end{eqnarray}
\endproclaim 

\demo{Proof} To prove $T_n (J/2) - T_n (J_0/2)$ is trace class, we need only show that 
$J^m - J_0^m =\sum_{j=1}^{m-1} J^j\,\mskip -1mu\delta J\,J^{m-1-j}$ is trace class, and that's obvious! 
Let $\widetilde{\delta J}_{n;F}$ be $\delta J_{n;F}$ extended to $\ell^2 (\bbZ_+)$ by 
setting it equal to the zero matrix on $\ell^2 (j\geq n)$. Let $\tilde J_{0,n}$ be $J_0$ with 
$a_{n+1}$ set equal to zero. Then
$$
\widetilde{\delta J}_{n;F} (\tilde J_{0,n}-E)^{-1} \to \delta J (J_0 -E)^{-1}
$$
in trace norm, which means that
\begin{equation} \label{2.58}
\det\biggl(\f{J_{n;F} -E(z)}{J_{0,n;F} - E(z)}\biggr) \to L(z;J).
\end{equation}
This convergence is uniform on a small circle about $z=0$, so the Taylor series coefficients 
converge. Thus \eqref{2.52}/\eqref{2.53} imply \eqref{2.54}/\eqref{2.55}. 
\enddemo

Next, we look at relations of $L(z;J)$ to certain critical functions beginning with the Jost 
function. As a preliminary, we note (recall $J^{(n)}$ is defined in \eqref{2.10a}), 

\proclaim{Proposition} \lb{P2.12} Let $\delta J$ be trace class. Then for each $z\in\bar D
\backslash\{-1,1\}${\rm ,}
\begin{equation} \label{2.59}
\lim_{n\to\infty}\, L(z; J^{(n)}) =1
\end{equation}
uniformly on compact subsets of $\bar D\backslash\{-1, 1\}$. If {\rm \eqref{2.32}} holds{\rm ,} 
{\rm \eqref{2.59}} holds uniformly in $z$ for all $z$ in $\bar D$.
\endproclaim

\demo{Proof} Use \eqref{2.19} and \eqref{2.25} with $B=0$ and the fact that $\|\delta J^{(n)}\|_1 
\to 0$ in the estimates above. 
\enddemo

Next, we note what is essentially the expansion of $\det (J-E(z))$ in minors in the 
first row:

\proclaim{Proposition} \lb{P2.13} Let $\delta J$ be trace class and $z\in\bar D\backslash 
\{-1, 1\}$. Then 
\begin{equation} \label{2.60}
L(z;J) = (E(z)-b_1) zL(z;J^{(1)}) - a_1^2 z^2 L(z; J^{(2)}).
\end{equation}
\endproclaim

\demo{Proof} Denote $(J^{(k)})_{n;F}$ by $J_{n;F}^{(k)}$, that is, the $n\times n$ matrix 
formed by rows and columns $k+1, \dots, k+n$ of $J$. Then expanding in minors,
\begin{equation} \label{2.61}
\det (E -J_{n;F}) = (E-b_1) \det (E-J_{n-1; F}^{(1)}) - a_1^2 \det 
(E-J_{n-2;F}^{(2)}).\hskip.25in
\end{equation}
Divide by $\det (E-J_{0; n;F})$ and take $n\to\infty$ using \eqref{2.58}. \eqref{2.60} 
follows if one notes
$$
\f{\det(E-J_{0;n-j;F})}{\det (E-J_{0; n;F})}\to z^j
$$
by \eqref{2.17}.
\enddemo

We now define for $z\in\bar D\backslash\{-1,1\}$ and $n=1,\dots, \infty$,
\begin{eqnarray} 
u_n (z;J) &= &\biggl(\, \prod_{j=n}^\infty a_j\biggr)^{-1} z^n L(z;J^{(n)}) \label{2.62} \\
u_0 (z;J) &= &\biggl(\, \prod_{j=1}^\infty a_j\biggr)^{-1} L(z;J). \lb{2.63}
\end{eqnarray}
$u_n$ is called the Jost solution and $u_0$ the Jost function. The infinite product 
of the $a$'s converges to a nonzero value since $a_j >0$ and $\sum_j \abs{a_j-1}<\infty$. 
We have:

\proclaim{Theorem} \lb{T2.14} The Jost solution{\rm ,} $u_n (z;J)${\rm ,} obeys 
\begin{equation} \label{2.64}
a_{n-1} u_{n-1} + (b_n -E(z)) u_n +a_n u_{n+1} =0, \quad n=1,2,\dots\hskip.5in
\end{equation}
where $a_0\equiv 1$. Moreover{\rm ,} 
\begin{equation} \label{2.65}
\lim_{n\to\infty}\, z^{-n} u_n (z;J) =1.
\end{equation}
\endproclaim 

\demo{Proof} \eqref{2.60} for $J$ replaced by $J^{(n)}$ reads
$$
L(z;J^{(n)})=(E(z)-b_{n+1})zL(z;J^{(n+1)}) - a_{n+1}^2 z^2 L(z;J^{(n+2)}), 
$$
from which \eqref{2.64} follows by multiplying by $z^n (\prod_{j=n+1}^\infty a_j)^{-1}$. 
Equation \eqref{2.65} is just a rewrite of
\eqref{2.59} because $\lim_{n\to\infty} \prod_{j=n}^\infty a_j =1$.
\enddemo

 {\it Remarks.} 1. If \eqref{2.32} holds, one can define $u_n$ for $z=\pm 1$.

\vglue4pt
2. By Wronskian methods, \eqref{2.64}/\eqref{2.65} uniquely determine $u_n (z;J)$.

\vglue8pt
Theorem~\ref{T2.14} lets us improve Theorem~\ref{T2.4}(iii) with an explicit estimate on 
the degree of $L(z;J)$.

\proclaim{Theorem} \lb{T2.15} Let $\delta J$ have range $n${\rm ,} that is{\rm ,} $a_j=1$ if $j\geq n${\rm ,} 
$b_j=0$ if $j>n$. Then $u_0(z;J)$ and so $L(z;J)$ is a polynomial in $z$ of degree at 
most $2n-1$. If $b_n\neq 0${\rm ,} then $L(z;J)$ has degree exactly $2n-1$. If $b_n=0$ but 
$a_{n-1}\neq 1${\rm ,} then $L(z;J)$ has degree $2n-2$.
\endproclaim

\demo{Proof} The difference equation \eqref{2.64} can be rewritten as 
\begin{eqnarray} 
\binom{u_{n-1}}{u_n} &= &\left( \begin{array}{ccccc} 
(E-b_n)/a_{n-1} & -a_n/a_{n-1}  \\ 1 & 0 \end{array}\right) \binom{u_n}{u_{n+1}} 
\label{2.65a}  \\
&= &\f{1}{za_{n-1}}\, A_n (z) \binom{u_n}{u_{n+1}}, \no
\end{eqnarray}
where
\begin{equation} \label{2.65b}
A_n(z)=\left( \begin{array}{ccccc} 
z^2 +1-b_n z & -a_n z \\ a_{n-1}z & 0 \end{array}\right).
\end{equation}
If $\delta J$ has range $n$, $J^{(n)}=J_0$ and $a_n =1$. Thus by \eqref{2.62}, 
$u_\ell (z;J) =z^\ell$ if $\ell\geq n$. Therefore by \eqref{2.65a}, \pagebreak

\begin{eqnarray} &&\no\\
\noalign{\vskip-30pt}
\binom{u_0}{u_1} &=& (a_1 \cdots a_{n-1})^{-1} z^{-n}\, A_1 (z) \cdots A_n(z) 
\binom{z^n}{z^{n+1}} \label{2.65c} \\
&=& (a_1 \cdots a_{n-1})^{-1} A_1 (z) \cdots A_n(z) \binom{1}{z} \no\\
&= &(a_1 \cdots a_{n-1})^{-1} A_1 (z) \cdots A_{n-1}(z) 
\binom{1-b_n z}{a_{n-1}z}. \no
\end{eqnarray}
Since $A_j(z)$ is a quadratic, \eqref{2.65c} implies $u_0$ is a polynomial of degree at 
most $2(n-1)+1=2n-1$. The top left component of $A_j$ contains $z^2$ while everything else
is of lower order.  Proceeding inductively, the top left component of $A_1 (z) \cdots 
A_{n-1}(z)$ is $z^{2n-2}+O(z^{2n-3})$. Thus if $b_n\neq 0$,
$$
u_0 =-(a_1\dots a_{n-1})^{-1} b_n z^{2n-1} + O(z^{2n-2}),
$$
proving $u_0$ has degree $2n-1$. If $b_n=0$, then
$$
A_{n-1}(z) \binom{1}{a_{n-1}z} = \binom{1+(1-a_{n-1}^2)z^2 - b_{n-1}z}{a_{n-2}z}
$$
so inductively, one sees that 
$$
u_0 = (a_1 \dots a_{n-1})^{-1} (1-a_{n-1}^2)z^{2n-2} + O(z^{2n-3})
$$
and $u_0$ has degree $2n-2$.
\enddemo

 {\it Remark.} Since the degree of $u$ is the number of its zeros (counting 
multiplicities), this can be viewed as a discrete analog of the Regge \cite{Regge}-Zworski 
\cite{Zwo} resonance counting theorem.

\vglue8pt
Recall the definitions \eqref{1.2} and \eqref{1.15a} of the $m$-function which we will 
denote for now by $M(z;J)=(E(z)-J)_{11}^{-1}$.

\proclaim{Theorem}\lb{T2.16} If $\delta J\in\calI_1$ then for $\abs{z}<1$ with $L(z;J)\neq 0${\rm ,} we have
\begin{eqnarray} 
M(z;J) &=&\f{zL(z;J^{(1)})}{L(z;J)} \label{2.66} \\ 
&=& \f{u_1 (z;J)}{u_0 (z;J)}\, . \lb{2.67}
\end{eqnarray}
\endproclaim 

\demo{Proof} \eqref{2.67} follows from \eqref{2.66} and \eqref{2.62}/\eqref{2.63}. 
\eqref{2.66} is essentially Cramer's rule. Explicitly, 
\begin{eqnarray*}
M(z;J) &=&\lim_{n\to\infty}\, (E(z) -J_{n;F})_{11}^{-1} \\
&=&\lim_{n\to\infty}\, \f{\det (E-J_{n-1; F}^{(1)})}{\det(E-J_{n;F})} \\
&=& \lim_{n\to\infty}\, w_n x_m y_n 
\end{eqnarray*}
where (by \eqref{2.58} and \eqref{2.17})
\begin{eqnarray*}
w_n &=& \f{\det(E-J_{n-1;F}^{(1)})}{\det(E-J_{0;n-1;F})}\to L(z;J^{(1)}) \\
x_n &= &\f{\det(E-J_{0;n;F})}{\det(E-J_{n;F})}\to L(z;J)^{-1} \\
y_n &=& \f{\det(E-J_{0;n-1; F})}{\det(E-J_{0;n;F})}\to z. \\
\noalign{\vskip-34pt}
\end{eqnarray*}
\enddemo
\vglue12pt
Theorem~\ref{T2.16} allows us to link $\abs{u_0}$ and $\abs{L}$ on $\abs{z}=1$ to $\Ima(M)$ 
there: 

\proclaim{Theorem} \lb{T2.17} Let $\delta J$ be trace class. Then for all $\theta\neq 0,\pi${\rm ,} 
the boundary value
$\lim_{r\uparrow 1} M (re^{i\theta};J)\equiv M(e^{i\theta};J)$ exists. Moreover{\rm ,} 
\begin{equation} \label{2.68}
\abs{u_0 (e^{i\theta};J)}^2 \Ima M(e^{i\theta}; J)=\sin\theta.
\end{equation}
Equivalently{\rm ,}
\begin{equation} \label{2.69}
\abs{L(e^{i\theta}; J)}^2 \Ima M(e^{i\theta}; J) =\biggl(\, \prod_{j=1}^\infty a_j^2 \biggr) 
\sin\theta.
\end{equation}
\endproclaim 

\demo{Proof} By \eqref{2.63}, \eqref{2.69} is equivalent to \eqref{2.68}. If $\abs{z}=1$, 
then $E(\bar z)=E(z)$ since $\bar z = z^{-1}$. Thus, $u_n (z; J)$ and $u_n (\bar z;J)$ 
solve the same difference equation. Since $z^{-n}u_n (z;J)\to 1$ and $a_n\to 1$, we have 
that
$$
a_n [u_n (\bar z; J) u_{n+1} (z; J) -u_n (z;J) u_{n+1} (\bar z; J)]\to z -z^{-1}.
$$
Since the Wronskian of two solutions is constant, if $z=e^{i\theta}$,
$$
a_n [u_n (e^{-i\theta}; J) u_{n+1} (e^{i\theta}; J) - u_n (e^{i\theta}; J) 
u_{n+1}(e^{-i\theta}; J)]=2i\sin\theta.
$$
Since $a_0=1$ and $u_n (\bar z; J) =\ol{u_n (z; J)}$, we have that 
\begin{equation} \label{2.70}
\Ima [\, \ol{u_0 (e^{i\theta}; J)}\, u_1 (e^{i\theta}; J)] =\sin\theta.
\end{equation}

\eqref{2.70} implies that $u_0 (e^{i\theta}; J)\neq 0$ if $\theta\neq 0,\pi$, so by 
\eqref{2.67}, $M(z;J)$ extends to $\bar D\backslash\{-1,1\}$. Since $u_1 (e^{i\theta}; J)
=u_0 (e^{i\theta}; J) M(e^{i\theta};J)$ (by \eqref{2.67}), \eqref{2.70} is the same as 
\eqref{2.68}. 
\enddemo

If $J$ has no eigenvalues in $\bbR\backslash [-2,2]$ and \eqref{2.32} holds so $u_0(z;J)$
has a continuation to $\bar D$, then 
\begin{eqnarray}
u_0(z;J) &=& \exp\biggl(\f{1}{2\pi} \int_0^{2\pi} \f{e^{i\theta}+z}{e^{i\theta}-z} 
\,\log \abs{u_0 (e^{i\theta}; J)}\, d\theta\biggr) \label{2.71} \\[4pt]
&= &\exp \biggl( -\f{1}{4\pi} \int_0^{2\pi} \f{e^{i\theta}+z}{e^{i\theta}-z}\, 
\log \biggl[ \f{\abs{\Ima M(e^{i\theta};J)}}{\abs{\sin\theta}}\biggr]\, d\theta 
\biggr) \lb{2.72} \\
&=& \exp \biggl( -\f{1}{4\pi} \int_0^{2\pi} \f{e^{i\theta}+z}{e^{i\theta}-z}\,
\log \biggl[ \f{\pi f(2\cos\theta)}{\abs{\sin\theta}}\biggr]\, d\theta
\biggr) \lb{2.72a}\\[4pt]
&= &(4\pi)^{-1/2}\,(1-z^2)\, D(z)^{-1}  \lb{2.73}
\end{eqnarray}
where $D$ is the Szeg\H{o} function defined by \eqref{1.33} and $f(E)=\frac{d\mu_{\ac}}{dE}$.
In the above, \eqref{2.71} 
is the Poisson-Jensen formula \cite{Rudin}.  It holds because under \eqref{2.32}, 
$u_0$ is bounded on $\bar D$ and by \eqref{2.70}, and the fact that $u_1$ is bounded, 
$\log(u_0)$ at worst has a logarithmic singularity at $\pm 1$. \eqref{2.72} follows 
from \eqref{2.68} and \eqref{2.72a} from \eqref{1.21}.  To obtain \eqref{2.73} we
use
$$
\f1{4}\, (1-z^2)^2 = \exp \biggl( \f{1}{2\pi} \int \f{e^{i\theta}+z}{e^{i\theta}-z}\, 
\log [  \sin^2\theta ]\, d\theta \biggr)
$$
which is the Poisson-Jensen formula for $\f1{2} (1-z^2)^2$ if we note that
$\abs{(1-e^{-2i\theta})^2}\break =4\sin^2\theta$.

As a final remark on perturbation theory and Jost functions, we note how easy they 
make Szeg\H{o} asymptotics for the polynomials:

\proclaim{Theorem} \lb{T2.18} Let $J$ be a Jacobi matrix with $\delta J$ trace class. 
Let $P_n(E)$ be an orthonormal polynomial associated to $J$. Then for $\abs{z}<1${\rm ,}
\begin{equation} \label{2.74}
\lim_{n\to\infty}\, z^n P_n (z+z^{-1}) = \f{u_0 (z;J)}{(1-z^2)}
\end{equation}
with convergence uniform on compact subsets of $D$.
\endproclaim 

 {\it Remarks.} 1. By looking at \eqref{2.74} near $z=0$, one gets results on the 
asymptotics of the leading coefficients of $P_n(E)$, that is, $a_{n,n-j}$ in $P_n(E) 
=\sum_{k=0}^n a_{n,k} E^k$; see Szeg\H{o} \cite{Szb}.

\vglue5pt
2. Alternatively, if $Q_n$ are the monic polynomials, 
\begin{equation} \label{2.74a}
\lim_{n\to\infty} z^n Q_n (z+z^{-1}) = \f{L(z;J)}{(1-z^2)}.
\end{equation}

\demo{Proof} This is essentially \eqref{2.58}. For let
\begin{equation} \label{2.75}
Q_n(E) = \det(E-J_{n;F}).
\end{equation}
Expanding in minors in the last rows shows 
\begin{equation} \label{2.76}
Q_n(E) =(E-b_n) Q_{n-1}(E) - a_{n-1}^2 Q_{n-2}(E)
\end{equation}
with $Q_0(E)=1$ and $Q_1(E)=E-b_1$. It follows $Q_n(E)$ is the monic orthogonal polynomial 
of degree $n$ (this is well known; see, e.g. \cite{Berez}).
Multiplying \eqref{2.75} by $(a_1, \dots, a_{n-1})^{-1}$, we see that
\begin{equation} \label{2.77}
P_n(E) = (a_1 \dots a_n)^{-1} Q_n(E)
\end{equation}
obeys \eqref{1.4} and so are the orthonormal polynomials. It follows then from \eqref{2.58} 
and \eqref{2.15} that
$$
L(z;J) = \lim_{n\to\infty}\, \f{z^{-1} -z}{z^{-(n+1)}}\, Q_n(z) 
= \lim_{n\to\infty}\, (1-z^2) z^n Q_n (z)
$$
which implies \eqref{2.74a} and, given \eqref{2.77} and $\lim_{n\to\infty} 
(a_1\cdots a_n)^{-1}$ exists, also \eqref{2.74}. 
\enddemo

\section{The sum rule: First proof}\label{S3}

Following Flaschka \cite{Fla2} and Case \cite{C1}, \cite{C2}, the Case sum rules follow from the 
construction of $L(z;J)$, the expansion \eqref{2.54} of $\log [L(z;J)]$ at $z=0$, the 
formula \eqref{2.69} for $\abs{L(e^{i\theta};J)}$, and the following standard result:

\proclaim{Proposition} \lb{P3.1} Let $f(z)$ be analytic in a neighborhood of $\bar D${\rm ,}
let $z_1, \dots, z_m$ be the zeros of $f$ in $D$ and suppose $f(0)\neq 0$. Then 
\begin{equation} \label{3.1}
\log \abs{f(0)} = \f{1}{2\pi}\int_0^{2\pi} \log \abs{f(e^{i\theta})}\, d\theta + 
\sum_{j=1}^m \log \abs{z_j}
\end{equation}
and for $n=1,2,\dots${\rm ,}
\begin{equation} \label{3.2}
\Real (\alpha_n) = \f{1}{\pi} \int_0^{2\pi} \log \abs{f(e^{i\theta})}\cos(n\theta)\, d\theta 
-\Real \biggl[\, \sum_{j=1}^m \f{z_j^{-n}- \bar z_j^n}{n}\biggr]
\end{equation}
where 
\begin{equation} \label{3.3}
\log \biggl[ \f{f(z)}{f(0)}\biggr] =  \sum_{n=1}^\infty \alpha_n z^n
\end{equation}
for $\abs{z}$ small.
\endproclaim

 {\it Remarks.} 1. Of course, \eqref{3.1} is  Jensen's formula. \eqref{3.2}
can be viewed as a 
derivative of the Poisson-Jensen formula, but the proof is so easy we give it.

\vglue4pt
2. In our applications, $\ol{f(z)} = f(\bar z)$ so $\alpha_n$ are real and the zeros are 
real or come in conjugate pairs. Therefore, $\Real$ can be dropped from both sides of \eqref{3.2} 
and the $\bar{\ }$ dropped from $\bar z_i$.

\demo{Proof} Define the Blaschke product, 
$$
B(z)=\prod_{j=1}^m \f{\abs{z_j}}{z_j}\, \f{z_j-z}{1-z\bar z_j}
$$
for which we have
\begin{eqnarray}
\log [B(z)] &= &\sum_{j=1}^m \log \abs{z_j} + \log \biggl[ \biggl( 1-\f{z}{z_j}\biggr)\biggr] 
- \log (1-z\bar z_j) \label{3.5}  \\
&=& \sum_{j=1}^m \log \abs{z_j} - \sum_{n=1}^\infty {z^n} \sum_{j=1}^m \, 
\frac{z_j^{-n} - \bar z_j^n}{n}\, . \no
\end{eqnarray}
By a limiting argument, we can suppose $f$ has no zeros on $\partial D$. Then $f(z)/B(z)$ 
is nonvanishing in a neighborhood of $\bar D$,  so $g(z)\equiv\log [f(z)/B(z)]$ is analytic
there and by \eqref{3.3}/\eqref{3.5}, its Taylor series
$$
g(z) = \sum_{n=0}^\infty c_n z^n
$$
\vglue-8pt\noindent 
has coefficients
\begin{eqnarray*}
c_0 &= &\log [f(0)] - \sum_{j=1}^m \log \abs{z_j} \\
c_n &= &\alpha_n + \sum_{j=1}^m \f{[z_j^{-n} - \bar z_j^n]}{n}\, .
\end{eqnarray*}
Substituting $d\theta = \f{dz}{iz}$ and $\cos(n\theta)=\f12 (z^n+ z^{-n})$ in the Cauchy
integral formula,
$$
\f{1}{2\pi i} \int_0^{2\pi} g(z) \,\f{dz}{z^{n+1}} = \left\{ \begin{array}{ll} 
c_n &\hbox{if } n\geq 0 \\
0 &\hbox{if } n\leq -1, \end{array}\right. 
$$
we get integral relations whose real part is \eqref{3.1} and \eqref{3.2}.  
\enddemo

While this suffices for the basic sum rule for finite range $\delta J$, which is the 
starting point of our analysis, we note three extensions:

\vglue6pt
(1) \ If $f(z)$ is meromorphic in a neighborhood of $\bar D$ with zeros $z_1, \dots, 
z_m$ and poles $p_1, \dots, p_k$, then \eqref{3.1} and \eqref{3.2} 
remain true so long as one makes the changes:
\begin{eqnarray}
\sum_{j=1}^m \log\abs{z_j}  &\mapsto& \sum_{j=1}^m \log \abs{z_j} 
- \sum_{j=1}^k \log \abs{p_j}  \label{3.6} \\
\sum_{j=1}^m \f{z_j^{-n} -\bar z_j^n}{n}  &\mapsto&\sum_{j=1}^m 
\f{z_j^{-n} -\bar z_j^n}{n} - \sum_{j=1}^k \f{p_j^{-n} - \bar p_j^n}{n} \lb{3.7} 
\end{eqnarray}
for we write $f(z) = f_1(z)/\prod_{j=1}^k (z-p_j)$ and apply Proposition~\ref{P3.1} 
to $f_1$ and to $\prod_{j=1}^k (z-p_j)$. We will use this extension in the next section. 

\vglue6pt
(2) \ If $f$ has continuous boundary values on $\partial D$, we know its zeros in $D$ 
obey $\sum_{j=1}^\infty (1-\abs{z_j})<\infty$ (so the Blaschke product converges) and we 
have some control on $-\log \abs{f(re^{i\theta})}$ as $r\uparrow 1$, one can prove 
\eqref{3.1}--\eqref{3.2} by a limiting argument. We could use this to extend the proof 
of Case's inequalities to the situation $\sum n[\abs{a_n-1}+\abs{b_n}]<\infty$. We first 
use a Bargmann bound (see \cite{CN}, \cite{Ger1}, \cite{Ger2}, \cite{HS}) to see there are only finitely many zeros 
for $L$ and \eqref{2.69} to see the only place $\log\abs{L}$ can be singular is at $\pm 1$. 
The argument in Section~\ref{S9} that $\sup_r \int [\log_- \abs{L(re^{i\theta})}]^2\,d\theta 
<\infty$ lets us control such potential singularities. Since Section~\ref{S9} will have a 
proof in the more general case of trace class $\delta J$, we do not provide the details. 
But we would like to emphasize that proving the sum rules in generality Case claims in 
\cite{C1}, \cite{C2} requires overcoming technical issues he never addresses.

\vglue6pt
(3) \ The final (one might say ultimate) form of \eqref{3.1}/\eqref{3.2} applies when $f$
is a Nevanlinna function, that is, $f$ is analytic in $D$ and
\begin{equation} \label{3.8}
\sup_{0<r<1}\,  \f{1}{2\pi} \int_0^{2\pi} \log_+ \abs{f(re^{i\theta})}\, d\theta <\infty,
\end{equation}
where $\log_+ (x)=\max(\log(x),0)$. If $f$ is Nevanlinna, then (\cite[pg.~311]{Rudin}; 
essentially one uses \eqref{3.1} for $f(z/r)$ with $r<1$), 
\begin{equation} \label{3.9}
\sum_{j=1}^\infty (1-\abs{z_j})< \infty
\end{equation}
and (\cite[pg.~310]{Rudin}) the Blaschke product converges. Moreover (see 
\cite[pp.~247, 346]{Rudin}), there is a finite real measure $d\mu^{(f)}$ on $\partial D$ so
\begin{equation} \label{3.10}
\log\abs{f(re^{i\theta})}\, d\theta\to d\mu^{(f)} (\theta)
\end{equation}
weakly, and for Lebesgue a.e.~$\theta$,
\begin{equation} \label{3.11}
\lim_{r\uparrow 1} \, \log\abs{f(re^{i\theta})}=\log\abs{f(e^{i\theta})}
\end{equation}
and
\begin{equation} \label{3.12}
d\mu^{(f)}(\theta) = \log \abs{f(e^{i\theta})}\, d\theta + d\mu_s^{(f)}(\theta)
\end{equation}
where $d\mu_s^{(f)}(\theta)$ is singular with respect to Lebesgue measure $d\theta$ on 
$\partial D$. $d\mu_s^{(f)}(\theta)$ is called the singular inner component.

By using \eqref{3.1}/\eqref{3.2} for $f(z/r)$ with $r\uparrow 1$ and \eqref{3.10}, we 
immediately have:

\proclaim{Theorem} \lb{T3.2} Let $f$ be a Nevanlinna function on $D$ and let $\{z_j\}_{j=1}^N$ 
{\rm{(}}$N=1,2,\dots${\rm ,} or $\infty${\rm{)}} be its zeros. Suppose $f(0)\neq 0$. Let $\log 
\abs{f(e^{i\theta})}$ be the {\rm a.e.}~boundary values of $f$ and $d\mu_s^{(f)}(\theta)$ 
the singular inner component. Then
\begin{equation} \label{3.13}
\log \abs{f(0)} =\f{1}{2\pi} \int_0^{2\pi} \log \abs{f(e^{i\theta})}\, d\theta 
+ \f{1}{2\pi} \int_0^{2\pi} d\mu_s^{(f)}(\theta) + \sum_{j=1}^N \log\abs{z_j}\qquad
\end{equation}
and for $n=1,2, \dots\, ${\rm ,}
\begin{eqnarray} \label{3.14}
\Real (\alpha_n) &=&\f{1}{\pi} \int_0^{2\pi} \log \abs{f(e^{i\theta})}\cos(n\theta)\, d\theta \\
&& +\ \f{1}{\pi} \int_0^{2\pi} \cos(n\theta)\, d\mu_s^{(f)}(\theta) 
 - \Real \biggl[ \sum_{j=1}^N \f{z_j^{-n} -\bar z_j^n}{n}\biggr]
\no
\end{eqnarray}
where $\alpha_n$ is given by {\rm \eqref{3.3}. }
\endproclaim 

We will use this form in Section~\ref{S9}. 
\vglue6pt
Now suppose that $\delta J$ has finite range, and apply Proposition~\ref{P3.1} to $L(z;J)$. 
Its zeros in $D$ are exactly the image under $E\to z$ of the (simple) eigenvalues of $J$ 
outside $[-2,2]$ (Theorem~\ref{T2.4}(ii)). The expansion of $\log [L(z;J)]$ at $z=0$ is 
given by Theorem~\ref{T2.12} and $\log \abs{L(e^{i\theta}; J)}$ is given by \eqref{2.69}. 
We have thus proven:

\proclaimtitle{Case's Sum Rules: Finite Rank Case}
\proclaim{Theorem} \lb{T3.3} Suppose $\delta J$ has finite 
rank. Then{\rm ,} with $\abs{\beta_1(J)}\geq\abs{\beta_2 (J)}\geq\cdots > 1$ defined so that
$\beta_j^{} + \beta_j^{-1}$ are the eigenvalues of $J$ outside $[-2,2]${\rm ,} we have
\begin{eqnarray} 
\qquad C_0:&  & \f{1}{4\pi} \int_0^{2\pi} \log \biggl( \f{\sin\theta}{\Ima M(e^{i\theta})} 
\biggr) \, d\theta = \sum_j \log\abs{\beta_j} - \sum_{n=1}^\infty 
\log(a_n)  \label{3.15} \\
\qquad C_n:& &  -\f{1}{2\pi} \int_0^{2\pi} \log \biggl( \f{\sin\theta}{\Ima M(e^{i\theta})}\biggr)
	\cos(n\theta)\, d\theta \lb{3.16} \\
& &\hskip.4in =  -\f{1}{n} \sum_j (\beta_j^n - \beta_j^{-n}) +\f{2}{n} \,
	\Tra \Bigl( T_n \bigl(\tfrac{1}{2}J\bigr) - T_n \bigl( \tfrac{1}{2}J_0 \bigr) \Bigr). \no 
\end{eqnarray}

In particular{\rm , }
\begin{equation} \label{3.17}
P_2: \f{1}{2\pi} \int_0^{2\pi} \log \biggl( \f{\sin\theta}{\Ima M}\biggr) \sin^2 \theta
\, d\theta + \sum_j F(e_j) = \f{1}{4} \sum_n b_n^2 + \f12 \sum_n G(a_n), \quad
\end{equation}
where
\begin{equation} \label{3.18}
G(a) = a^2 - 1 - \log(a^2)
\end{equation}
and
\begin{equation} \label{3.19}
F(e) =\tfrac14 (\beta^2 - \beta^{-2}-\log \abs{\beta}^4); \qquad e=\beta + \beta^{-1},\ |\beta|>1.\hskip.5in
\end{equation}
\endproclaim  

{\it Remarks.} 1. Actually, when $\delta J$ is finite rank all eigenvalues must
lie outside $[-2,2]$ ---it is easily checked that the corresponding difference equation has no
(nonzero) square summable solutions.
While eigenvalues may occur at $-2$ or $2$ when \pagebreak $\delta J\in \calI_1$,
there are none in $(-2,2)$.  This follows from the fact that
$\lim_{r\uparrow 1} M(re^{i\theta};J)$ exists for $\theta\in(0,\pi)$ (see Theorem~\ref{T2.17}) 
or alternately from the fact that one can construct two independent solutions $u_n 
(e^{\pm i\theta}, J)$ whose linear combinations are all non-$L^2$. 

\vglue6pt
2. In \eqref{2.69},
$$
\log\abs{L} = \f12 \,\log \biggl| \f{\sin\theta}{\Ima M}\biggr| + \sum_{n=1}^\infty 
\log a_n,
$$
the $\sum_{n=1}^\infty \log a_n$ term is constant and so contributes only to $C_0$ because\break
$\int_0^{2\pi} \cos(n\theta)\, d\theta =0$.

\vglue6pt
3. As noted, $P_2$ is $C_0 + \f12 C_2$.

\vglue6pt
4. We have looked at the combinations of sum rules that give $\sin^4 \theta$ and $\sin^6 
\theta$ hoping for another miracle like the one below that for $\sin^2\theta$, the function 
$G$ and $F$ that result are positive. But we have not found anything but a mess of 
complicated terms that are not in general positive.

\vglue6pt
$P_2$ is especially useful because of the properties of $G$ and $F$:  

\proclaim{Proposition} \lb{P3.4} The function $G(a)=a^2-1-2\log(a)$ for 
$a\in (0,\infty)$ is nonnegative and vanishes only at $a=1$. For $a-1$ small{\rm ,}
\begin{equation} \label{3.20}
G(a) = 2(a-1)^2 + O((a-1)^3).
\end{equation}
\endproclaim

\demo{Proof} By direct calculations, $G(1)=G'(1)=0$ and
$$
G''(a) = \f{2(1+a^2)}{a^2} \geq 2
$$
so $G(a)\geq (a-1)^2$ (since $G(1)=G'(1)=0$). 
\eqref{3.20} follows from $G''(1)=4$. 
\enddemo

\proclaim{Proposition} \lb{P3.5} The function $F(e)$ given by {\rm \eqref{3.19}} is positive
throughout its domain{\rm ,} $\{\abs{e}>2\}$.
It is even{\rm ,} increases with increasing $|e|${\rm ,} and  for $\abs{e}-2$ small{\rm ,}
\begin{equation} \label{3.21}
F(e) = \tfrac23 (\abs{e}-2)^{3/2} + O((\abs{e}-2)^2).
\end{equation}
In addition{\rm ,}
\begin{equation} \label{3.22}
F(e) \leq \tfrac23 (e^2 -4)^{3/2}.
\end{equation}
\endproclaim

\demo{Proof} Let $R(\beta) =\f14 (\beta^2 - \beta^{-2} -\log \abs{\beta}^4)$ for $\beta\geq 1$ 
and compute
$$
R'(\beta) =\f12 \biggl(\beta + \beta^{-3} -\f{2}{\beta}\biggr) =\f12\biggl(\f{\beta+1}{\beta} 
\biggr)^2 \f{1}{\beta}\, (\beta-1)^2.
$$
This shows that $R(\beta)$ is increasing.  It also follows that
$$
R'(\beta) = 2(\beta-2)^2 + O((\beta-1)^3)
$$
and since $\beta\geq 1$, $(\beta +1)/\beta \leq 2$ and $\beta^{-1} \leq 1$ so
$$
R'(\beta)\leq 2(\beta-1)^2.
$$
As $R(1)=0$, we have
\begin{equation} \label{3.23}
R(\beta) \leq \tfrac23\, (\beta -1)^3
\end{equation}
and
\begin{equation} \label{3.24}
R(\beta) =\tfrac23\, (\beta-1)^3 + O((\beta-1)^4).
\end{equation}
Because $F(-e)=F(e)$, which is simple to check, we can suppose $e>2$ so $\beta >1$.
As $\beta=\f12 [e+\sqrt{e^2-4}\,]$ is an increasing function of $e$, $F(e)=R(\beta)$ is an
increasing function of $e>2$.  Moreover,
$\beta -1=(e-2)^{1/2} + O(e-2)$ and so \eqref{3.24} implies \eqref{3.21}. Lastly,
$$
(\beta -1) \leq \beta - \beta^{-1} =\sqrt{e^2-4}\, ,
$$
so \eqref{3.23} implies \eqref{3.22}.
\enddemo

\section{The sum rule: Second proof}\label{S4}

In this section, we will provide a second proof of the sum rules that never mentions a 
perturbation determinant or a Jost function explicitly. We do this not only because it 
is nice to have another proof, but because this proof works in a situation where we 
{\it a~priori} know the $m$-function is analytic in a neighborhood of $\bar D$ and the other 
proof does not apply.  And this is a situation we will meet in proving Theorem~6. On the 
other hand, while we could prove Theorems~1, 2, 3, 5, 6 without Jost functions, we 
definitely need them in our proof in Section~\ref{S9} of the $C_0$-sum rule for the 
trace class case.

The second proof of the sum rules is based on the continued fraction expansion of $m$ 
\eqref{1.5}. Explicitly, we need,
\begin{equation} \label{4.1}
-M(z;J)^{-1} = -(z+z^{-1}) + b_1 + a_1^2 M(z;J^{(1)})
\end{equation}
which one obtains either from the Weyl solution method of looking at $M$ (see \cite{GS}, \cite{S270}) 
or by writing $M$ as a limit of ratio of determinants
\begin{equation} \label{4.2}
M(z;J)=\lim_{n\to\infty}\, \f{\det (E(z) - J_{n-1;F}^{(1)})}{\det(E(z) - J_{n;F})}
\end{equation}
and expanding the denominator in minors in the first row. For any $J$, \eqref{4.1} holds 
for $z\in D$. Suppose that we know $M$ has a meromorphic continuation to a neighborhood 
of $\bar D$ and consider \eqref{4.1} with $z=e^{i\theta}$:
\begin{equation} \label{4.2a}
-M(e^{i\theta}; J)^{-1} = -2\cos\theta + b_1 + a_1^2 M(e^{i\theta}; J^{(1)}).
\end{equation}
Taking imaginary parts of both sides,
\begin{equation} \label{4.3}
\f{\Ima M(e^{i\theta}; J)}{\abs{M(e^{i\theta}; J)}^2} = a_1^2 \Ima M(e^{i\theta}; J^{(1)}) 
\end{equation}
or, letting
$$
g(z;J) = \f{M(z;J)}{z}
$$
({\it Note}: because 
\begin{equation} \label{4.3a}
M(z;J) = (z+z^{-1} -J)_{11}^{-1} = z(1+z^2 - zJ)_{11}^{-1} = z+O(z^2)\hskip.4in
\end{equation}
near zero, $g$ is analytic in $D$), we have
\begin{equation}  \lb{4.3b}
\f12\biggl[\log\biggl(\f{\Ima M(e^{i\theta};J)}{\sin\theta}\biggr) - 
\log \biggl( \f{\Ima M(e^{i\theta}; J^{(1)})}{\sin\theta}\biggr)\biggr] = 
\log a_1 +\log \abs{g(e^{i\theta};J)}.
\end{equation}

To see where this is heading, 

\proclaim{Theorem} \lb{T4.1} \hskip-8pt Suppose $M(z;J)$ is meromorphic in a neighborhood of~$\bar D$. Then $J$ and
$J^{(1)}$ have finitely many eigenvalues outside $[-2,2]$ and if 
\begin{equation} \label{4.4}
C_0(J) =\f{1}{4\pi}\int_0^{2\pi} \log\biggl( \f{\sin\theta}{\Ima M(e^{i\theta}; J)} \biggr)
d\theta - \sum_{j=1}^N \log \abs{\beta_j (J)}
\end{equation}
{\rm{(}}with $\beta_j$ as in Theorem~{\rm{\ref{T3.3})},} then
\begin{equation} \label{4.5}
C_0 (J) =-\log(a_1) + C_0 (J^{(1)}).
\end{equation}
In particular{\rm ,} if $\delta J$ is finite rank{\rm ,} then the $C_0$ sum rule holds\/{\rm :}
\begin{equation} \label{4.6}
C_0(J) =-\sum_{n=1}^\infty \log (a_n).
\end{equation}
\endproclaim 

\demo{Proof} The eigenvalues, $E_j$, of $J$ outside $[-2,2]$ are precisely the poles of 
$m(E;J)$ and so the poles of $M(z; J)$ under $E_j^{}=z_j^{} + z_j^{-1}$. By \eqref{4.1}, 
the poles of $M(z;J^{(1)})$ are exactly the zeros of $M(z;J)$. Thus $\{\beta_j(J)^{-1}\}$ 
are the poles of $M(z;J)$ and $\{\beta_j (J^{(1)})^{-1}\}$ are its zeros. Since $g(0;J)
=1$ by \eqref{4.3a}, \eqref{3.1}/\eqref{3.6} becomes
$$
\f1{2\pi} \int \log (\abs{g(e^{i\theta}, J)}\,d\theta = -\sum_j \log (\abs{\beta_j(J)}) 
+\sum_j \log (\abs{\beta_j (J^{(1)})}).
$$
\eqref{4.3b} and this formula imply \eqref{4.5}. By \eqref{4.2a}, if $M(z;J)$ is meromorphic 
in a neighborhood of $\bar D$, so is $M(z;J^{(1)})$. So we can iterate \eqref{4.5}. The free 
$M$ function is
\begin{equation} \label{4.7}
M(z;J_0)=z
\end{equation}
(e.g., by \eqref{2.13} with $m=n=1$), so $C_0 (J_0)=0$ and thus, if $\delta J$ is finite rank, 
the remainder is zero after finitely many steps. 
\enddemo

To get the higher-order sum rules, we need to compute the power series for $\log (g(z;J))$ 
about $z=0$. For low-order, we can do this by hand. Indeed, by \eqref{4.1} and \eqref{4.3a} 
for $J^{(1)}$,
\begin{eqnarray*}
g(z;J) &=& (z[(z+z^{-1}) -b_1 - a_1^2 z+O(z^2)])^{-1} \\
&=& (1-b_1z -(a_1^2 -1)z^2 + O(z^3))^{-1} \\
&= &1+b_1 z + ((a_1^2 -1) + b_1^2)z^2 + O(z^3)
\end{eqnarray*}
so since $\log (1+w)=w - \frac{1}{2}w^2 + O(w^3)$,
\begin{equation} \label{4.8}
\log (g(z;J)) = b_1z + ({\textstyle \frac{1}{2}}\, b_1^2 + a_1^2 -1) z^2 + O(z^3).
\end{equation}
Therefore, by mimicking the proof of Theorem~\ref{T4.1}, but using \eqref{3.2}/\eqref{3.7} 
in place of \eqref{3.1}/\eqref{3.6}, we have

\proclaim{Theorem} \lb{T4.2} \hskip-8pt Suppose $M(z;J)$ is meromorphic in a neighborhood of~$\bar D$. 
Let
\begin{equation} \label{4.9}
C_n (J) =-\f{1}{2\pi}\int_0^{2\pi} \log \biggl( \f{\sin\theta}{\Ima M(e^{i\theta})}\biggr) 
\cos(n\theta) + \f{1}{n} \biggl[\, \sum_{j} \beta_j(J)^n -\beta_j(J)^{-n}\biggr].
\end{equation}
Then
\begin{eqnarray}
C_1(J) &=& b_1 + C_1 (J^{(1)})  \label{4.10} \\
C_2(J) &=& [{\textstyle \frac{1}{2}} \, b_1^2 + (a_1^2 -1)\ + C_2 (J^{(1)})]. \lb{4.11}
\end{eqnarray}
If 
\begin{equation} \label{4.12}
P_2 (J) =\f{1}{2\pi} \int_0^{2\pi} \log \biggl(\f{\sin\theta}{\Ima M (e^{i\theta})}\biggr) 
\sin^2\theta\, d\theta +\sum_j F(e_j(J)) \hskip.5in
\end{equation}
with $F$ given by {\rm \eqref{3.19},} then writing $G(a)=a^2-1-2\log(a)$ as in {\rm \eqref{3.18}}
\begin{equation} \label{4.13}
P_2 (J) = \tfrac14\, b_1^2 + {\textstyle \frac{1}{2}}\, G(a_1) + P_2 (J^{(1)}).
\end{equation}
In particular{\rm ,} if $\delta J$ is finite rank{\rm ,} we have the sum rules $C_1, C_2, P_2$ of 
{\rm \eqref{3.16}/\eqref{3.17}.}
\endproclaim 

To go to order larger than two, we expand $\log(g(z;J))$ systematically as follows: 
We begin by noting that by \eqref{4.2} (Cramer's rule),
\begin{equation} \lb{4.17} 
g(z;J) = \lim_{n\to\infty}\, g_n (z;J) 
\end{equation}
where
\begin{eqnarray}
g_n (z;J) &= &\f{z^{-1} \det (z+z^{-1} - J_{n-1;F}^{(1)})}{\det(z+z^{-1} - J_{n;F})} \lb{4.18} \\
&=& \f{1}{1+z^2} \, \f{\det(1-E(z)^{-1} J_{n-1;F}^{(1)})}{\det(1-E(z)^{-1} J_{n;F})} \lb{4.19} 
\end{eqnarray}
where we used $z(E(z))=1+z^2$ and the fact that because the numerator has a matrix of order 
one less than the denominator, we get an extra factor of $E(z)$. We now use Lemma~\ref{L2.10},
writing $F_j(x)$ for $\tfrac{2}{j}[T_j(0)-T_j(x/2)]$,
\begin{eqnarray}   &&\lb{4.20} \\
\noalign{\vskip-6pt}
\log g_n (z;J) &\nhs=\nhs&-\log (1+z^2) + \sum_{j=1}^\infty z^j \Big[\Tra \big(F_j(J_{n-1;F}^{(1)})\big) - 
\Tra\big(F_j (J_{n;F})\big)\Big]  \no\\
&\nhs=\nhs&-\log (1+z^2) - \sum_{j=1}^\infty \f{z^{2j}}{j}\, (-1)^j  \lb{4.21}   \\
&\nhs\nhs& + \ \sum_{j=1}^\infty \f{2z^j}{j}\, \Bigl[ \Tra \Bigl(T_j\big(\tfrac{1}{2}J_{n;F}\big)\Bigr)
	- \Tra \Bigl( T_j\big(\tfrac{1}{2}J_{n-1;F}^{(1)}\big) \Bigr)\Bigr] \no
\end{eqnarray}
where we picked up the first sum because $J_{n;F}$ has dimension one greater than 
$J_{n-1;F}^{(1)}$ so the $T_j(0)$ terms in $F_j(J_{n;F})$ and $J_{n-1;F}^{(1)}$ contribute
differently.  Notice 
$$
\sum_{j=1}^\infty \f{z^{2j}}{j}\, (-1)^j = -\log (1+z^2)
$$
so the first two terms cancel! Since $g_n (z;J)$ converges to $g(z;J)$ in a neighborhood 
of $z=0$, its Taylor coefficients converge. Thus

\proclaim{Proposition} \lb{P4.3} For each $j${\rm ,} 
\begin{equation} \lb{4.22} 
\alpha_j (J,J^{(1)}) =\lim_{n\to\infty} \Bigl[ \Tra \Bigl(T_j \bigl(\tfrac{1}{2} J_{n;F} 
\bigr) \Bigr) - \Tra \Bigl( T_j\bigl(\tfrac{1}{2}J_{n-1;F}^{(1)}\bigr) \Bigr)\Bigl] \hskip.5in
\end{equation}
exists{\rm ,} and for $z$ small{\rm ,}
\begin{equation} \lb{4.23} 
\log g(z;J) = \sum_{j=1}^\infty \f{2z^j}{j}\, \alpha_j (J,J^{(1)}).
\end{equation}
\endproclaim

 {\it Remark.} Since
$$
(J^{(1)\ell})_{mm} = (J^\ell)_{m+1\, m+1}
$$
if $m\geq \ell$, the difference of traces on the right side of \eqref{4.22} is constant for 
$n >j$, so one need not take the limit.
\vglue8pt 

Plugging this into the machine that gives Theorem~\ref{T4.1} and Theorem~\ref{T4.2}, we obtain 

\proclaim{Theorem} \lb{T4.4} \hskip-8pt Suppose $M(z;J)$ is meromorphic in a neighborhood of~$\bar D$.
 Let $C_n (J)$ be given by {\rm \eqref{4.9}} and $\alpha$ by {\rm \eqref{4.22}. }
Then
\begin{equation} \lb{4.24} 
C_n (J) = \f{2}{n}\, \alpha_n (J,J^{(1)}) + C_n (J^{(1)}).
\end{equation}
In particular{\rm ,} if $\delta J$ is finite rank{\rm ,} we have the sum rule $C_n$ of {\rm \eqref{3.16}.} 
\endproclaim 

\demo{Proof} The only remaining point is why if $\delta J$ is finite rank, we have recovered 
the same sum rule as in \eqref{3.16}. Iterating \eqref{4.24} when $J$ has rank $m$ gives 
\begin{eqnarray} 
C_n (J) &= &\f{2}{n} \sum_{j=1}^m \alpha_n (J^{(j-1)}, J^{(j)})  \lb{4.25}  \\[6pt]
&=& \lim_{\ell\to\infty} \f{2}{n} \Bigl[ \Tra \bigl[ T_n \bigl( \tfrac{1}{2} J_{\ell;F}\bigr) 
-T_n \bigl( \tfrac{1}{2} J_{0,\ell-m;F}\bigr) \bigr] \Bigr]\no
\end{eqnarray}
while \eqref{3.16} reads 
\begin{eqnarray} 
C_n(J) &=&\f{2}{n} \Tra \Bigl[ T_n \bigl( \tfrac{1}{2} J\bigr) - T_n \bigl(\tfrac{1}{2}J_0\bigr) 
\Bigr] \lb{4.26}  \\[6pt]
&=&\lim_{\ell\to\infty} \f{2}{n} \Bigl[ \Tra\bigl[ T_n\bigl( \tfrac{1}{2}J_{\ell;F}\bigr)\bigr] - 
\Tra\bigl[ T_n\bigl( \tfrac{1}{2} J_{0,\ell;F} \bigr)\bigr] \Bigr]. \no
\end{eqnarray}
That \eqref{4.25} and \eqref{4.26} are the same is a consequence of Proposition~\ref{P2.3a}. 
\enddemo

\vglue-8pt
\section{Entropy and lower semicontinuity \\ of the Szeg\H{o} and 
	quasi-Szeg\H{o} terms}\label{S5}
\vglue-4pt

In the sum rules $C_0$ and $P_2$ of most interest to us, there appear two terms involving 
integrals of logarithms:
\begin{equation} \label{5.1}
Z(J) =\f{1}{4\pi} \int_0^{2\pi} \log \biggl( \f{\sin\theta}{\Ima M(e^{i\theta}, J)}
\biggr)\, d\theta
\end{equation}
and
\begin{equation} \label{5.2}
Q(J) =\f{1}{2\pi} \int_0^{2\pi} \log \biggl( \f{\sin\theta}{\Ima M(e^{i\theta}, J)} 
\biggr) \sin^2\theta\, d\theta.
\end{equation}

One should think of $M$ as related to the original spectral measure on $\sigma (J)\supset 
[-2,2]$ as
\begin{equation} \label{5.3}
\Ima M(e^{i\theta}) =  \pi\,\f{d\mu_{\ac}}{dE}\, (2\cos\theta)
\end{equation}
in which case, \eqref{5.1}, \eqref{5.2} can be rewritten
\begin{equation} \label{5.4}
Z(J) = \f{1}{2\pi} \int_{-2}^2 \log \biggl( \f{\sqrt{4-E^2}}{2\pi\, d\mu_{\ac}/dE}\biggr) 
\f{dE}{\sqrt{4-E^2}}
\end{equation}
and
\begin{equation} \label{5.4a}
Q(J) = \f{1}{4\pi} \int_{-2}^2 \log \biggl( \f{\sqrt{4-E^2}}{2\pi\, d\mu_{\ac}/dE} 
\biggr)\, \sqrt{4-E^2}\, dE.
\end{equation}
Our main result in this section is to view $Z$ and $Q$ as functions of $\mu$  \pagebreak and to 
prove if $\mu_n\to \mu$ weakly, then $Z(\mu_n)$ (resp.~$Q(\mu_n)$) obeys
\begin{equation} \label{5.5}
Z(\mu) \leq \liminf Z(\mu_n); \qquad Q(\mu) \leq \liminf Q(\mu_n),
\end{equation}
that is, that $Z$ and $Q$ are weakly lower semicontinuous. This will let us prove sum 
rule-type inequalities in great generality.

The basic idea of the proof will be to write variational principles for $Z$ and $Q$ as 
suprema of weakly continuous functions. Indeed, as Totik has pointed out to us, 
Szeg\H{o}'s theorem (as extended to the general, not only a.c., case \cite{Akh}, \cite{Garnett})
gives what is essentially $Z(J)$ by a variational principle; explicitly,
\begin{equation} \label{5.6}
\exp\biggl\{ \f{1}{2\pi} \int_0^{2\pi} \log \biggl( \f{d\mu_{\ac}}{d\theta}\biggr)
d\theta\biggr\} = \inf_P \biggl[\f{1}{2\pi}\int_{-\pi}^\pi \abs{P(e^{i\theta})}^2\, 
d\mu(\theta)\biggr]
\end{equation}
where $P$ runs through all polynomials with $P(0)=1$, which can be used to prove the 
semicontinuity we need for $Z$. It is an interesting question of what is the relation 
between \eqref{5.6} and the variational principle \eqref{5.14a} below. It also would be 
interesting to know if there is an analog of \eqref{5.6} to prove semicontinuity of $Q$.

We will deduce the semicontinuity by providing a variational principle. We originally found 
the variational principle based on the theory of Legendre transforms, then realized that 
the result was reminiscent of the inverse Gibbs variation principle for entropy (see 
\cite[pg.~271]{SSM} for historical remarks; the principle was first written down by 
Lanford-Robinson \cite{LR}) and then realized that the quantities of interest to us 
aren't merely reminiscent of entropy, they are exactly relative entropies where $\mu$ 
is the second variable rather than the first one that is usually varied. We have located 
the upper semicontinuity of the relative entropy in the second variable in the literature 
(see, e.g., \cite{CKZ}, \cite{KL}, \cite{OP}), but not in the generality we need it, so especially since 
the proof is easy, we provide it below. We use the notation 
\begin{equation} \label{5.7}
\log_\pm (x) = \max(\pm\log(x),0).
\end{equation}

\vglue12pt
\demo{{D}efinition} Let $\mu,\nu$ be finite Borel measures on a compact Hausdorff 
space, $X$. We define the entropy of $\mu$ relative to $\nu$, $S(\mu\mid \nu)$, by 
\begin{equation} \lb{5.8}
S(\mu\mid\nu) = \left\{ \begin{array}{ll} 
-\infty &\hbox{if $\mu$ is not $\nu$-ac} \\
-\int \log (\f{d\mu}{d\nu}) d\mu &\hbox{if $\mu$ is $\nu$-ac}.
\end{array}\right. 
\end{equation}
\enddemo

{\it Remarks.} 1. Since $\log_- (x) = \log_+ (x^{-1})\leq x^{-1}$ and 
$$
\int\biggl(\f{d\mu}{d\nu}\biggr)^{-1} d\mu = \nu \biggl( \biggl\{ x\biggm| \f{d\mu}{d\nu} 
\neq 0 \biggr\}\biggr) \leq \nu(X) <\infty,
$$
the integral \pagebreak in \eqref{5.8} can only diverge to $-\infty$, not to $+\infty$.

2. If $d\mu = f\, d\nu$, then
\begin{equation} \label{5.9}
S(\mu\mid\nu) = -\int f\log(f)\, d\nu,
\end{equation}
the more usual formula for entropy.

\proclaim{Lemma}\lb{L5.1} Let $\mu$ be a probability measure. Then   
\begin{equation} \label{5.10}
S(\mu\mid\nu) \leq \log\nu(X).
\end{equation}
In particular{\rm ,} if $\nu$ is also a probability measure{\rm ,}
\begin{equation} \label{5.11}
S(\mu\mid\nu) \leq 0.
\end{equation}
Equality holds in {\rm \eqref{5.11}} if and only if $\mu=\nu$.
\endproclaim

\demo{Proof} If $\mu$ is not $\nu$-ac, \eqref{5.10}/\eqref{5.11} is trivial, so suppose 
$\mu=f\, d\nu$ and let
\begin{equation} \label{5.11a}
d\tilde\nu = \chi_{\{x\mid f(x)\neq 0\}}\, d\nu
\end{equation}
so $\tilde\nu$ and $\mu$ are mutually ac. Then, 
\begin{eqnarray}
S(\mu\mid\nu) &= &\int \log \biggl( \f{d\tilde \nu}{d\mu}\biggr) \, d\mu   \lb{5.12} \\
&\leq& \log \biggl( \int \biggl( \f{d\tilde\nu}{d\mu}\biggr) d\mu\biggr)\no \\
&= &\log \tilde\nu (X) \lb{5.13} \\
&\leq& \log \nu(X) \no
\end{eqnarray}
where we used Jensen's inequality for the concave function $\log (x)$. For equality to 
hold in \eqref{5.11}, we need equality in \eqref{5.13} (which says $\nu=\tilde\nu$) and 
in \eqref{5.12}, which says, since $\log$ is strictly convex, that $d\nu/d\mu$ is a 
constant. When $\nu(X)=\mu(X)=1$, this says $\nu=\mu$. 
\enddemo

\proclaim{Theorem} \lb{T5.2} For all $\mu,\nu${\rm ,} 
\begin{equation} \label{5.14a}
S(\mu\mid\nu) =\inf \biggl[ \int F(x)\, d\nu - \int (1+\log F)\, d\mu(x)\biggr] 
\end{equation}
where the $\inf$ is taken over all real\/{\rm -}\/valued continuous functions $F$ with  
$\min_{x\in X} F(x)\break >0$.
\endproclaim 

\demo{Proof} Let us use the notation 
$$
\calG (F, \mu,\nu) = \int F(x)\, d\nu - \int (1+\log F)\, d\mu(x)
$$
for any nonnegative function $F$ with $F\in L^1(d\nu)$ and $\log F\in L^1 (d\mu)$.

Suppose first that $\mu$ is $\nu$-ac with $d\mu=f\,d\nu$ and $F$ is positive and continuous.
Let $A=\{x\mid f(x) \neq 0\}$ and define $\tilde\nu$ by \eqref{5.11a}.
As $\log (a)$ is concave, $\log(a) \leq a-1$ so for $a,b >0$,
\begin{equation} \label{5.13a}
ab^{-1} \geq 1 + \log (ab^{-1}) = 1 + \log (a)- \log b.
\end{equation}
Thus for $x\in A$, 
$$
F(x) f(x)^{-1} \geq 1 + \log F(x) -\log f(x).
$$
Integrating with $d\mu$ and using 
$$
\int F(x)\, d\nu \geq \int F(x)\, d\tilde\nu = \int_A F(x) f(x)^{-1}\, d\mu,
$$
we have that
$$
\int F(x)\, d\nu \geq \int (1+\log F(x))\, d\mu(x) + S(\mu\mid\nu)
$$
or
\begin{equation} \label{5.14}
S(\mu\mid\nu) \leq \calG (F,\mu,\nu).
\end{equation}

To get equality in \eqref{5.14a}, take $F=f$ so $\int d\mu$ and $\int F\,d\nu$ cancel. Of course, 
$f$ may not be continuous or strictly positive, so we need an approximation argument. Given 
$N,\veps$, let
$$
f_{N,\veps}(x) = \left\{ \begin{array}{ll} 
N &\hbox{ if } f(x) \geq N \\
f(x) &\hbox{ if } \veps \leq f(x) \leq N \\
\veps &\hbox{ if } f(x) \leq \veps. 
\end{array}\right. 
$$
Let $f_{\ell, N,\veps}(x)$ be continuous functions with $\veps\leq f_{\ell, N,\veps}\leq N$ 
so that as $\ell\to\infty$, $f_{\ell, N,\veps}\to f_{N,\veps}$ in $L^1 (X, d\mu+d\nu)$. 
For $N>1$, $ff_{N,\veps}^{-1} \leq 1 + f$, so we have 
\begin{eqnarray*} 
-\int \log (f_{N,\veps})\, d\mu &=& \mu(X) \int \log (f_{N,\veps}^{-1})\, 
\f{d\mu}{\mu(X)} \\
&\leq &\mu (X) \log \biggl[\int ff_{N,\veps}^{-1} \f{d\nu}{\mu(X)}\biggr] \\
&\leq &\mu(X) \log \bigg[ 1 + \f{\nu(X)}{\mu(X)}\biggr] < \infty
\end{eqnarray*}
and thus, since $-\log (f_{N,\veps})$ increases as $\veps\downarrow 0$, $f_{N,\veps=0}
\equiv\lim_{\veps\downarrow 0} f_{N,\veps}$ has\break $\log f_{N,\veps=0}\in L^1 (d\mu)$ and the
integrals converge. It follows that as $\ell\to\infty$ and then $\veps\downarrow 0$, 
$$
\calG (f_{\ell, N,\veps}, \mu,\nu) \to \calG(f_{N,\veps}, \mu,\nu)\to \calG (f_{N,\veps=0}, 
\mu,\nu).
$$
We now take $N\to\infty$. By monotonicity, $-\int \log f_{N,\veps=0}\, d\mu$ converges to\break $-\int 
\log f\,d\mu$ which may be infinite. In addition, $\int f_{N,\veps=0}\, d\nu - \int d\mu\to 0$ so 
$\calG(f_{N,\veps=0},\mu,\nu)\to S(\mu\mid\nu)$, and we have proven \eqref{5.14}.

Next, suppose $\mu$ is not $\nu$-ac. Thus, there is a Borel subset $A\subset X$ with $\mu(A) 
>0$ and $\nu(A)=0$. By regularity of measures, we can find $K\subset A$ compact and for any 
$\veps$, $U_\veps$ open so $K\subset A\subset U_\veps$ and
\begin{equation} \label{5.16}
\mu(K)>0 \qquad \nu(U_\veps)<\veps.
\end{equation}
By Urysohn's lemma, find $F_\veps$ continuous with 
\begin{equation} \label{5.17}
1\leq F_\veps (x) \leq \veps^{-1} \hbox{ all }x, \qquad F_\veps \equiv \veps^{-1} 
\hbox{ on } K, \, F_\veps \equiv 1 \hbox{ on } X\backslash U_\veps.\hskip.35in
\end{equation}
Then
$$
\int F_\veps\, d\nu \leq \nu (X\backslash U_\veps) + \veps^{-1} \nu(U_\veps) \leq 
\nu (X) +1
$$
while 
$$
\int (1+\log F_\veps)\, d\mu \geq \log (\veps^{-1}) \mu(K)
$$
so
$$
\calG (F_\veps, \mu,\nu) \leq \nu(X)+1 - \log (\veps^{-1})\mu(K)\to-\infty
$$
as $\veps\downarrow 0$, proving the right side of \eqref{5.14a} is $-\infty$. 
\enddemo

As an infimum of continuous functions is upper semicontinuous, we have

\proclaim{{C}orollary} \lb{C5.3} $S(\mu\mid\nu)$ is jointly weakly upper semicontinuous in 
$\mu$ and $\nu${\rm ,} that is{\rm ,} if $\mu_n \stackrel{w}{\longrightarrow} \mu$ and $\nu_n
\stackrel{w}{\longrightarrow}\nu${\rm ,} 
then 
$$
S(\mu\mid\nu) \geq \limsup_n S(\mu_n\mid\nu_n).
$$
\endproclaim

 {\it Remarks.} 1. In our applications, $\mu_n$ will be fixed.

\vglue5pt
2. This proof can handle functions other than $\log$. If $\int \log ((d\nu/d\mu)^{-1})\,
d\mu$ is replaced by $\int G((d\nu/d\mu))\,d\mu$ where $G$ is an arbitrary increasing 
concave function with $\lim_{y\downarrow 0} G(y) =\infty$, there is a variational 
principle where $1+\log F$ in \eqref{5.14} is replaced by $H(F(x))$ with $H(y)=\inf_x 
(xy-G(x))$.

\vglue8pt
To apply this to $Z$ and $Q$, we note

\proclaim{Proposition} \lb{P5.4} 
 
 {\rm{(a)}}  Let
\begin{equation} \label{5.14x}
d\mu_0(E) = \f{1}{2\pi} \,\sqrt{4-E^2}\, dE.
\end{equation}
Then
\begin{equation} \label{5.15x}
Q(J) = -{\textstyle \frac{1}{2}}\, S(\mu_0\mid\mu_J).
\end{equation}

{\rm{(b)}}  Let
\begin{equation} \label{5.16x}
d\mu_1 (E) = \f{1}{\pi}\, \f{dE}{\sqrt{4-E^2}}\,.
\end{equation}
Then
\begin{equation} \label{5.17x}
Z(J) = -{\textstyle \frac{1}{2}}\, \log(2)-{\textstyle \frac{1}{2}}\, S(\mu_1\mid\mu_J).
\end{equation}
\endproclaim

 {\it Remarks.} 1. Both $\mu_0$ and $\mu_1$ are probability measures, as is easily 
checked by setting $E=2\cos\theta$.

\vglue5pt
2. $d\mu_0$ is the spectral measure for $J_0$. For $M(z;J_0)=z$ and thus $\Ima M(e^{i\theta};
J_0)=\sin\theta$ so $m(E;J_0)=\f12\sqrt{4-E^2}$ and $\f{1}{\pi} \Ima m\,dE=d\mu_0$.

\vglue5pt
3. $d\mu_1$ is the spectral measure for the whole-line free Jacobi matrix and also for the 
half-line matrix with $b_n=0$, $a_1=\sqrt{2}$, $a_2=a_3=\cdots = 1$. An easy way to see this 
is to note that after $E=2\cos\theta$, $d\mu_1 (\theta)=\f{1}{\pi}\,d\theta$ and so the orthogonal 
polynomials are precisely the normalized scaled Chebyshev polynomials of the first kind 
that have the given values of $a_j$.

\demo{Proof} (a) \ Follows immediately from \eqref{5.14a} if we note that
$$
\f{d\mu_0}{d\mu} = \f{d\mu_0}{dE}\bigg/ \f{d\mu_{\ac}}{dE} = \f{\sqrt{4-E^2}}
{2\pi\, d\mu_{\ac}/dE}\,.
$$

\vglue5pt
(b) \ As above,
$$
\f{d\mu_1}{d\mu} = 2(4-E^2)^{-1} \,\f{\sqrt{4-E^2}}{2\pi\,d\mu_{\ac}/dE}\,.
$$
Thus
$$
Z(J) = c -{\textstyle \frac{1}{2}}\, S(\mu_1\mid \mu_J) ,
$$
where
\begin{eqnarray*}
c&=& -\f{1}{2\pi} \int_{-2}^2 \log \biggl[ \f{2}{4-E^2}\biggr] \sqrt{4-E^2}\, dE \\
&= &\f{1}{4\pi} \int_0^{2\pi} \log [2\sin^2\theta]\, d\theta \\
&= &\f{1}{2}\, \log(2) + \f{1}{2\pi} \int_0^{2\pi} \log \abs{\sin\theta}\, d\theta \\
&=& \f{1}{2}\, \log(2) + \f{1}{2\pi} \int_0^{2\pi} \log \biggl| 
\f{1-e^{i\theta}}{2}\biggr|\, d\theta \\
&=& \f{1}{2}\, \log(2) + \log \biggl(\f12\biggr) = -\f12\, \log(2) 
\end{eqnarray*}
where we used Jensen's formula for $f(z) =\f12 (1-z^2)$ to do the integral. 
\enddemo

 {\it Remark.} As a check on our arithmetic, consider the Jacobi matrix $\tilde J$ 
with $a_1 = \sqrt2$ and all other $a$'s and $b$'s the same as for $J_0$ so $d\mu_{\tilde J}$ 
is $d\mu_1$. The sum rule, $C_0$, for this case says that 
$$
Z(\tilde J) = -\log (\sqrt2) = -{\textstyle \frac{1}{2}} \,\log 2
$$
since there are no eigenvalues and $a_1 =\sqrt2$. But $\mu_1 = \mu_J$, so $S(\mu_1\mid\mu_J)\break
=0$. This shows once again that $c=-\f12\log 2$ (actually, it is essentially the calculation 
we did---done the long way around!).

Given this proposition, Lemma~\ref{L5.1}, and Corollary~\ref{C5.3}, we have 

\proclaim{Theorem} \lb{T5.5}  For any Jacobi matrix{\rm ,}
\begin{equation} \label{5.19}
Q(J) \geq 0
\end{equation}
and
\begin{equation} \label{5.20}
Z(J) \geq -{\textstyle \frac{1}{2}}\, \log(2).
\end{equation}
If $\mu_{J_n}\to \mu_J$ weakly{\rm ,} then
\begin{equation} \label{5.21}
Z(J)\leq \liminf \,Z(J_n).
\end{equation}
and
\begin{equation} \label{5.22}
Q(J) \leq \liminf\, Q(J_n).
\end{equation}
\endproclaim 

We will call \eqref{5.21} and \eqref{5.22} lower semicontinuity of $Z$ and $Q$.

\section{Fun and games with eigenvalues}\label{S6}

Recall that $J_n$ denotes the Jacobi matrix with truncated perturbation, as given by 
\eqref{2.10}. In trying to get sum rules, we will approximate $J$ by $J_n$ and need to 
estimate eigenvalues of $J_n$ in terms of eigenvalues of $J$. Throughout this section, $X$ 
denotes a continuous function on $\bbR$ with $X(x) = X(-x)$, $X(x)=0$ if $\abs{x}\leq 2$, 
and $X$ is monotone increasing in $[2,\infty)$. Our goal is to prove:

\proclaim{Theorem} \lb{T6.1} For any $J$ and all $n${\rm ,} we have $N^\pm (J_n) \leq N^\pm (J)+1$ 
and 
\vglue4pt
{\rm (i)} $|E_1^\pm (J_n)| \leq |E_1^\pm (J)| +1,$
\vglue6pt
{\rm (ii)} $|E_{k+1}^\pm (J_n)| \leq |E_k^\pm (J)|$.
\vglue4pt\noindent 
In particular{\rm ,} for any function $X$ of the type described above{\rm ,} 
\begin{equation} \label{6.1}
\sum_{j=1}^{N^\pm (J_n)} X(E_j^\pm (J_n)) \leq X(E_1^\pm (J) +1) + \sum_{j=1}^{N^\pm (J)} 
X(E_j^\pm (J)).
\end{equation}
\endproclaim 

\proclaim{Theorem} \lb{T6.2} If $J-J_0$ is compact{\rm ,} then 
\begin{equation} \label{6.2}
\lim_{n\to\infty} \, \sum_{j=1}^{N^\pm (J_n)} X(E_j^\pm (J_n)) = \sum_{j=1}^{N^\pm (J)} 
X(E_j^\pm (J)).
\end{equation}
This quantity may be infinite. 
\endproclaim 

\demo{Proof  of Theorem~{\rm \ref{T6.1}}} To prove these results, we pass from $J$ to 
$J_n$ in several intermediate steps.
\begin{itemize}
\item[(1)] We pass from $J$ to $J_{n;F}$.
\item[(2)] We pass from $J_{n;F}$ to $J_{n;F}\pm d_{n,n} \equiv J_{n;F}^\pm$ where 
$d_{n,n}$ is the matrix with $1$ in the $n,n$ place and zero elsewhere.
\item[(3)] We take a direct sum of $J_{n;F}^\pm$ and $J_0 \pm d_{1,1}$.
\item[(4)] We pass from this direct sum to $J_n$. 
\end{itemize}

 \vglue-6pt
{\it Step} 1.   $J_{n;F}$ is just a restriction of $J$ (to $\ell^2 (\{1,\dots, 
n\})$). The min-max principle \cite{RS4} implies that under restrictions, the most positive 
eigenvalues become less positive and the most negative, less negative. It follows that
\begin{eqnarray}
N^\pm (J_{n;F}) &\leq &N^\pm (J)  \label{6.2a} \\
\pm E_j^\pm (J_{n;F}) &\leq &\pm E_j^\pm (J). \lb{6.2b} 
\end{eqnarray}
\vglue2pt
 {\it Step} 2.  To study $E_j^+$, we add $d_{n,n}$, and to study $E_j^-$, 
we subtract $d_{n,n}$. The added operator $d_{n,n}$ has two critical properties: 
It is rank one and its norm is one. From the norm condition, we see
\begin{equation} \label{6.3}
\abs{E_1^\pm (J_{n;F}^\pm) - E_1^\pm (J_{n;F})}\leq 1
\end{equation}
so
\begin{eqnarray} 
E_1^+ (J_{n;F}^+) &\leq& E_1^+ (J_{n;F}) + 1\label{6.4} \\
&\leq& E_1^+ (J) + 1. \no
\end{eqnarray}
(Note \eqref{6.3} and \eqref{6.4} hold for all indices $j$, not just $j=1$, but we only 
need $j=1$.) Because $d_{n,n}$ is rank 1, and positive, we have 
$$
E_{m+1}^+ (J_{n;F}) \leq E_{m+1}^+ (J_{n;F}^+) \leq E_m^+ (J_{n;F})
$$
and so, by \eqref{6.2b},
\begin{equation} \label{6.5}
E_{m+1}^+ (J_{n;F}^+) \leq E_m^+(J)
\end{equation}
and thus also
\begin{equation} \label{6.6}
N^\pm (J_{n;F}^\pm) \leq N^\pm (J) +1.
\end{equation}
\vglue2pt
{\it Step} 3.  Take the direct sum of $J_{n;F}^\pm$ and $J_0\pm d_{11}$.  This
should be interpreted as a matrix with entries 
$$
\big[J_{n;F}^\pm\oplus (J_0\pm d_{11})\big]_{k,\ell}=\left\{ \begin{array}{ll}
(J_{n;F}^\pm)_{k,\ell} 			& k,\ell\leq n \\
(J_0\pm d_{11})_{k-n,\ell-n}	& k,\ell >n \\
0				& \hbox{otherwise}. 
\end{array}\right. 
$$ 
Since $J_0\pm d_{11}$ has no eigenvalues, \eqref{6.5} and \eqref{6.6} still hold.
\pagebreak

{\it Step} 4. Go from the direct sum to $J_n$. In the $+$ case, we add the 
$2\times 2$ matrix in sites $n, n+1$:
$$
dJ^+ = \left( 
\begin{array}{rr}
 -1 & 1 \\ 1 & -1 \end{array} \right)
$$
and, in the $-$ case,
$$
dJ^- = 
\left( 
\begin{array}{cc} 1 & 1 \\ 1 & 1 \end{array} \right)
$$
$dJ^+$ is negative, so it moves eigenvalues down, while $dJ^-$ is positive. Thus
$$
E_{m+1}^+ (J_n) \leq E_{m+1}^+ (J_{n;F}^+) \leq E_m^+ (J)
$$
and
\vglue6pt
\hfill ${\displaystyle
N^\pm (J_n) \leq N^\pm (J_{n;F}^\pm) \leq N^\pm (J) + 1.
}$
\enddemo
\vglue12pt
\demo{Proof  of Theorem~{\rm \ref{T6.2}}} We have, since $J-J_0$ is compact, that 
$$
\|J_n -J\| \leq \sup_{m\geq n+1}\, \abs{b_m} + 2\, \sup_{m\geq n}\, \abs{a_m}\to 0.
$$
Thus
\begin{equation} \label{6.7}
\abs{E_j^\pm (J_n) - E_j^\pm (J)} \leq \|J_n -J\|\to 0.
\end{equation}
If $\sum_{j=1}^{N^\pm (J)} X(E_j^\pm (J))=\infty$, then, by \eqref{6.7}, for all fixed 
$m$, 
\begin{eqnarray*}
\liminf\, \sum_{j=1}^{N^\pm (J_n)} X(E_j^\pm (J_n)) &\geq& \liminf\, \sum_{j=1}^m 
X(E_j^\pm (J_n))\\
&= &\sum_{j=1}^m X(E_j^\pm (J))
\end{eqnarray*}
so taking $m$ to infinity, \eqref{6.2} results.

If the sum is finite, \eqref{6.7}, dominated convergence and \eqref{6.1} imply \eqref{6.2}. 
\enddemo

\section{Jacobi data dominate spectral data in $P_2$}\label{S7}

Our goal in this section is to prove Theorem~5. Explicitly, for a Jacobi matrix, $J$, let 
\begin{equation} \label{7.1}
D_2 (J) = \tfrac14 \sum_{j=1}^\infty b_j^2 + {\textstyle \frac{1}{2}} \sum_{j=1}^\infty G(a_j)
\end{equation}
with $G=a^2-1-2\log(a)$ as in \eqref{3.18}. For a probability measure, $\mu$ on $\bbR$,
define 
\begin{equation} \label{7.2}
P_2 (\mu) =\f{1}{2\pi} \int_{-\pi}^\pi \log \biggl( \f{\sin\theta}{\Ima M_\mu(e^{i\theta})}
\biggr) \sin^2 \theta\, d\theta + \sum_j F(E_j) 
\end{equation}
where $E_j$ are the mass points of $\mu$ outside $[-2,2]$ and $F$ is given by \eqref{3.19}. 
Recall that $\Ima M_\mu (e^{i\theta})\equiv \pi\,d\mu_{\ac}/dE$ at $E=2\cos\theta$. We will let $\mu_J$ 
be the measure associated with $J$ by the spectral theorem and $J_\mu$ the Jacobi matrix 
associated to $\mu$.

Then Theorem~5 says

\proclaim{Theorem}\lb{T7.1} If $J-J_0$ is Hilbert\/{\rm -}\/Schmidt so that $D_2 (J)<\infty${\rm ,} then 
$P_2(\mu_J)<\infty$ and
\begin{equation} \label{7.3}
P_2 (\mu_J) \leq D_2 (J).
\end{equation}
\endproclaim 

\demo{Proof} Let $J_n$ be a truncation of $J$ given by \eqref{2.10}. Then $D_2(J_n)$ is 
monotone increasing with limit $D_2(J)$.  This is finite because $J-J_0$ is Hilbert-Schmidt.
By the definition 
\eqref{5.4a}, 
\begin{eqnarray}
 P_2(\mu) &=& Q(J) + \sum_j F(E_j) \label{7.4}  \\
&=&-\tfrac 12\, S(\mu_0, \mu) + \sum_j F(E_j) \no
\end{eqnarray}
by \eqref{5.15x}. Since $Q\geq 0$ \eqref{5.19} and $F>0$ (Proposition~\ref{P3.5}), 
\eqref{7.4} is a sum of positive terms. Moreover, by Theorem~\ref{T6.2}, $\sum_j 
F(E_j(J_n)) \to \sum_j F(E_j(J))$ even if the right side 
is infinite.  As $J_n\to J$ in Hilbert-Schmidt sense, $(J_n-E)^{-1}$ converges (in norm) to
$(J-E)^{-1}$ for all $E\in\bbC\setminus\bbR$.  This implies that $\mu_{J_n}$ converges
weakly to $\mu$ and so by \eqref{5.22}, $Q(J)\leq \limsup\, Q(J_n)$. It follows that 
\begin{eqnarray*}
P_2 (J_\mu) & \leq& \limsup \Bigl[Q(J_n) + \sum F\big(E_j(J_n)\big)\Bigr] \\
&=&\limsup\, D_2 (J_n) \qquad \hbox{(by Theorem~\ref{T3.3})} \\
&=& D_2(J).
\end{eqnarray*}
Thus $P_2 (\mu_J)<\infty$ and \eqref{7.3} holds. 
\enddemo

The result in this section is essentially a quantitative version of the main result in 
Deift-Killip \cite{DK}.

\section{Spectral data dominate Jacobi data in $P_2$}\label{S8}

Our goal in this section is to prove the following, which is essentially Theorem~6:

\proclaim{Theorem}\lb{T8.1} If $\mu$ is a probability measure with $P_2 (\mu)<\infty${\rm ,} then
\begin{equation} \label{8.1}
D_2 (J_\mu) \leq P_2 (\mu)
\end{equation}
and so $J_\mu$ is Hilbert-Schmidt.
\endproclaim  

The idea of the proof is to start with a case where we have the sum rule and then pass to 
successively more general cases where we can prove an inequality of the form \eqref{8.1}. 
There will be three steps: 
\vglue6pt
\noindent\hskip8pt\hangindent=27pt\hangafter=1
 (1) Prove the inequality in the case $M_\mu$ is meromorphic in a neighborhood of 
$\bar D$.

\vglue6pt
\noindent\hskip8pt\hangindent=27pt\hangafter=1(2) \hskip2pt Prove the inequality in the case $\mu\geq \delta
\mu_0$ where
$\delta$ is a  positive real number and $\mu_0$ is the free Jacobi measure \eqref{5.14}.

\vglue6pt
\noindent\hskip8pt\hangindent=25pt\hangafter=1(3) \hskip1pt  Prove the inequality in the case $P_2 (\mu)<\infty$.
 
\proclaim{Proposition} \lb{P8.2} Let $J$ be a Jacobi matrix for which $M_\mu$ 
has a meromorphic continuation to a neighborhood of $\bar D$. Then
\begin{equation} \label{8.2}
D_2(J) \leq P_2 (J).
\end{equation}
\endproclaim

\demo{Proof} By Theorem~\ref{T4.2}, 
$$
P_2(J) =\tfrac14\, b_1^2 + {\textstyle \frac{1}{2}}\, G(a_1) + P_2 (J^{(1)}).
$$
so iterating,
\begin{eqnarray*}
P_2 (J) &=& \tfrac14 \sum_{j=1}^m b_j^2 + {\textstyle \frac{1}{2}} \sum_{j=1}^m G(a_j) + P_2 (J^{(m)}) \\
&\geq& \tfrac14 \sum_{j=1}^m b_j^2 + {\textstyle \frac{1}{2}} \sum_{j=0}^m G(a_j)
\end{eqnarray*}
since $P_2 (J^{(m)})\geq 0$. Now $G\geq 0$, so we can take $m\to\infty$ and obtain \eqref{8.2}. 
\enddemo

 {\it Remark.} If $M_\mu$ has a meromorphic continuation into $\{z\mid\,\abs{z}
<\eta\}$ for some $\eta >1$, then by a theorem of Geronimo \cite{Ger3}, $\sum \abs{a_n -1}
\rho^n + \abs{b_n}\rho^n <\infty$ for all $\rho <\eta$, so the sum rule also follows from 
the methods of Section~\ref{S3}. We prefer to avoid the use of Geronimo's theorem. 

Given any $J$ and associated $M$-function $M(z;J)$, there is a natural approximating 
family of $M$-functions meromorphic in a neighborhood of $\bar D$.

\proclaim{Lemma} \lb{L8.3} Let $M_\mu$ be the $M$\/{\rm -}\/function of a probability measure $\mu$ 
obeying the Blumenthal\/{\rm -}\/Weyl condition{\rm ,} and define
\begin{equation} \label{8.3}
M^{(r)}(z) = r^{-1} M_\mu (rz)
\end{equation}
for $0<r<1$. Then{\rm ,} there is a set of probability measures $\mu^{(r)}$ so that $M^{(r)}
=M_{\mu^{(r)}}$.
\endproclaim

\demo{Proof} Return to the $E$ variable. Since $M^{(r)}(z)$ is meromorphic in a 
neighborhood of $\bar D$ with $\Ima M^{(r)}(z)>0$ if $\Ima z>0$, 
$$
m^{(r)}(E) = -M^{(r)}(z(E)) 
$$
(where $z(E) + z(E)^{-1} =E$ with $\abs{z}<1$) is meromorphic on $\bbC\backslash [-2,2]$ and 
Herglotz. It follows that it is the Borel transform of a measure $\mu^{(r)}$ of total 
weight $\lim_{E\to\infty} -E m^{(r)} (E)= \lim_{z\downarrow 0} z^{-1}M_\mu^{(r)}(z)=1$. 
\enddemo

\proclaim{Proposition} \lb{P8.4} Let $\mu$ be a probability measure obeying the Blumenthal\/{\rm -}\/Weyl 
condition and 
\begin{equation} \label{8.4}
\mu\geq \delta \mu_0 
\end{equation}
where $\mu_0$ is the free Jacobi measure {\rm{(}}\/the measure with $M_{\mu_0}(z)=z${\rm{)}} 
and $\delta >0$. 
Then
\begin{equation} \label{8.4a}
D_2 (J_\mu) \leq P_2 (\mu).
\end{equation}
\endproclaim

\demo{Proof} We claim that
\begin{equation} \label{8.5}
\limsup_{r\uparrow 1}\, \int -\log\abs{\Ima M_{\mu^{(r)}}(e^{i\theta})}\, d\theta \leq 
\int -\log\abs{\Ima M_\mu (e^{i\theta})}\, d\theta. \hskip.5in
\end{equation}
Accepting \eqref{8.5} for the moment, let us complete the proof. The eigenvalues of 
$\mu^{(r)}$ that lie outside $[-2,2]$ correspond to $\beta$'s of the form 
$$
\beta_k (J_{\mu^{(r)}}) =\f{\beta_k(J)}{r}
$$
for those $k$ with $\abs{\beta_k (J)}<r$. Thus $\sum F(E_k^\pm(J_{\mu^{(r)}}))$ is 
monotone increasing to $\sum F(E_k^\pm (J_\mu))$, so \eqref{8.5} shows that
\begin{equation} \label{8.6}
P_2 (\mu) \geq \limsup\, P_2 (\mu^{(r)}).
\end{equation}

Moreover, $M_{\mu^{(r)}}(z)\to M_\mu(z)$ uniformly on compact subsets of $D$ which means 
that the continued fraction parameters for $m^{(r)}(E)$, which are the Jacobi coefficients, 
must converge. Thus for any $N$, 
\begin{eqnarray*}
\tfrac14 \sum_{j=1}^N b_j^2 + {\textstyle \frac{1}{2}} \sum_{j=1}^{N-1} G(a_j) 
&=&\lim_{r\uparrow 1} \, \tfrac14 \sum_{j=1}^N \big(b_j^{(r)}\big)^2 + {\textstyle \frac{1}{2}} \sum_{j=1}^{N-1} 
G(a_j^{(r)}) \\
&\leq &\liminf\, D_2 (J_{\mu^{(r)}}) \\
&\leq& \liminf\, P_2 (\mu^{(r)}) \qquad \hbox{(by Proposition~\ref{P8.2})} \\
&\leq& P_2 (\mu) \qquad \hbox{(by \eqref{8.6})}
\end{eqnarray*}
so \eqref{8.4a} follows by taking $N\to\infty$.

Thus, we need only prove \eqref{8.5}. Since $M_{\mu^{(r)}}(\theta) =r^{-1} M_\mu 
(re^{i\theta})\to M_\mu (e^{i\theta})$ for a.e.~$\theta$, Fatou's lemma implies that 
\begin{equation} \label{8.7}
\liminf_{r\uparrow 1} \int \log_+ \abs{\Ima M_{\mu^{(r)}}(e^{i\theta})} \,d\theta \geq 
\int \log_+ \abs{\Ima M_{\mu}(\theta)}\, d\theta.
\end{equation}
On the other hand, \eqref{8.4} implies $\abs{\Ima M_\mu(z)} \geq\delta \abs{\Ima z}$, so 
$\abs{\Ima M_{\mu^{(r)}}(z)} \geq \delta\abs{\Ima z}$. Thus uniformly in $r$, 
\begin{equation} \label{8.8}
\abs{\Ima M_{\mu^{(r)}}(e^{i\theta})} \geq\delta\abs{\sin\theta}.
\end{equation}
Thus 
$$
\log_- \abs{\Ima M_{\mu^{(r)}}(e^{i\theta})} \leq -\log\delta - \log\abs{\sin\theta},
$$
so, by the dominated convergence theorem, 
$$
\lim\int\log_- (\abs{\Ima M_{\mu^{(r)}}})\, d\theta = \int \log_- (\abs{\Ima M_\mu (\theta)}) 
\, d\theta.
$$
This, together with \eqref{8.7} and $-\log (x) = -\log_{+}(x) + \log_- (x)$ implies 
\eqref{8.5}. 
\enddemo

 {\it Remark.} Semicontinuity of the entropy and \eqref{8.8} actually imply one 
has equality for the limit in \eqref{8.5} rather than inequality for the $\limsup$. 

\demo{Proof  of Theorem~{\rm \ref{T8.1}}} For each $\delta\in (0,1)$, let $\mu_\delta =
(1-\delta)\mu + \delta\mu_0$. Since $\mu_\delta$ obeys \eqref{8.4} and the Blumenthal-Weyl
criterion, 
\begin{equation} \label{8.9}
D_2 (J_{\mu_\delta}) \leq P_2 (\mu_\delta).
\end{equation}
Let $M_\delta \equiv M_{\mu_\delta}$ and note that 
$$
\Ima M_\delta (e^{i\theta}) = (1-\delta) \Ima M(e^{i\theta})+\delta\sin\theta
$$
so
$$
\log\abs{\Ima M_\delta (e^{i\theta})} = \log (1-\delta)+ \log\biggl|\Ima M 
(e^{i\theta})  + \f{\delta}{1-\delta}\, \sin\theta\biggr|.
$$
We see that up to the convergent $\log (1-\delta)$ factor, $\log\abs{\Ima M_\delta 
(e^{i\theta})}$ is monotone in $\delta$, so by the monotone convergence theorem,
\begin{equation} \label{8.10}
P_2(\mu) = \lim_{\delta\downarrow 0}\, P_2 (\mu_\delta)
\end{equation}
(the eigenvalue terms are constant in $\delta$, since the point masses of $\mu_\delta$ have the
same positions as those of $\mu$!).

On the other hand, since $\mu_\delta\to\mu$ weakly, as in the last proof, 
\begin{equation} \label{8.11}
D_2 (J_\mu) \leq \liminf\, D_2 (J_{\mu_\delta}).
\end{equation}
\eqref{8.9}--\eqref{8.11} imply \pagebreak \eqref{8.1}. 
\enddemo

\section{Consequences of the $C_0$ sum rule}\label{S9}

In this section, we will study the $C_0$ sum rule and, in particular, we will prove 
Nevai's conjecture (Theorem~2) and several results showing that control of the eigenvalues
can have strong consequences for $J$ and $\mu_J$, specifically Theorems~4$^\prime$ and 7. 
While Nevai's conjecture will be easy, the more complex results will involve some machinery, 
so we provide this overview: 
\begin{itemize}
\item[(1)] By employing semicontinuity of the Szeg\H{o} term, we easily get a  $C_0$-in\-equality 
that implies Theorems~2 and 7 and the part of Theorem~4$^\prime$ that says $J-J_0$ is 
Hilbert-Schmidt.
\item[(2)] We prove Theorem~\ref{T4.1} under great generality when there are no eigenvalues 
and use that to prove a semicontinuity in the other direction, and thereby show that the 
Szeg\H{o} condition implies a $C_0$-equality when there are no eigenvalues, including 
conditional convergence of $\sum_n (a_n-1)$.
\item[(3)] We use the existence of a $C_0$-equality to prove a $C_1$-equality, and thereby 
conditional convergence of $\sum_n b_n$.
\item[(4)] Returning to the trace class case, we prove that the perturbation determinant is 
a Nevanlinna function with no singular inner part, and thereby prove a sum rule in the Nevai 
conjecture situation.
\end{itemize}

\proclaimtitle{$\equiv$ Theorem~3}
\proclaim{Theorem}
\lb{T9.1} Let $J$ be a Jacobi matrix with $\sigma_{\ess} 
(J)\subset [-2,2]$ and 
\begin{equation} \label{9.1}
\sum_k e_k (J)^{1/2} <\infty,
\end{equation}
\vglue-18pt
\begin{equation} \label{9.2}
\limsup_{N\to\infty} \sum_{j=1}^N \log (a_j) >-\infty.
\end{equation}
Then
\begin{itemize}
\ritem{(i)} $\sigma_{\ess}(J) = [-2,2]$. 
\ritem{(ii)} The Szeg\H{o} condition holds{\rm ;} that is{\rm ,}
$$
Z(J)<\infty
$$
with $Z$ given by {\rm \eqref{5.1}.}
\ritem{(iii)} $\sigma_{\ac}(J)=[-2,2];$ indeed{\rm ,} the essential support of $\sigma_{\ac}$ 
is $[-2,2]$. 
\end{itemize}
\endproclaim 

 {\it Remarks.} 1. We emphasize \eqref{9.2} says $>-\infty$, not $<\infty$, that is,
it is a condition which prevents the $a_n$'s from being too small (on average). 

\vglue4pt
2. We will see below that \eqref{9.1} and \eqref{9.2} also imply
$\abs{a_j -1}\to 0$ and $\abs{b_j}\to 0$ and that at least inequality holds for the $C_0$
sum rule:
\begin{equation} \label{9.4x}
Z(J) \leq \sum_k \log \abs{\beta_k(J)} - \limsup_N \sum_{j=1}^N \log (a_j)
\end{equation}
holds.

\demo{Proof} Pick $N_1, N_2, \dots$ (tending to $\infty$) so that 
\begin{equation} \label{9.3}
 \inf_\ell \biggl(\,\sum_{j=1}^{N_\ell} \log (a_j) \biggr) >-\infty 
\end{equation}
and let $J_{N_\ell}$ be given by \eqref{2.10}. By Theorem~\ref{T3.3}, 
\begin{eqnarray} 
\quad Z(J_{N_\ell}) &\leq &-\sum_{j=1}^{N_\ell} \log (a_j) + \sum \log (\abs{\beta_k (J_{N_\ell})}) 
 \label{9.4}  \\
&\leq& - \inf_\ell \sum_{j=1}^{N_\ell} \log (a_j) + \sum \log (\abs{\beta_k(J)}) 
+ 2\log (\abs{\beta_1 (J)}+2) \no
\end{eqnarray}
where in \eqref{9.4} we used Theorem~\ref{T6.1} and the fact that the $\tilde\beta$ solving 
$e_1(J) + 1=\tilde\beta + \tilde\beta^{-1}$ (i.e., $1 +\beta_1^{} + \beta_1^{-1} = \tilde\beta
+ \tilde\beta^{-1}$) has $\tilde\beta \leq \beta_1 (J)+2$. For later purposes, we note 
that if $\abs{b_n(J)}+ \abs{a_n (J) -1}\to 0$, Theorem~\ref{T6.2} implies we can drop the last 
term in the limit.

Now use \eqref{5.21} and \eqref{9.4} to see that  
\begin{eqnarray*}
Z(J) &\leq \liminf Z(J_{N_\ell}) <\infty.
\end{eqnarray*}

This proves (ii). But (ii) implies $\f{d\mu_{\ac}}{dE}> 0$ a.e.~on $E\in [-2,2]$, that is, 
$[-2,2]$ is the essential support of $\mu_{\ac}$. That proves (iii). (i) is then immediate. 
\enddemo

\demo{{P}roof of Theorem~{\rm 2} {\rm{(}}Nevai\/{\rm '}\/s conjecture{\rm{)}}} We need only check 
that $J-J_0$ trace class implies \eqref{9.1} and \eqref{9.2}. The finiteness of
\eqref{9.1} follows from a 
bound of Hundertmark-Simon \cite{HS},
$$
\sum [\,\abs{e_k(J)}\, \abs{e_k(J)+4}\,]^{1/2} \leq \sum_n \abs{b_n} + 2\abs{a_n-1}
$$
where $e_k(J) = \abs{E^\pm}-2$ so $\abs{e}\,\abs{e+4}=(E^\pm)^2 -4$.

Condition \eqref{9.2} is immediate for, as is well-known, $a_j>0$ and\break $\sum(\abs{a_j}-1)<\infty$ 
implies $\prod a_j$ is absolutely convergent, that is, $\sum\abs{\log(a_j)}<\infty$.  \phantom{lotsofwind}
\hfill\qed\enddemo

\proclaimtitle{$\equiv$ Theorem~7}
\proclaim{{C}orollary} \lb{C9.2} A discrete half\/{\rm -}\/line Schr{\rm \"{\it o}}dinger operator 
{\rm{(}}\/i.e.{\rm ,} $a_n\equiv 1${\rm{)}} with $\sigma_{\ess}(J)\subset [-2,2]$ and $\sum 
e_n(J)^{1/2}<\infty$ has $\sigma_{\ac}=[-2,2]$.
\endproclaim

This is, of course, a special case of Theorem~\ref{T9.1} but a striking one discussed 
further in Section~\ref{S10}. In particular, if $a_n\equiv 1$ and $b_n=n^{-\alpha} w_n$ 
where $\alpha <\f12$ and $w_n$ are identically distributed independent random variables 
with distribution $g(\lambda)\, d\lambda$ with $g\in L^\infty$ and $\supp(g)$ bounded, 
then it is known that $[-2,2]$ is dense pure point spectrum (see Simon \cite{S157}). It 
follows that $J$ must also have infinitely many eigenvalues outside $[-2,2]$, indeed, 
enough that $\sum e_n (J)^{1/2}\break =\infty$.

Next, we deduce some additional aspects of Theorem~4$^\prime$:

\proclaim{{C}orollary} \lb{C9.3} \hskip-8pt If $\sigma_{\ess}(J)\!\subset\! [-2,2]$ and {\rm \eqref{9.1}, }
{\rm \eqref{9.2}} hold{\rm ,} then $J\! - J_0\!\in \calI_2${\rm ,} that is{\rm ,}
\begin{equation} \label{9.5}
\sum b_n^2 + \sum (a_n -1)^2 <\infty.
\end{equation}
\endproclaim

\demo{Proof} By Theorem~6, \eqref{9.5} holds if $\sum_k e_k(J)^{3/2}<\infty$, and $Q(J)$ 
(given by \eqref{5.14x}) is finite. By \eqref{9.1} and $e_k(J)^{3/2}\leq e_1 (J) e_k 
(J)^{1/2}$, we have that $\sum e_k(J)^{3/2}<\infty$. Moreover, $Z(J)<\infty$ (i.e., 
Theorem~\ref{T9.1}) implies $Q(J)<\infty$. For, in any event, $\int \Ima M\,d\theta
<\infty$ implies 
$$
\int_0^{2\pi} \log_- \biggl( \f{\sin\theta}{\Ima M}\biggr) \sin^2 (\theta)\,d\theta < \infty 
\quad\hbox{and}\quad
\int_0^{2\pi} \log_- \biggl( \f{\sin\theta}{\Ima M}\biggr)\, d\theta <\infty.
$$
Thus
\begin{eqnarray*}
Z(J)<\infty &\Rightarrow& \int_0^{2\pi} \log_+ \biggl( \f{\sin\theta}{\Ima M}\biggr)\, 
d\theta <\infty \\
&\Rightarrow& \int_0^{2\pi} \log_+ \biggl( \f{\sin\theta}{\Ima M}\biggr) 
\sin^2 \theta\, d\theta < \infty \\ 
&\Rightarrow& Q(J) <\infty. \\
\noalign{\vskip-36pt}
\end{eqnarray*}
\enddemo
\vglue12pt
What remains to be shown of Theorem~4$^\prime$ is the existence of the conditional sums.
We will start with $\sum (a_n-1)$. Because $\sum (a_n-1)^2 <\infty$, it is easy to see 
that $\sum (a_n-1)$ is conditionally convergent if and only if $\sum \log(a_n)$ is 
conditionally convergent. By \eqref{9.4} and the fact that $J-J_0$ is compact, we have: 

\proclaim{Proposition} \lb{P9.4} If {\rm \eqref{9.2}} holds and $\sigma(J)\subset [-2,2]${\rm ,} that is{\rm ,} 
no eigenvalues outside $[-2,2]${\rm ,} then
\begin{equation} \label{9.6}
Z(J) \leq -\limsup \biggl[\, \sum_{j=1}^N \log (a_j)\biggr].
\end{equation}
\endproclaim

We are heading towards a proof that
\begin{equation} \label{9.7}
Z(J) \geq -\liminf \biggl[ \,\sum_{j=1}^N \log (a_j)\biggr]
\end{equation}
from which it follows that the limit exists and equals $Z(J)$.

\proclaim{Lemma} \lb{L9.5} If $\sigma(J)\subset [-2,2]${\rm ,} then $\log [z^{-1} M(z;J)]$ lies in 
every $H^p(D)$ space for $p<\infty$. In particular{\rm ,} $z^{-1} M(z;J)$ is a Nevanlinna 
function with no singular inner part.
\endproclaim

\demo{Proof} In $D\backslash (-1,0)$, we can define $\Arg\, M(z;J)\subset (-\pi,\pi)$ 
and $\Arg\, z\subset (-\pi,\pi)$ since $\Ima M(z;J)/\Ima z >0$. Thus $g(z;J)=z^{-1}M(z;J)$ 
in the same region has argument in $(-\pi,\pi)$. But $\Arg \,g$ is single-valued and 
continuous across $(-1,0)$ since $M$ has no poles and precisely one zero at $z=0$. Thus 
$\Arg\, g\in L^\infty$. It follows by Riesz's theorem on conjugate functions 
(\cite[pg.~351]{Rudin}) that $\log (g)\in H^p (D)$ for any $p<\infty$. Since it lies in 
$H^1$, $g$ is Nevanlinna. Since for $p>1$, any $H^p$ function, $F$, has boundary values 
$F(re^{i\theta})\to F(e^{i\theta})$ in $L^p$, $\log (g)$ has no singular part in its 
boundary value. 
\enddemo

\proclaim{Proposition} \lb{P9.6} Let $\sigma(J)\subset [-2,2]$. Suppose $Z(J)<\infty$. 
Let $C_0, C_n$ be given by {\rm \eqref{4.7}} and {\rm \eqref{4.12}} {\rm{(}}\/where the $\beta(J)$ 
terms are absent\/{\rm{)}}. Then the step\/{\rm -}\/by\/{\rm -}\/step sum rules{\rm , \eqref{4.5}, \eqref{4.10}, 
\eqref{4.11}, \eqref{4.24}} hold. In particular{\rm ,} 
\begin{eqnarray}
Z(J)& =&-\log (a_1) + Z(J^{(1)})  \label{9.8} \\[4pt]
C_1 (J)& =& b_1 + C_1 (J^{(1)}). \lb{9.9}
\end{eqnarray}
\endproclaim

\demo{Proof} \eqref{4.3} and therefore \eqref{4.1} hold. Thus, we only need apply 
Theorem~\ref{T3.2} to $g$, noting that we have just proven that $g$ has no singular 
inner part. 
\enddemo

\proclaim{Theorem} \lb{T9.7} If $J$ is such that $Z(J)<\infty$ and $\sigma(J)\subset 
[-2,2]${\rm ,} then 
\begin{itemize}
\item[{\rm{(i)}}] $\lim_{N\to\infty} \sum_{j=1}^N \log (a_j)$ exists.
\item[{\rm{(ii)}}] The limit in {\rm{(i)}} is $-Z(J)$.
\item[{\rm{(iii)}}] 
\end{itemize}
\vglue-24pt
\begin{equation} \label{9.9a}
\lim_{n\to\infty} Z(J^{(n)}) =0 \qquad (=Z(J_0))
\end{equation}
\endproclaim 

\demo{Proof} By \eqref{9.8}, 
\begin{equation} \label{9.10}
Z(J) + \sum_{j=1}^n \log (a_j) =Z(J^{(n)}).
\end{equation}
Since $J-J_0\in \ell_2$, $\mu_{J^{(n)}}\to \mu_{J_0}$ weakly, and so, by \eqref{5.21}, 
$\liminf Z(J^{(n)})\geq 0$, or by \eqref{9.10},
\begin{equation} \label{9.11}
\liminf \biggl[ \, \sum_{j=1}^n \log (a_j)\biggr] \geq -Z(J).
\end{equation}
But \eqref{9.6} says 
$$
\limsup \biggl[ \, \sum_{j=1}^n \log (a_j)\biggr] \leq -Z(J). 
$$
Thus the limit exists and equals $Z(J)$, proving (i) and (ii). Moreover, by \eqref{9.10}, 
(i) and (ii) imply (iii). 
\enddemo

If $Z(\dott)$ had a positive integrand, \eqref{9.9a} would immediately imply that $C_1 (J^{(n)}) 
\to 0$ as $n\to\infty$, in which case, iterating \eqref{9.9} would imply that $\sum_{j=1}^n 
b_j$ is conditionally convergent. $Z(\dott)$ does not have a positive integrand but a theme 
is that concavity often lets us treat it as if it does. Our goal is to use \eqref{9.9a} 
and the related $\lim_{n\to\infty} Q(J^{(n)})=0$ (which follows from Theorem~5) to still 
prove that $C_1 (J^{(n)})\to 0$. We begin with

\proclaim{Lemma} \lb{L9.8} Let $d\mu$ be a probability measure and suppose $f_n \geq 0${\rm ,} 
$\int f_n\, d\mu\leq 1${\rm ,} and
\begin{equation} \label{9.12}
\lim_{n\to\infty} \int \log (f_n)\, d\mu =0.
\end{equation}
Then
\begin{equation} \label{9.13}
\int \abs{\log (f_n)}\, d\mu + \int \abs{f_n -1}\, d\mu\to 0.
\end{equation}
\endproclaim

\demo{Proof} Let 
\begin{equation} \label{9.14}
H(y) =-\log (y) -1+y.
\end{equation}
Then 
\begin{itemize}
\item[(i)] $H(y)\geq 0$ for all $y$.
\item[(ii)] $\inf_{\abs{y-1}\geq\veps} H(y) >0.$
\item[(iii)] $H(y) \geq \f12 y$ if $y>8$.
\end{itemize}

 (i) is concavity of $\log (y)$, (ii) is strict concavity, and (iii) holds because $-\log 
y - 1+\f12 y$ is monotone on $(2,\infty)$ and $>0$ at $y=8$ since $\log(8)$ is slightly 
more than $2$.

Since $\int (f_n -1)\,d\mu \leq 0$, \eqref{9.12} and (i) implies that
\begin{equation} \label{9.14a}
\int f_n (x)\, d\mu (x)\to 1
\end{equation} 
and 
\begin{equation} \label{9.15}
\lim_{n\to\infty} \int H(f_n (x))\, d\mu(x) \to 0.
\end{equation}
Since $H\geq 0$, (ii) and the above imply $f_n\to1$ in measure:
\begin{equation} \label{9.16}
\mu (\{x\mid\, \abs{f_n (x)-1}>\veps\})\to 0.
\end{equation}
By (i), (iii) and \eqref{9.15}, 
\begin{equation} \label{9.17}
\int_{f_n(x)>8} \abs{f_n(x)}\, d\mu \to 0.
\end{equation}
Now \eqref{9.16}/\eqref{9.17} imply that
$$
\int \abs{f_n (x)-1}\, d\mu(x)\to 0
$$
and this together with \eqref{9.15} implies $\int \abs{\log (f_n)}\, d\mu =0$. 
\enddemo

\proclaim{Proposition} \lb{P9.9} Suppose $Z(J)<\infty$ and $\sigma(J)\subset [-2,2]$. Then 
\begin{equation} \label{9.18}
\lim_{n\to\infty} \int_{-\pi}^\pi \biggl| \log \biggl( \f{\sin\theta}
{\Ima M(e^{i\theta}, J^{(n)})}\biggr)\biggr|\, d\theta =0.
\end{equation}
\endproclaim

\demo{Proof} By \eqref{9.9a}, the result is true if $\abs{\dott}$ is dropped. Thus it 
suffices to show 
$$
\lim_{n\to\infty} \int_{-\pi}^\pi \log_- \biggl( \f{\sin\theta}{\Ima M(e^{i\theta}, J^{(n)})} 
\biggr) \, d\theta =0
$$
or equivalently, 
\begin{equation} \label{9.19}
\lim_{n\to\infty} \int_{-\pi}^\pi \log_+ \biggl( \f{\Ima M(e^{i\theta}, J^{(n)})}{\sin\theta} 
\biggr)\, d\theta =0.
\end{equation}

Now, let $d\mu_0(\theta) = \f{1}{\pi} \sin^2\theta\, d\theta$ and $f_n(\theta) =
(\sin\theta)^{-1} \Ima M(e^{i\theta}, J^{(n)})$. By \eqref{1.22}, 
\begin{equation} \label{9.19a}
\int_{-\pi}^\pi f_n (\theta)\, d\mu_0 (\theta)\leq 1
\end{equation}
and by Theorem~5 (and Corollary~\ref{C9.3}, which implies $\|J^{(n)}-J_0\|_2^2\to 0$), 
$$
\int\log (f_n(\theta))\, d\mu_0 (\theta)\to 0
$$
so, by Lemma~\ref{L9.8}, we control $\abs{\log}$ and so $\log_+$; that is, 
\begin{equation} \label{9.20}
\lim_{n\to\infty} \int_{-\pi}^\pi \log_+ \biggl(\f{\Ima M(e^{i\theta}, J^{(n)})}{\sin\theta} 
\biggr) \sin^2\theta\, d\theta =0.
\end{equation}
Thus, to prove \eqref{9.19}, we need only prove
\begin{equation} \label{9.21}
\lim_{\veps\downarrow 0} \, \limsup_{n\to\infty} 
\int_{{ \abs{\theta}<\veps \atop \hbox{or} \\ \abs{\pi -\theta}<\veps }} 
\log_+ \biggl( \f{\Ima M(e^{i\theta}, J^{(n)})}{\sin\theta}\biggr)\, d\theta =0.
\end{equation}

To do this, use
\begin{eqnarray*}
\log_+ \biggl(\f{a}{b}\biggr) &\leq &\log_+ (a) + \log_- (b) = 2\log_+ (a^{1/2}) + \log_- (b) \\
&\leq& 2a^{1/2} + \log_-(b)
\end{eqnarray*}
with $a=\sin\theta \Ima M(e^{i\theta}, J^{(n)})$ and $b=\sin^2\theta$. The contribution of 
$\log_-(b)$ in \eqref{9.21} is integrable and $n$-independent, and so goes to zero as 
$\veps\downarrow 0$. The contribution of the $2a^{1/2}$ term is, by the Schwartz inequality, 
bounded by 
$$
(4\veps)^{1/2} \biggl( 4\int_{-\pi}^\pi f_n(\theta)\, d\mu_0 (\theta)\biggr)^{1/2}
$$
also goes to zero as $\veps\downarrow 0$. Thus \eqref{9.21} is proven. 
\enddemo

The following concludes the proofs of Theorems~4 and 4$^\prime$.

\proclaim{Theorem} \lb{T9.10} If $Z(J)<\infty$ and $\sigma(J)\subset [-2,2]${\rm ,} then
\begin{equation} \lb{9.22}
\lim_{N\to\infty} \sum_{j=1}^N b_j \hbox{ exists and equals } -\f{1}{2\pi} \int_0^{2\pi} 
\log \biggl( \f{\sin\theta}{\Ima M(e^{i\theta})}\biggr)\cos(\theta)\, d\theta. \quad
\end{equation}
\endproclaim 

\demo{Proof} By Proposition~\ref{P9.9}, $C_1 (J^{(n)})\to 0$ and, by \eqref{9.9}, 
\vglue12pt
\hfill ${\displaystyle
C_1 (J) = \sum_{j=1}^n b_j+ C_1 (J^{(n)}). 
}$
\enddemo
\vglue12pt

As a final topic in this section, we return to the general trace class case where we want to 
prove that the $C_0$ (and other) sum rules hold; that is, we want to improve the inequality 
\eqref{9.4} to an equality. The key will be to show that in this case, the perturbation 
determinant is a Nevanlinna function with vanishing inner singular part.

\proclaim{Proposition} \lb{P9.11} Let $J-J_0$ be trace class. Then{\rm ,} the perturbation 
determinant $L(z;J)$ is in Nevanlinna class.
\endproclaim

\demo{Proof} By \eqref{2.22}, if $J_n$ is given by \eqref{2.10}, then 
\begin{equation} \label{9.23}
L(z;J_n) \to L(z;J)
\end{equation}
uniformly on compact subsets of $D$. Thus
\begin{eqnarray}  && \label{9.24}\\
\noalign{\vskip-8pt}
\sup_{0<r<1} \int_0^{2\pi} \log_+ \abs{L(re^{i\theta}; J)} \, \f{d\theta}{2\pi} 
&\leq &\sup_n \, \sup_{0<r<1} \int_0^{2\pi} \log_+ \abs{L(re^{i\theta}; J_n)}\, 
\f{d\theta}{2\pi}\no \\
&= &\sup_n \int_0^{2\pi} \log_+ \abs{L(e^{i\theta}; J_n)}\, \f{d\theta}{2\pi} 
\no
\end{eqnarray}
where \eqref{9.24} follows from the monotonicity of the integral in $r$ 
(see \cite[pg.~336]{Rudin}) and the fact that $L(z;J_n)$ is a polynomial.
\vglue2pt

In \eqref{9.24}, write $\log_+ \abs{L} = \log \abs{L}+\log_-\abs{L}$. By Jensen's formula, 
\eqref{3.1}, and $L(0;J)=1$, 
$$
\int_0^{2\pi} \log \abs{L(e^{i\theta}; J_n)}\, \f{d\theta}{2\pi} = -\sum_{j=1}^{N_n} 
\log \abs{\beta_j(J_n)}
$$
and this is uniformly bounded in $n$ by the $\f12$ Lieb-Thirring inequality of\break 
Hundertmark-Simon \cite{HS}, together with Theorem~\ref{T6.1}. On the other hand, by 
\eqref{2.69}, 
\begin{eqnarray} 
& & 2\log_- \abs{L(e^{i\theta}; J_n)} = \log_- \biggl( \, \prod_{j=1}^{n-1} a_j^2
	\, \f{\sin\theta}{\Ima M(e^{i\theta}; J_n)}\biggr) \label{9.24a} \\
&&\qquad \leq 2\sum_{j=1}^{n-1} \log_- (a_j) + 2 \log (\sin\theta) + \log_+\big( 
\Ima M(e^{i\theta}; J_n) \sin \theta \big) \no
\end{eqnarray}
since $\log_- \abs{ab/c} = [\log(a) + \log(b) - \log(c)]_- \leq \log_- (a) + \log_-(b) 
+\log_+ (c)$.

The first term in \eqref{9.24a} is $\theta$-independent and uniformly bounded in $n$ since 
$\sum_{j=1}^\infty \abs{a_j-1}<\infty$. The second term is integrable. For the final 
term, we note that $\log_+ (y) \leq y$ so by \eqref{1.22}, the integral over $\theta$ 
is uniformly bounded. 
\enddemo

 {\it Remark.} Our proof that $L$ is Nevanlinna used $\sum_k (e_k (J))^{1/2} 
<\infty$ as input. If we could find a proof that did not use this {\it a~priori}, we would 
have, as a consequence, a new proof that $\sum_k e_k (J)^{1/2}<\infty $ since $\sum 
[1-\beta_k (J)^{-1}]<\infty$ is a general property of Nevanlinna functions.

\proclaim{Proposition} \lb{P9.12} If $\delta J\in\calI_1${\rm ,} the singular inner part of $L(z;J)${\rm ,} 
if any{\rm ,} is a positive point mass at $z=1$ and/or at $z=-1$.
\endproclaim 

\demo{Proof} By Theorem~\ref{T2.7}, $L(z;J)$ is continuous on $\bar D\backslash \{-1,1\}$ 
and by \eqref{2.69}, it is nonvanishing on $\{e^{i\theta}\mid \theta\neq 0,\pi\}$. It 
follows that on any closed interval, $I\subset (0,\pi)\cup (\pi,2\pi)$, $\log 
\abs{L(re^{i\theta}, J)}\,d\theta$ converges to an absolutely continuous measure, so the 
support of the singular inner part is $\{\pm 1\}$.

Returning to \eqref{9.24a} and using $\log_+(x) = 2\log_+ (x^{1/2})\leq 2x^{1/2}$, we 
see that $\log_- \abs{L}$ lies in $L^2$; that is,
$$
\sup_n \, \sup_{0<r<1}\int [\log_- \abs{L(re^{i\theta}, J_n)}]^2\, \f{d\theta}{2\pi} <\infty
$$
and this implies $\log_- \abs{L(re^{i\theta}, J)}$ has an a.c.~measure as its boundary 
value. Thus $\pm 1$ can only be positive \pagebreak pure points. 
\enddemo

 {\it Remark.} We will shortly prove $L$ has no singular inner part. However, we can 
ask a closely related question. If $\abs{\sum \log(a_n)}<\infty$ and $\sum e_k(J)^{1/2} < 
\infty$ so $Z(J)<\infty$, does the sum rule always hold or is there potentially a positive 
singular part in some suitable object?

\vglue8pt

Theorem~\ref{T2.9} will be the key to proving that $L(x;J)$ has no pure point singular part. 
The issue is whether the Blaschke product can mask the polar singularity, since, if not, 
\eqref{2.40} says there is no polar singularity in $L$ which combines the singular inner 
part, outer factor, and Blaschke product. Experts that we have consulted  tell us that 
the idea that Blaschke products cannot mask poles goes back to Littlewood and is known to 
experts, although our approach in the next lemma seems to be a new and interesting way of 
discussing this:

\proclaim{Lemma} \lb{L9.13} Let $f(z)$ be a Nevanlinna function on $D$. Then for any $\theta_0 
\in \partial D${\rm ,} 
\begin{equation}\label{9.25}
\lim_{r\uparrow 1} \; [\log(1-r)^{-1}]^{-1} \int_0^r \log \abs{f(ye^{i\theta_0})}\, dy = 
2\mu_s (\{\theta_0\}).
\end{equation}
\endproclaim

\demo{Proof} Let $B$ be a Blaschke product for $f$. Then (\cite[pg.~346]{Rudin}),
\begin{equation} \label{9.26}
\log\abs{f(z)} =\log \abs{B(z)} + \int_{-\pi}^\pi P(z,\theta)\, d\mu(\theta) 
\end{equation}
where $P$ is the Poisson kernel 
$$
P(re^{i\phi},\theta) =\f{(1-r^2)}{(1 + r^2 - 2r\cos (\theta-\varphi))}
$$
and $d\mu (\theta)$ is the boundary value of $\log \abs{f(re^{i\theta})}\, d\theta$, that is, 
outer plus singular inner piece. By an elementary estimate, 
$$
\sup_{r,\theta, \varphi}\, (1-r) P(re^{i\theta}, \varphi)<\infty
$$
and
$$
\lim_{r\uparrow 1} \, (1-r) P(re^{i\theta}, \varphi) = 
\left\{ \begin{array}{ll} 0 & \theta\neq\varphi \\
2 & \theta=\varphi
\end{array}\right. 
$$
and thus 
$$ 
(1-r) \int_{-\pi}^\pi P(re^{i\varphi}, \theta)\, d\mu(\theta)\to 2\mu_s (\{\varphi\}).
$$

This means \eqref{9.25} is equivalent to
\begin{equation} \label{9.27}
\lim_{r\uparrow 1}\, [\log (1-r)^{-1}]^{-1} \int_0^r \log \abs{B(ye^{i\theta_0})}\, dy =0
\end{equation}
for any Blaschke product. Without loss, we can take $\theta_0=0$ in \eqref{9.27}. Now let 
$$
b_\alpha (z) = \f{\abs{\alpha}}{\alpha}\, \f{\alpha-z}{1-\bar\alpha z} 
$$
so
$$
B(z) = \prod_{z_i} b_{z_i}(z)
$$
and note that for $0<x<1$ and any $\alpha\in D$, 
$$
1 > \abs{b_\alpha (x)} \geq \abs{b_{\abs{\alpha}}(x)},
$$
so 
\begin{equation} \label{9.27a}
0 < -\log \abs{B(x)} \leq \sum_{z_i} -\log \abs{b_{\abs{z_i}}(x)}.
\end{equation}
Thus, also without loss, we can suppose all the zeros $z_i$ lie on $(0,1)$.

If $\alpha\in (0,1)$, a straightforward calculation (or Maple!) shows 
\begin{equation} \label{9.28}
\int_0^1 - \log \abs{b_\alpha (x)}\, dx = \alpha\log\biggl( \f{1}{\alpha}\biggr)+ 
\f{1-\alpha^2}{\alpha}\, \log \biggl( \f{1}{1-\alpha}\biggr). \hskip.5in
\end{equation}
We claim that for a universal constant $C$ and $r>\f34$, $\alpha >\f12$, 
\begin{equation} \label{9.29}
-\int_0^r \log \abs{b_\alpha(x)}\,dx \leq C(1-\alpha) \log (1-r)^{-1}.
\end{equation}
Accepting \eqref{9.29} for the moment, by \eqref{9.27a}, we have for $r>\f34$, 
$$
-\int_0^r \log \abs{B(x)}\, dx \leq \sum_{j=1}^n \eta(\alpha_j) + 
C \biggl( \sum_{j=n+1}^\infty (1-\alpha_j)\biggr) \log(1-r)^{-1} 
$$
where $\eta(\alpha)$ is the right side of \eqref{9.28}. Dividing by $\log (1-r)^{-1}$ 
and using $\eta (\alpha)<\infty$, we see 
$$
\limsup \biggl[\frac{1}{\log (1-r)^{-1}}\biggl\{ -\int_0^r \log \abs{B(x)}\, dx\biggr\}\biggr] \leq 
C \sum_{j=n+1}^\infty (1-\alpha_j).
$$
Taking $n\to\infty$, we see that the $\limsup$ is $0$. Since $-\log \abs{B(x)}>0$, 
the limit is $0$ as required by \eqref{9.27}. Thus, the proof is reduced to establishing 
\eqref{9.29}.

Note first that if $1>\alpha >\f12$, $\f{1}{\alpha} (1+\alpha)<4$. Moreover, if $g(\alpha) 
=\alpha\log (\f{1}{\alpha})$, then $g''(\alpha) =-\f{1}{\alpha^2}<0$, so $g(\alpha)\leq 
(1-\alpha)$. Thus, if $\eta(\alpha)$ is the right side of \eqref{9.28} and $\alpha >\f12$, 
then $\phantom{\sum^\int}$
\begin{equation} \label{9.30}
\eta(\alpha) \leq 1-\alpha + 4(1-\alpha) \log \biggl( \f{1}{1-\alpha}\biggr).
\end{equation}
Suppose now 
\begin{equation} \label{9.31}
1 -\alpha \geq (1-r)^2. \pagebreak
\end{equation}
Then 
\begin{eqnarray} 
-\int_0^r \log \abs{b_\alpha (x)}\, dx &\leq& \eta(\alpha) \leq (1-\alpha) 
\biggl[ 1 + 4 \log \biggl( \f{1}{1-\alpha}\biggr) \biggr]\label{9.32}  \\
&\leq& (1-\alpha) \biggl[ 1 + 8\log \biggl( \f{1}{1-r}\biggr) \biggr] \no
\end{eqnarray} 
by \eqref{9.31}. 

On the other hand, suppose 
\begin{equation} \label{9.33}
(1-\alpha)\leq (1-r)^2.
\end{equation}
By an elementary estimate (see, e.g., \cite[pg.~310]{Rudin}), 
\begin{equation} \label{9.34}
\abs{1-b_\alpha (x)} \leq \f{2}{1-x}\, (1-\alpha).
\end{equation}
If \eqref{9.33} holds and $x<r$, then, by \eqref{9.34}, 
\begin{equation} \label{9.35}
\abs{1- b_\alpha (x)} \leq \f{2(1-r)^2}{1-x} \leq 2(1-r) < \f12
\end{equation}
since $r$ is supposed larger than $\f34$. If $u\in (\f12, 1)$, then 
$$
-\log u =\int_u^1 \f{dy}{y}\leq 2(1-u)
$$
so if \eqref{9.35} holds, 
$$
-\log (b_\alpha (x)) \leq 2 (1-b_\alpha (x)) \leq \f{4(1-\alpha)}{1-x}
$$
and so 
\begin{equation} \label{9.36}
\int_0^r -\log (b_\alpha (x))\, dx \leq 4 (1-\alpha) \log (1-r)^{-1}.
\end{equation}
We have thus proven \eqref{9.32} if \eqref{9.31} holds and \eqref{9.36} if \eqref{9.33} 
holds. Together this proves \eqref{9.29}. 
\enddemo

\proclaim{Theorem} \lb{T9.14} Let $J-J_0$ be trace class. Then the Nevanlinna function 
$L(z;J)$ has a vanishing singular inner component and all the sum rules $C_0, C_1, \dots$ 
hold with no singular term. 
\endproclaim 

\demo{Proof} By Proposition~\ref{P9.12}, the only possible singular parts are positive 
points at $\pm 1$. By \eqref{9.25} and the estimate \eqref{2.43}, these point masses are 
absent. Thus the singular part vanishes and the sum rules hold by \pagebreak Theorem~\ref{T3.2}. 
\enddemo

\section{Whole-line Schr\"odinger operators   with no bound states}\label{S10}

 \vglue-8pt

Our goal in this section is to prove Theorem~8 that the only whole-line Schr\"odinger 
operator with $\sigma (W)\subset [-2,2]$ is $W_0$, the free operator. We do this here 
because it illustrates two themes: that absence of bound states is a strong assertion 
and that sum rules can be very powerful tools.

Given two sequences of real numbers $\{a_n\}_{n=-\infty}^\infty$, $\{b_n\}_{n=-\infty}^\infty$
with $a_n>0$, we will denote by $W$ the operator on $\ell^2 (\bbZ)$ defined by 
\begin{equation} \label{10.1}
(Wu)_n = a_{n-1} u_{n-1} + b_n u_n + a_{n+1} u_{n+1}.
\end{equation}
$W_0$ is the operator with $a_n \equiv 1$, $b_n \equiv 0$. The result we will prove is:

\proclaim{Theorem} \lb{T10.1} Let $W$ be a whole\/{\rm -}\/line operator with $a_n \equiv 1$ and $\sigma 
(W)\subset [-2,2]$. Then $W=W_0${\rm ,} that is{\rm ,} $b_n\equiv 0$.
\endproclaim 

The proof works if 
$$
\limsup_{{ n\to\infty \atop m\to\infty }} \biggl[\, \sum_{j=-n}^m \log (a_j)\biggr] 
\geq 0.
$$
The strategy of the proof will be to establish analogs of the $C_0$ and $C_2$ sum rules. 
Unlike the half-line case, the integrand inside the Szeg\H{o}-like integral will be 
nonnegative. The $C_0$ sum rule will then imply this integrand is zero and the 
$C_2$ sum will therefore yield $\sum_n b_n^2=0$. As a preliminary, we note:

\proclaim{Proposition} \lb{P10.2} If $a_n \equiv 1$ and $\sigma(W)\subset [-2,2]${\rm ,} then 
$\sum_n b_n^2 <\infty$. In particular{\rm ,} $b_n\to 0$ as $\abs{n}\to\infty$. 
\endproclaim

\demo{Proof} Let $J$ be a Jacobi matrix obtained by restricting to $\{1,2, \dots\}$. By 
the min-max principle \cite{RS4}, $\sigma(J)\subset [-2,2]$. By Corollary~\ref{C9.3}, 
$\sum_{n=1}^\infty b_n^2 <\infty$. Similarly, by restricting to $\{0, -1, -2, \dots, \}$, 
we obtain $\sum_{n=-\infty}^0 b_n^2 <\infty$. 
\enddemo

Let $W^{(n)}$ for $n=1,2,\dots$ be the operator with
\begin{equation} \label{10.1a}
\left\{ \begin{array}{ll}
a_j^{(n)}\equiv 1 \\
b_j^{(n)} =b_j &\hbox{if } \abs{j}\leq n \\
b_j^{(n)} =0 &\hbox{if } \abs{j} >n.
\end{array}\right. 
\end{equation}
Then, Proposition~\ref{P10.2} and the proofs of Theorems~\ref{T6.1} and \ref{T6.2} immediately 
imply:

\proclaim{Theorem} \lb{T10.3} If $a_n \equiv 1$ and $\sigma (W)\subset [-2,2]${\rm ,} then $W^{(n)}$ 
has at most four eigenvalues in $\bbR\backslash [-2,2]$ {\rm{(}}\/up to two in each of
$(-\infty, -2)$ and $(2, \infty)${\rm{)}} and for $j=1, \dots, 4${\rm ,} 
\begin{equation} \label{10.2}
\abs{e_j (W^{(n)})}\to 0
\end{equation}
as $n\to\infty$. 
\endproclaim 

 {\it Note}. As in the Jacobi case, $e_j(W)$ is a relabeling of $\abs{E_j^\pm (W)} 
-2$ in decreasing order.

\vglue8pt
To get the sum rules, we need to study whole-line perturbation determinants. We will use the 
same notation as for the half-line, allowing the context to distinguish the two cases. So, 
let $\delta W=W-W_0$ be trace class and define
\begin{eqnarray}
L(z;W) &= &\det ((W-E(z)(W_0 -E(z))^{-1})  \label{10.3} \\
&=&\det (1+\delta W(W_0-E(z))^{-1}) \lb{10.4} 
\end{eqnarray} 
where as usual, $E(z) = z + z^{-1}$.

The calculation of the perturbation series for $L$ is algebraic and so immediately extends to 
imply:

\proclaim{Proposition} \lb{P10.4} If $\delta W$ is trace class{\rm ,} for each $n${\rm ,} $T_n(W/2) -T_n 
(W_0/2)$ is trace class. Moreover{\rm ,} for $\abs{z}$ small{\rm ,} 
\begin{equation} \label{10.5}
\log [L(z;W)] = \sum_{n=1}^\infty c_n (W) z^n 
\end{equation}
where $c_n (W)$ is 
\begin{equation} \label{10.6}
c_n (W) = -\f{2}{n}\, \Tra \Bigl( T_n \bigl( \tfrac{1}{2} W\bigr) - T_n \bigl( \tfrac{1}{2} W_0 
\bigr)\Bigr).
\end{equation}
In particular{\rm , }
\begin{equation} \label{10.7}
c_2 (W) = -\f12 \sum_{m=1}^\infty b_m^2 + 2(a_m^2 -1).
\end{equation}
\endproclaim

The free resolvent, $(W_0 - E(z))^{-1}$, has matrix elements that we can compute as we did
to get \eqref{2.15}, 
$$
(W_0 -E(z))_{nm}^{-1} = -(z^{-1} -z)^{-1} z^{\abs{m-n}}
$$
which has poles at $z=\pm 1$. We immediately get

\proclaim{Proposition} \lb{P10.5} If $\delta W$ is finite rank{\rm ,} $L(z;J)$ is a rational 
function on $\bbC$ with possible singularities only at $z=\pm 1$.
\endproclaim

 {\it Remarks.} 1. If $\delta W$ has $b_0=1$, all other elements zero, then 
$$
L(z;W) = 1-(z^{-1}-z)^{-1} = \f{(1-z-z^2)}{(1-z^2)}
$$
has poles at $\pm 1$, so poles can occur.

\vglue4pt
2. The rank one operator
$$
R(z) = -(z^{-1} -z)^{-1} z^{m+n}
$$
is such that if $\delta W=C^{1/2} UC^{1/2}$ with $C$ finite rank, then 
$$
C^{1/2} [(W_0 -E(z)^{-1} - R(z)] C^{1/2}
$$
is entire. Using this, one can see $L(z;J)$ has a pole of order at most $1$ when $\delta W$ 
is finite rank. We will see this below in another way.

If $z\in\bar D\backslash\{-1,1\}$, we can define a Jost solution $u_n^+ (z;W)$ so that 
\eqref{2.64} holds for all $n\in\bbZ$ and 
\begin{equation} \label{10.8x}
\lim_{n\to\infty} z^{-n} u_n^+ (z;W) =1.
\end{equation}
Moreover, if $\delta W$ has finite rank, $u_n^+$ is a polynomial in $z$ for each $n\geq 0$. 
Moreover, for $n<0$, $z^{-n}u_n^+$ is a polynomial in $z$ by using \eqref{2.65a}.

Similarly, we can construct $u_n^-$ solving \eqref{2.64} for all $n\in\bbZ$ with 
$$
\lim_{n\to -\infty} z^n u_n^- (z;W)=1.
$$
As above, if $\delta W$ is finite rank, $u_n^-$ is a polynomial in $z$ if $n\leq 0$ and for 
$n>0$, $z^n u_n^-$ is a polynomial.

\proclaim{Proposition} \lb{P10.6} Let $\delta W$ be trace class. Then for $z\in\bar D\backslash 
\{-1, 1\}$ and all $n\in \bbZ${\rm ,} 
\begin{eqnarray} \label{10.8}
\quad L(z;W) &=& (z^{-1} -z)^{-1} \biggl(\, \prod_{j=-\infty}^\infty a_j \biggr) \\
&& a_n [u_n^+ (z;W) u_{n+1}^- (z;W) - u_n^- (z;W) u_{n+1}^+ (z;W)].\no
\end{eqnarray}
\endproclaim

\demo{Proof} Both sides of \eqref{10.8} are continuous in $W$, so we need only prove the 
result when $\delta W$ is finite rank. Moreover, by constancy of the Wronskian, the right 
side of \eqref{10.8} is independent of $n$ so we need only prove \eqref{10.8} when 
$\abs{z}<1$ and $n$ is very negative--so negative it is to the left of the support 
of $\delta W$, that is, choose $R$ so $a_n=1$, $b_n=0$ if $n<-R$, and we will 
prove that \eqref{10.8} holds for $n<-R$. 

For $n<-R$, $z^n$ and $z^{-n}$ are two solutions of \eqref{2.64} so in that region
we have 
\begin{equation} \label{10.9}
u_n^+ = \alpha_\ell z^n + \beta_\ell z^{-n}.
\end{equation}
Taking the Wronskian of $u_n^+$ given by \eqref{10.9} and $u_n^-= z^{-n}$ at some point
$n<-R$, we see 
\begin{equation} \label{10.10}
\hbox{RHS of \eqref{10.8}} = \alpha_\ell \biggl( \,\prod_{j=-\infty}^\infty a_j  \biggr).
\end{equation}
Let $W_n, W_{0;n}$ be the Jacobi matrices on $\ell^2 (\{n+1, n+2, \dots, \})$
obtained by truncation. On the one hand, as with the proof of 
\eqref{2.70}, for $\abs{z}<1$, 
\begin{equation} \label{10.11}
L(z;W) =\lim_{n\to -\infty} \, \det((W_n -E(z))(W_{n;0}-E(z))^{-1}) \hskip.5in
\end{equation}
and on the other hand, for $n<-R$, by \eqref{2.63}, 
$$
\hbox{RHS of \eqref{10.11}} = \biggl(\, \prod_n a_n \biggr) z^{-n} u_n^+ (z; W).
$$
Thus 
\begin{eqnarray}
L(z;W) &=& \biggl( \prod_n a_n \biggr) \lim_{n\to -\infty} \, z^{-n} u_n^+ (z;W) \lb{10.12}  \\
&= &\biggl( \prod_n a_n\biggr) \lim_{n\to -\infty}\, (\alpha_\ell + \beta_\ell z^{-2n}) \no  \\
&= &\biggl( \prod_n a_n\biggr) \alpha_\ell \no
\end{eqnarray}
since $\abs{z}<1$ and $n\to -\infty$. Comparing \eqref{10.10} and \eqref{10.12} yields 
\eqref{10.8}. 
\enddemo

 {\it Note}. In \eqref{10.9}, $\alpha_\ell, \beta_\ell$ use ``$\ell$" for ``left" 
since they are related to scattering from the left. 

\proclaim{{C}orollary} \lb{C10.7} If $\delta W$ is finite rank{\rm ,} then $(1-z^2) L(z;W)$ is a 
polynomial and{\rm ,} in particular{\rm ,} $L(z;W)$ is a rational function. 
\endproclaim

\demo{Proof} By \eqref{10.8}, this is equivalent to $z(u_0^+ u_1^- - u_0^- u_1^+)$ being a 
polynomial. But $u_0^+$, $zu_1^-$, $u_0^-$, and $u_1^+$ are all polynomials. 
\enddemo

Let $\delta W$ be finite rank. Since $L$ is meromorphic in a neighborhood of $\bar D$ and 
analytic in $D$, Proposition~\ref{P3.1} immediately implies the following sum rule:

\proclaim{Theorem} \lb{T10.8} If $\delta W$ is finite rank{\rm ,} then
\begin{eqnarray}
\qquad C_0:&& \nhs\nhs\f{1}{2\pi} \int_0^{2\pi} \log \abs{L(e^{i\theta};W)}\, d\theta = 
\sum_{j=1}^{N(W)} \log \abs{\beta_j (W)} \label{10.13} \\
\qquad C_n : && \nhs\nhs\f{1}{\pi} \int_0^{2\pi} \log \abs{L(e^{i\theta}; W)} \cos(n\theta)\, d\theta \lb{10.14} \\
\qquad &&\qquad  =\f{1}{n} \sum_{j=1}^{N(W)} [\beta_j^n - \beta_j^{-n}] - \f{2}{n}\, 
\Tra \Bigl( T_n \bigl( \tfrac{1}{2} W \bigr) - T_n \bigl(\tfrac{1}{2} W_0 \bigr) \Bigr)\no
\end{eqnarray} 
for $n\geq 1$.
\endproclaim 

The final element of our proof is an inequality for $L(e^{i\theta};W)$ that depends on 
what a physicist would call conservation of probability.

\proclaim{Proposition} \lb{P10.9} Let $\delta W$ be trace class. Then for all $\theta\neq 0,\pi${\rm ,} 
\begin{equation} \label{10.15}
\abs{L(e^{i\theta}; W)} \geq \prod_{j=-\infty}^\infty a_j.
\end{equation}
\endproclaim

\demo{Proof} As above, we can suppose that $\delta W$ is finite range. Choose $R$ so that
all nonzero matrix elements of $\delta W$ have indices lying within $(-R,R)$. By 
\eqref{10.10}, \eqref{10.15} is equivalent to
\begin{equation} \label{10.16}
\abs{\alpha_\ell} \geq 1 
\end{equation}
where $\alpha_\ell$ is given by \eqref{10.9}. 

Since $u_n^+ (z;W)$ is real for $z$ real, we have
$$
u_n^+ (\bar z; W) = \ol{u_n^+ (z;W)}.
$$
Thus for $z=e^{i\theta}$, $\theta\neq 0,\pi$, and $n<-R$, 
\begin{eqnarray*}
u_n^+ (e^{i\theta}; W) &=& \alpha_\ell (e^{i\theta}) e^{in\theta} + \beta_\ell (e^{i\theta})
e^{-in\theta} \\
u_n^+ (e^{-i\theta}; W) &= &\ol{\alpha_\ell (e^{i\theta})}\, e^{-in\theta} + 
\ol{\beta_\ell (e^{i\theta})}\, e^{+in\theta}.
\end{eqnarray*}
Computing the Wronskian of the left-hand sides for $n>R$, where $u_n^+ =z^n$ and then the 
Wronskian of the right-hand sides for $n<-R$, we find
$$
i(\sin\theta) = i(\sin\theta) [\abs{\alpha_\ell}^2 - \abs{\beta_\ell}^2 ]
$$
or, since $\theta\neq 0,\pi$,
\begin{equation} \label{10.17}
\abs{\alpha_\ell}^2 = 1 + \abs{\beta_\ell}^2
\end{equation}
from which \eqref{10.16} is obvious. 
\enddemo

 {\it Remark.} In terms of the transmission and reflection coefficients of scattering 
theory \cite{Tes}, $\alpha_\ell =1/t$, $\beta_\ell = r/t$, \eqref{10.17} is $\abs{r}^2 + \abs{t}^2 
=1$ and \eqref{10.16} is $\abs{t}\leq 1$. 
\vglue6pt
We are now ready for

\demo{Proof of Theorem~{\rm \ref{T10.1}}} Let $W^{(n)}$ be given by \eqref{10.1a}. Then, 
by \eqref{10.2} and $C_0$ \eqref{10.13},
\begin{equation} \label{10.18}
\lim_{n\to\infty}\, \f{1}{2\pi} \int_0^{2\pi} \log \abs{L(e^{i\theta}; W^{(n)})}\, 
d\theta =0.
\end{equation}
Since $a_n\equiv 1$, \eqref{10.15} implies $\log \abs{L(e^{i\theta}; W^{(n)})} \geq 0$, 
and so \eqref{10.18} implies 
\begin{equation} \label{10.19}
\lim_{n\to\infty} \, \f{1}{2\pi} \int_0^{2\pi} \cos(2\theta) \log \abs{L(e^{i\theta}; 
W^{(n)})}\, d\theta =0.
\end{equation}
By \eqref{10.7}, $a_m\equiv 1$, $C_2$, and \eqref{10.2}, we see 
$$
\lim_{n\to\infty}\, \sum_{\abs{j}<n} b_j^2 =0
$$
which implies $b\equiv 0$. 
\enddemo

Finally, a remark on why this result holds that could provide a second proof (without sum 
rules) if one worked out some messy details. Here is a part of the idea:

\proclaim{Proposition} \lb{P10.10} Let $\{b_n\}$ be a bounded sequence and $W$ the associated 
whole\/{\rm -}\/line Schr{\rm \"{\it o}}dinger operator {\rm{(}}\/with $a_n\equiv 1${\rm{)}}. Let 
\begin{equation} \label{10.20}
A(\alpha) =\sum b_n e^{-\alpha\abs{n}}.
\end{equation}
If 
\begin{equation} \label{10.21}
\limsup_{\alpha\downarrow 0}\, A(\alpha) > 0,
\end{equation}
$W$ has spectrum in $(2,\infty)${\rm ,} and if 
\begin{equation} \label{10.22}
\liminf_{\alpha\downarrow 0}\, A(\alpha) <0,
\end{equation}
then $W$ has spectrum in $(-\infty, -2)$. 
\endproclaim

\demo{Proof} Let \eqref{10.21} hold and set $\varphi_\alpha (n) =e^{-\alpha\abs{n}/2}$. 
Then
$$
(W_0\varphi_\alpha)(n) = \left\{ \begin{array}{ll} 
2\cosh (\f{\alpha}{2}) &\hbox{if } n\neq 0 \\
{[} 2\cosh (\f{\alpha}{2})-2\sinh (\f{\alpha}{2}){]}\varphi_\alpha(n) &\hbox{if } n=0.
\end{array}\right. 
$$
It follows that 
\begin{eqnarray} 
&& \lb{10.23}\\
\noalign{\vskip-8pt}
(\varphi_\alpha, W\varphi_\alpha)(n) & \nhs=\nhs &2\cosh \biggl( \f{\alpha}{2}\biggr) 
\|\varphi_\alpha\|^2 +A(\alpha) - 2\sinh \biggl( \f{\alpha}{2}\biggr) \no  \\ 
&\nhs=\nhs& 2 \|\varphi_\alpha \|^2 + 2 \biggl[ \cosh \biggl( \f{\alpha}{2}\biggr) -1 \biggr] 
 \|\varphi_\alpha\|^2 + A(\alpha) - 2\sinh \biggl(\f{\alpha}{2}\biggr) .\no
\end{eqnarray}
Now, $\sinh(\alpha/2)\to 0$ as $\alpha\downarrow 0$ and since $\|\varphi_\alpha\|^2 
= O(\alpha^{-1})$ and $\cosh (\alpha/2)-1 =O(\alpha^2)$, $2[\cosh (\alpha/2) -1] 
\|\varphi_\alpha\|^2 \to 0$ as $\alpha\downarrow 0$. If there is a sequence with $\lim A
(\alpha_n) >0$, for $n$ large, $(\varphi_{\alpha_n}, (W-2)\varphi_{\alpha_n})>0$ which 
implies there is spectrum in $(2,\infty)$.

If \eqref{10.22} holds, use $\varphi_\alpha (n) =(-1)^n e^{-\alpha\abs{n}/2}$ and a 
similar calculation to deduce $(\varphi_{\alpha_n}, (W+2)\varphi_{\alpha_n})<0$. 
\enddemo

This proof is essentially a variant of the weak coupling theory of Simon \cite{S70}. Those 
ideas immediately show that if 
\begin{equation} \label{10.23a}
\sum n\abs{b_n} <\infty
\end{equation}
and $\sum b_n =0$ (so Proposition\eqref{P10.10} does not apply), then $W$ has eigenvalues 
in {\it both} $(2,\infty)$ and $(-\infty, 2)$ unless $b\equiv 0$. This reproves 
Theorem~\ref{T10.1} when \eqref{10.23} holds by providing explicit eigenvalues outside 
$[-2,2]$. It is likely using these ideas as extended in \cite{S80}, \cite{Kl}, one can provide an 
alternate proof of Theorem~\ref{T10.1}. In any event, the result is illuminated.


\vglue-18pt 
{\it Added Notes}. During the refereeing process, several results were obtained
which relate to this paper.  In connection with Theorem~10.1, D.~Damanik,
D.~Hundertmark, R.~Killip, and B.~Simon (to appear) have proved that if
the essential spectrum of a whole- or half-line Schr\"odinger operator is
contained in $[-2,2]$, then it is a compact perturbation of the free operator.  
B.~Simon and A.~Zlato\v{s} (to appear) have studied when the $C_0$ sum rule holds, have
simplified the proofs of Theorems~7.1 and 9.14, and have extended Theorem~$4'$
to the case where one assumes (\ref{1.12}) rather than that there is no 
discrete spectrum.

\vglue6pt
\centerline{\ninerm (Received December 13, 2001)}
\end{document}